\documentclass[%
 aip,
 amsmath,amssymb,
 reprint,%
]{revtex4-1}

\usepackage{graphicx}% Include figure files
\usepackage{dcolumn}% Align table columns on decimal point
\usepackage{bm}% bold math
\usepackage[utf8]{inputenc}
\usepackage[T1]{fontenc}
\usepackage{mathptmx}
\usepackage{etoolbox}

\usepackage{listings}

\makeatletter
\def\@email#1#2{%
 \endgroup
 \patchcmd{\titleblock@produce}
  {\frontmatter@RRAPformat}
  {\frontmatter@RRAPformat{\produce@RRAP{*#1\href{mailto:#2}{#2}}}\frontmatter@RRAPformat}
  {}{}
}%
\makeatother
\begin{document}

\preprint{AIP/123-QED}

\title[Black-box optimization using factorization and Ising machines]{Black-box optimization using factorization and Ising machines}
% Force line breaks with \\

\author{Ryo Tamura}
\email{tamura.ryo@nims.go.jp}
\altaffiliation[Also at ]{Graduate School of Frontier Sciences, The University of Tokyo, 5-1-5 Kashiwanoha, Kashiwa 2778561, Japan}
\altaffiliation[Also at ]{RIKEN Center for Advanced Intelligence Project, 1-4-1 Nihonbashi, Chuo-ku, Tokyo 103-0027, Japan}
\affiliation{Center for Basic Research on Materials, National Institute for Materials Science, 1-1 Namiki, Tsukuba, Ibaraki 305-0044, Japan}

\author{Yuya Seki}
\affiliation{Graduate School of Science and Technology, Keio University, 3-14-1 Hiyoshi, Kouhoku-ku, Yokohama, Kanagawa 223-8522, Japan}

\author{Yuki Minamoto}
\affiliation{Fixstars Amplify Corporation, 3-1-1 Shibaura, Minato-ku, Tokyo 108-0023, Japan}

\author{Koki Kitai}
\affiliation{Unprecedented-scale Data Analytics Center, Tohoku University, 6-3 Aoba, Aramakiaza, Aoba-ku, Sendai, Miyagi 980-8578, Japan}

\author{Yoshiki Matsuda}
\altaffiliation[Also at ]{Fixstars Corporation, 3-1-1 Shibaura, Minato-ku, Tokyo 108-0023, Japan}
\affiliation{Fixstars Amplify Corporation, 3-1-1 Shibaura, Minato-ku, Tokyo 108-0023, Japan}

\author{Shu Tanaka}
\email{shu.tanaka@keio.jp}
\altaffiliation[Also at ]{Department of Applied Physics and Physico-Informatics, Keio University, 3-14-1 Hiyoshi, Kohoku-ku, Yokohama-shi, Kanagawa 223-8522, Japan}
\altaffiliation[Also at ]{Keio University Sustainable Quantum Artificial Intelligence Center (KSQAIC), Keio University, Tokyo 108-8345, Japan}
\altaffiliation[Also at ]{Human Biology-Microbiome-Quantum Research Center (WPI-Bio2Q), Keio University, Tokyo 108-8345, Japan}
\affiliation{Graduate School of Science and Technology, Keio University, 3-14-1 Hiyoshi, Kouhoku-ku, Yokohama, Kanagawa 223-8522, Japan}

\author{Koji Tsuda}
\email{tsuda@k.u-tokyo.ac.jp}
\altaffiliation[Also at ]{Center for Basic Research on Materials, National Institute for Materials Science, 1-1 Namiki, Tsukuba, Ibaraki 305-0044, Japan}
\altaffiliation[Also at ]{RIKEN Center for Advanced Intelligence Project, 1-4-1 Nihonbashi, Chuo-ku, Tokyo 103-0027, Japan}
\affiliation{Graduate School of Frontier Sciences, The University of Tokyo, 5-1-5 Kashiwanoha, Kashiwa 277-8561, Japan}

\date{\today}% It is always \today, today,

\begin{abstract}
Black-box optimization (BBO) is used in materials design, drug discovery, and hyperparameter tuning in machine learning. 
The world is experiencing several of these problems.
In this review, a factorization machine with quantum annealing or with quadratic-optimization annealing (FMQA) algorithm to realize fast computations of BBO using Ising machines (IMs) is discussed.
The FMQA algorithm uses a factorization machine (FM) as a surrogate model for BBO.
The FM model can be directly transformed into a quadratic unconstrained binary optimization model that can be solved using IMs.
This makes it possible to optimize the acquisition function in BBO, which is a difficult task using conventional methods without IMs.
Consequently, it has the advantage of handling large BBO problems.
To be able to perform BBO with the FMQA algorithm immediately,
we introduce the FMQA algorithm along with Python packages to run it.
In addition, we review examples of applications of the FMQA algorithm in various fields, including physics, chemistry, materials science, and social sciences.
These successful examples include binary and integer optimization problems, as well as more general optimization problems involving graphs, networks, and strings, using a binary variational autoencoder.
We believe that BBO using the FMQA algorithm will become a key technology in IMs including quantum annealers.
\end{abstract}

\maketitle

\tableofcontents

\section{\label{sec:level1} Introduction}

Black-box optimization (BBO) methods\cite{10.1145/3097983.3098043,ALARIE2021100011,Terayama:2021aa,Kumagai2023} can provide better solutions to black-box functions\cite{Belevitch1962SummaryOT,10.7551/mitpress/11810.001.0001}, 
where the output value is determined only from the input value,
without knowing the functional form of the function.
Examples of such optimization problems include the optimization of experimental conditions, simulation parameters, and hyperparameters in the machine learning (ML) models, which appear everywhere in the basic and social sciences.
As a BBO method, the most famous one is the Bayesian optimization\cite{frazier2018tutorialbayesianoptimization,NIPS2012_05311655,7352306,10.1145/3582078}, which has solved several optimization problems in physics\cite{PhysRevLett.115.205901,PhysRevX.7.021024,PhysRevLett.126.104801,Yu:2020aa}, chemistry\cite{Hase:2018aa,Homma:2020aa,Fang:2021aa,WANG2022100728}, and materials science\cite{10.1063/1.5123019,Zhang:2020aa,TAMURA2021109290,Liang:2021aa}.

In BBO methods, ML surrogate models are prepared from training data, where both the input and output are already known by black-box functions,
such that the output value can be correctly predicted from any inputs using the surrogate models.
By optimizing the surrogate model, 
the best solution to the surrogate model is obtained.
If the surrogate model has a better prediction accuracy, then the best solution to the surrogate model will also be a better solution to the black-box function.
This is the idea behind BBO.

Let us consider the optimization of the surrogate models.
When the input is a continuous variable, continuous optimization methods, 
such as gradient descent and Newton's method, are used to optimize the surrogate models.
However,
obtaining globally optimal solutions is difficult, because these algorithms tend to be trapped in local minima.
In addition, when the input variables are discrete,
the optimization of surrogate models is a combinatorial optimization problem,
and determining the best solution is significantly more difficult.
Of course,
because the computation time required to obtain the output value from the input using the surrogate models is quite short,
exhaustive search is useful and an optimal solution is obtained,
when the number of candidate inputs is small.
However, combinatorial optimization problems are NP-hard\cite{10.3389/fphy.2014.00005},
and as the number of candidate inputs increases, 
the optimization of a surrogate model using an exhaustive search is impossible.
Consequently, effectively searching for better surrogate model solutions is difficult when huge number of candidate inputs are present.

From another perspective,
the world is full of combinatorial optimization problems such as scheduling, matching, and route search problems\cite{Combinatorial_book}.
To address these problems, Ising machines (IMs)\cite{Mohseni:2022aa} have been developed that can effectively explore the ground state of the Ising models\cite{10.5555/3159044,doi:10.7566/JPSJ.88.061010,Yarkoni_2022}.
The Ising model was introduced to address critical phenomena in statistical physics\cite{RevModPhys.47.773,PELISSETTO2002549}.
Nearly all combinatorial optimization problems can be expressed using the Ising model.
The ground state of the Ising model corresponds to the optimal solution of the combinatorial optimization problem.
The Hamiltonian of Ising model is defined as follows:
\begin{eqnarray}
H_{\rm Ising} = \sum_{i = 1}^N h_i s_i + \sum_{1 \le i < j \le N } J_{ij} s_i s_j, 
\label{eq:E_Ising}
\end{eqnarray}
where $s_i = \pm 1$ denotes the classical spin variables and $N$ denotes the number of variables.
Parameters $h_i$ and $J_{ij}$ can have real values known as the magnetic field and magnetic interaction, respectively.
Because the number of states is $2^N$, determining the optimal state with the minimum energy of the Ising model is difficult when $N$ increases.
This model can be individually translated into
\begin{eqnarray}
H_{\rm QUBO} = \sum_{1 \le i \le j \le N } Q_{ij} x_i x_j,
\label{eq:E_QUBO}
\end{eqnarray}
where $x_i$ is a 0/1 binary variable, and this type of model is known as quadratic unconstrained binary optimization (QUBO)\cite{Kochenberger:2014aa}.
To effectively obtain the minimum energy states in Eqs.~(\ref{eq:E_Ising}) and (\ref{eq:E_QUBO}), that is, the optimal solution to the combinatorial optimization problem,
IMs implement classical annealing\cite{7350099,10.1007/978-3-319-61566-0_39,Goto:2016aa,doi:10.1126/sciadv.aav2372}, quantum annealing\cite{Johnson:2011aa,Easttom2024}, or dynamic evolution\cite{doi:10.1126/science.aah4243,doi:10.1126/sciadv.abh0952} as hardware.
In addition, an algorithm for solving QUBO by gated quantum computing, known as quantum approximate optimization algorithm (QAOA), has also been developed.\cite{QAOA}

\begin{figure}
  \begin{center}
    \includegraphics[scale = 0.7]{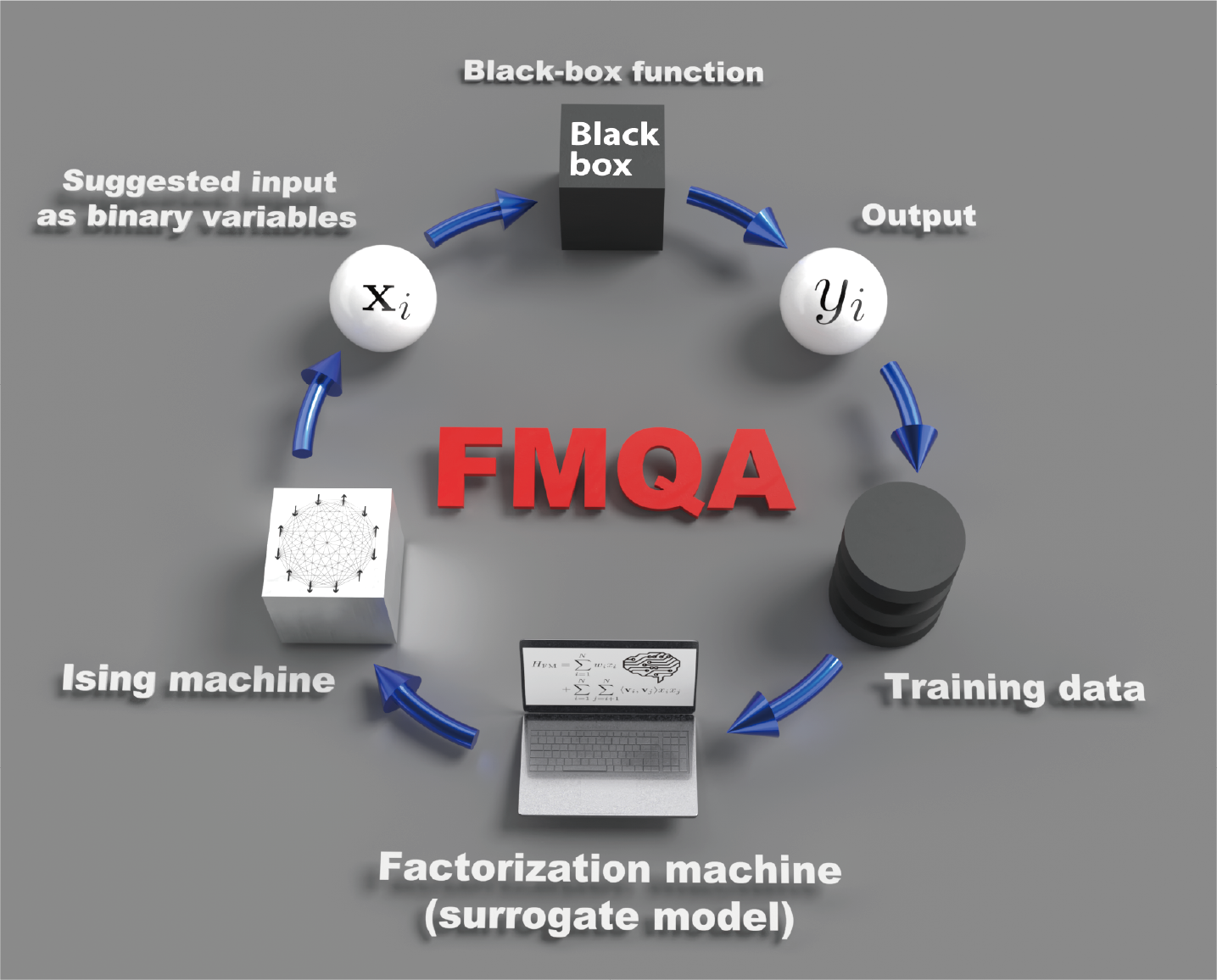}
  \end{center}
  \caption{
  Flow of the FMQA algorithm that performs BBO using FM and IMs.
  } 
  \label{fig:flow}
\end{figure}

The use of IMs can solve the difficulties of BBO using discrete variables\cite{PhysRevResearch.2.013319}.
In other words, the IM was used to solve the combinatorial optimization problem based on surrogate model optimization with discrete variables, 
and better inputs for black-box functions can be identified at high speeds.
The attempt to use IMs for discrete BBO is roughly categorized according to the surrogate model used: factorization machine (FM)\cite{5694074}, Bayesian optimization of combinatorial structures (BOCS)\cite{baptista2018bayesianoptimizationcombinatorialstructures,doi:10.7566/JPSJ.90.064001,10113307,doi:10.7566/JPSJ.92.123801}, and the Ising model Hamiltonian itself\cite{https://doi.org/10.1002/aisy.202000209}.
Because these surrogate models can be directly converted to QUBO defined by Eq.~(\ref{eq:E_QUBO}), 
better solutions to the surrogate models were instantly obtained by using IMs.
This review focuses on the case using an FM that has been used in several applications.
For the first time,
the use of an FM for BBO was proposed by Kitai et al.\cite{PhysRevResearch.2.013319},
and a D-Wave quantum annealer was used to solve the FM model.
The proposed algorithm is known as FMQA (factorization machine with quantum annealing).
In this review, the BBO algorithm using an FM is collectively referred to as FMQA (factorization machine with quadratic-optimization annealing), 
if the IMs do not use quantum technologies.
This process is summarized in Fig.~\ref{fig:flow}.

\begin{figure}
  \begin{center}
    \includegraphics[scale=0.3]{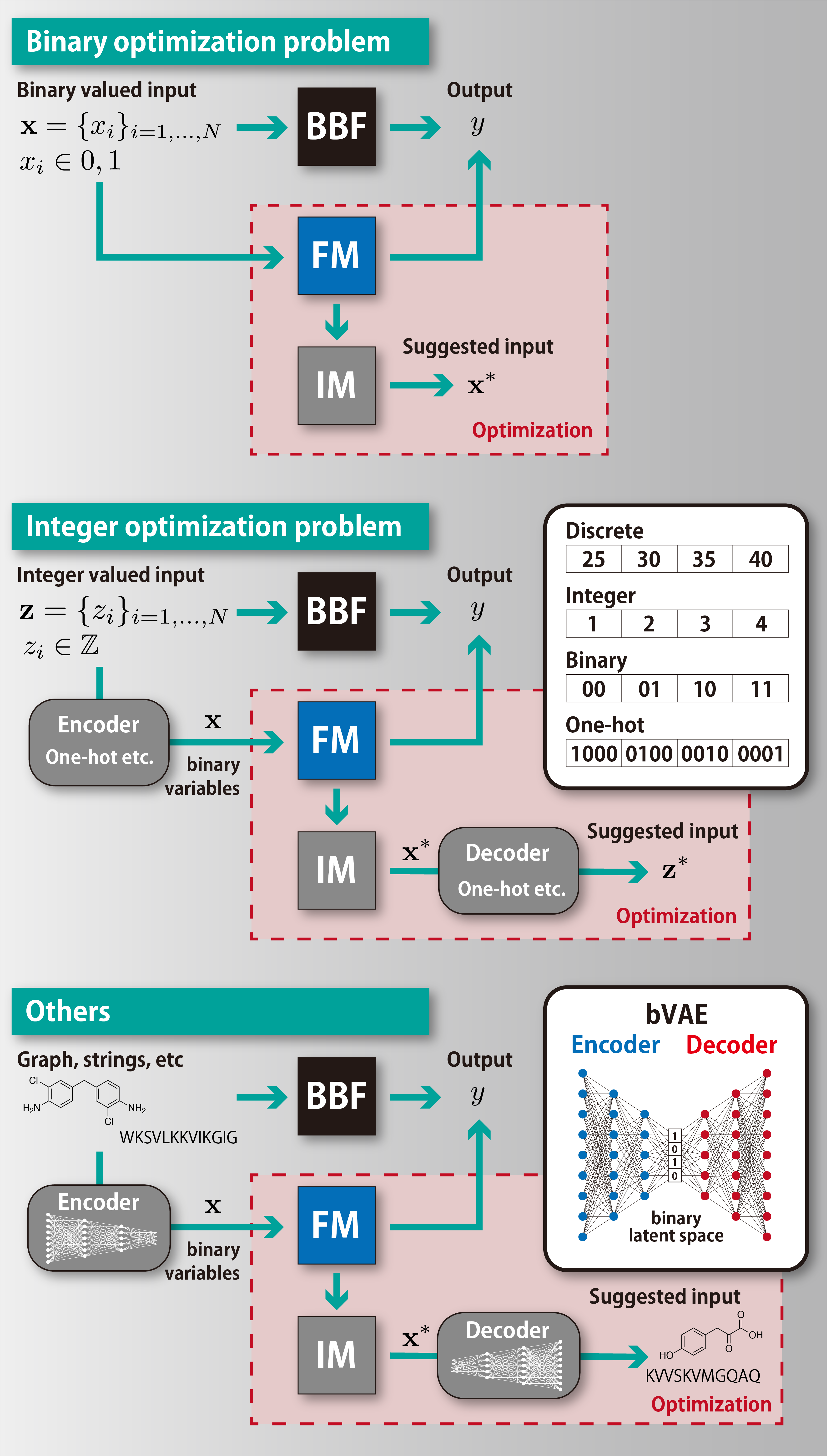}
  \end{center}
  \caption{
Three types of optimization problems depending on the type of input to the black-box function.
When the input is binary variables, 
preparing the encoder and decoder is not necessary.
For integer optimization problems,
integer variables should be converted into binary variables using an encoder.
To perform optimization problems with variables other than binary or integer values, a bVAE is useful for the encoder and decoder.
  } 
  \label{fig:types}
\end{figure}

In Section~\ref{sec:methods}, the details of the FMQA algorithm are described, along with the elemental techniques used, that is, FM and IM.
In Section~\ref{sec:software},
we present several open-source implementations of the FMQA algorithm
In Section~\ref{sec:applications}, application studies of FMQA for the basic and social sciences and engineering are introduced.
For instance, materials designs using computational materials science methods have been actively performed\cite{PhysRevResearch.2.013319,Kim:2022aa,PhysRevApplied.20.024044}.
Furthermore, the crystal structure prediction was considered using FMQA\cite{QALO,couzinie2024machinelearningsupportedannealing,Lin}.
In addition, peptide design in biotechnology\cite{Tucs:2023aa}, optimization of resonance avoidance in engineering\cite{Kadowaki:2022aa}, and traffic signal control in social science\cite{fixtraffic} have also been reported.
These application examples can be classified into three types depending on the type of input to the black-box function, as follows (see Fig.~\ref{fig:types}).

\begin{itemize}

\item[] {\bf Binary optimization problem}: 
For this category,
the input is given by binary variables.
The FM can be used as a surrogate model for the black-box function, 
and the obtained solution can be used as an input to the black-box function as it is.

\item[] {\bf Integer optimization problem}: 
For this category,
the input is given by integer variables.
By converting a single integer variable into multiple binary variables using an encoder, the FM can be trained using binary variables as inputs.
The optimal solution to an FM using binary variables should be decoded into integer variables that are input to the black-box function.
To ensure that the solution with binary variables of an FM can be correctly decoded into integer variables for the black-box function, 
it is necessary to search for the better solutions to the FM under appropriate constraints by IMs.
The continuous input can be treated as an integer optimization problem, if the variables are discretized.

\item[] {\bf Others}:
For this category,
the input for the black-box function is not binary and integer variables such as graphs, networks, and strings.
The binary variational autoencoder (bVAE)\cite{10.1007/978-3-030-33904-3_12} is useful to convert these inputs into binary variables.
Using the converted binary variables, FM is trained, and the optimal solution to FM is selected by IMs.
The solution with binary variables is converted back to inputs for the black-box function using the decoder.

\end{itemize}
Section~\ref{sec:outlook} discusses the future prospects of BBO methods using IMs.

\section{FMQA Algorithm} \label{sec:methods}

In this section, the strategy for combining FM and IMs to effectively solve BBO problems is presented.
First, the general problem of BBO is defined.
Thereafter, the FM, which is used as a surrogate model, is introduced as an elementary technique, and the FMQA algorithm is described.
Subsequently, encoding methods from integer variables and others to binary variables are introduced to solve the general optimization problem using the FMQA algorithm.
In addition, some IMs that are solvers of the surrogate model in the FMQA algorithm are introduced.
Finally, the implementation of FMQA is explained.

\subsection{Black-box optimization}

BBO is a method for optimizing black-box functions.
The input to the black-box function can be any vector of integer and real values, graphs, networks, or strings, and the input is expressed by $\mathbf{x}$.
A black-box function returns only the function value $y (\mathbf{x})$ when $\mathbf{x}$ is inputted.
In other words, it is a function for which no information useful for optimization computations, such as function forms and gradients, can be obtained.
Simulations and experimental results in basic science are generally considered to be black-box functions.
The prediction accuracy of neural networks is also a black-box function, where the input is hyperparameters.
Owing to the lack of useful information for optimization,
BBO uses a surrogate model that approximates the black-box function.
As the surrogate model only needs to predict the output from the inputs, 
various ML models can be used.
The BBO procedure is summarized below.

\begin{itemize}

\item[Step 1] For some inputs, the outputs of the black-box functions are prepared as initial training data.
If the number of initial data is $M$, the training data are expressed as $D = \{ ({\bf x}_m, y_m) \}_{m=1,...,M}$.

\item[Step 2] Using $D$, the surrogate model is trained to correctly predict the output value from the inputs.

\item[Step 3] The optimal input of the surrogate model is explored. 
In this step, the acquisition function defined by the surrogate model is often used instead of surrogate model.

\item[Step 4] For the selected input, the output value is obtained via a black-box function.
Then, the number of training data is increased as $D = \{ ({\bf x}_m, y_m) \}_{m=1,...,M+1}$.

\item[Step 5] Repeats from Steps 2 to 4.

\end{itemize}

The most well-known BBO method is Bayesian optimization, which uses Gaussian process regression as a surrogate model.
Because Gaussian process regression is a stochastic model, the predictions and their variances, which correspond to the uncertainty of the predictions, can be evaluated.
By selecting candidates with better predictions, better candidates for the black-box function are obtained; this is known as ``exploitation.''
In contrast, 
by selecting candidates with a large variance,
candidates far from the training data are selected.
This is known as ``exploration.''
In Bayesian optimization, an acquisition function is introduced to simultaneously perform exploitation and exploration with a single function using both the predictions and variances obtained from the Gaussian process.
Rather than optimizing the surrogate model in Step 3, Bayesian optimization selects the input that maximizes the acquisition function as the next input for the black-box function.
Python packages can be used to perform Bayesian optimization, such as Optuna\cite{optuna_2019}, GpyOpt\cite{HPyOpt}, Summit\cite{https://doi.org/10.1002/cmtd.202000051},
BoTorch\cite{balandat2020botorchframeworkefficientmontecarlo,optuna_2019},
and PHYSBO\cite{UENO201618,MOTOYAMA2022108405}.
However, as explained in the introduction, 
optimizing the acquisition function is a time-consuming task.

\subsection{Factorization machine (FM)}

In the FMQA algorithm, the FM is used as a surrogate model.
The ML prediction model up to the quadratic term is the FM proposed by Rendle\cite{rendle:tist2012}.
The FM is defined as
\begin{eqnarray}
\bar{y} (\mathbf{x}) = w_0 + \sum_{i=1}^N w_i x_i + \sum_{i=1}^N \sum_{j=i+1}^N \langle \mathbf{v}_i, \mathbf{v}_j \rangle x_i x_j := H_{\rm FM}, \label{eq:FM}
\end{eqnarray}
where $\langle \mathbf{v}_i, \mathbf{v}_j \rangle$ is defined as
\begin{eqnarray}
\langle \mathbf{v}_i, \mathbf{v}_j \rangle = \sum_{k=1}^K v_{ik} v_{jk}.
\end{eqnarray}
Here, $w_0$, $\{ w_i \}_{i=1,...,N}$ and $\{ v_{ik} \}_{i=1,...,N, k=1,...,K}$ are the model parameters,
and are determined such that $\bar{y} (\mathbf{x})$ can correctly predict $y$ by $\mathbf{x}$.
In this model, only a single hyperparameter $K$ determines the dimensionality of the model parameters.
In other words, the number of model parameters is $NK + N + 1$.
An important aspect of the FM is that the computation time of the third term in Eq.~(\ref{eq:FM}) exhibits linear complexity.
In addition, the training of the model parameters can be easily performed using gradient descent methods with linear complexity.
For instance, the root mean square loss is considered, which is defined as follows:
\begin{eqnarray}
L = \sum_{m=1,...,M}{\left( \bar{y} (\mathbf{x}_m) - y_m \right)^2},
\end{eqnarray}
when the training data are $D = \{ ({\bf x}_i, y_i) \}_{i=1,...,M}$.
To minimize $L$, the gradients of $y (\mathbf{x}_i)$, which depend on the model parameters, are necessary,
and can be easily evaluated as follows:
\begin{eqnarray}
\frac{\partial L}{\partial w_0} &=& 1, \\
\frac{\partial L}{\partial w_i} &=& x_i, \\
\frac{\partial L}{\partial v_{ik}} &=& x_i \sum_{j=1}^N v_{jk} - v_{ik} x_i^2,
\end{eqnarray}
The computational order for these gradient calculations is $\mathcal{O} (1)$.
Thus, all the parameter updates can be performed using $\mathcal{O} (NK)$,
and fast training of the FM is guaranteed.
The computation time increases linearly with the amount of training data $M$.
Some packages are available for training FMs, such as libFM\cite{rendle:tist2012}, fastFM\cite{JMLR:v17:15-355}, PyTorch\cite{NEURIPS2019_9015}, and FMQA\cite{FMQA_git}.

Considering the binary variables for $x_i = \pm 1$, the model in Eq.~(\ref{eq:FM}) can be easily converted into QUBO.
Because the constant term has no meaning for QUBO, except for $w_0$, the following relationships are established between the FM and QUBO:
\begin{eqnarray}
Q_{ij} = \begin{cases} w_i & (i=j) \\  \langle \mathbf{v}_i, \mathbf{v}_j \rangle & (i \neq j) \end{cases},
\end{eqnarray}
where $x_i^2 = x_i$.
Subsequently, the FM can be solved effectively using IMs.

\subsection{FMQA algorithm for binary optimization problem}

Using the FM as a surrogate model for BBO, an IM can be used to quickly search for the optimal input for the surrogate model.
This is known as the FMQA algorithm, and the procedure when the input is binary is as follows (see also Fig.~\ref{fig:flow}).

\begin{itemize}

\item[Step 1] For some inputs that are expressed by binary variables as ${\bf x} = (x_1, ..., x_N)$, the outputs of black-box functions are prepared as initial training data.
When the number of initial data is $M$, the training data are expressed as $D = \{ ({\bf x}_m, y_m) \}_{m=1,...,M}$.

\item[Step 2] Using $D$, the FM is trained to correctly predict the output value from the inputs.

\item[Step 3] The optimal input of FM defined by Eq.~(\ref{eq:FM}) is explored by the IM. 

\item[Step 4] For the selected input, the output value is obtained via a black-box function.
Then, the number of training data is increased as $D = \{ ({\bf x}_i, y_i) \}_{i=1,...,M+1}$.

\item[Step 5] Repeats from Steps 2 to 4.

\end{itemize}

In Step 3, the input already included in $D$ may become optimal.
In this case, it is necessary to select a solution that is not included in $D$ by randomly generating, slightly changing the optimal input, or adjusting the parameters of the IM.
In Ref.\cite{PhysRevResearch.2.013319}, which proposes the FMQA algorithm, the case of random generation is suggested.
This will work roughly as ``exploitation'' in Bayesian optimization.

\subsection{Encoding method to binary variables}

The FMQA algorithm is a specialized optimization method for binary variables.
However, optimization problems expressed in terms of integers and other variables can also be treated by converting them into binary variables using an appropriate encoding method.
This section describes encoding methods for integer variables and the use of a bVAE.

\subsubsection{Encoding integer-variables}

Here, we introduce three types of representative encoding methods, from integers to binary variables,
that is, binary, one-hot, and domain-wall.
In Ref.\cite{seki2022blackboxoptimizationintegervariableproblems}, the optimization performances using three encoding methods were compared for benchmark problems.
In addition,
other encoding methods such as the Gray code\cite{1362904} are also used.
Note that if continuous values are discretized, they can be treated as integer variables.
Thus, the encoding methods introduced here can also handle continuous values.

{\it Binary encoding}: A positive integer variable $z$ is expressed by the binary variables $\{ x_i \}_{i=1,...,N}$ where $x_i =$ 0 or 1, and $N$ denotes the length of the binary variables,
as follows
\begin{eqnarray}
z = 1 + x_1 + 2 x_2 + ... + 2^{(N -1)} x_N.
\end{eqnarray}
The integer variable $z$ ranges from 1 to $2^N$.
Depending on the range of the integer, the dimension of $N$ should be given.
If $L$ integer variables are required to express the optimization problem,
$L \times N$-dimensional binary variables are used as inputs for BBF.
Therefore, the FM can be expressed as
 \begin{eqnarray}
H_{\rm FM} = w_0 + \sum_{i=1}^{LN} w_i x_i + \sum_{i=1}^{LN} \sum_{j=i+1}^{LN} \langle \mathbf{v}_i, \mathbf{v}_j \rangle x_i x_j.
\end{eqnarray}
Thus, by solving $H_{\rm FM}$ using IMs as it is, the better solution to the integer optimization problem can be obtained.

{\it One-hot encoding}: In addition to the integer encoding, 
a positive integer variable $z$ is expressed by the binary variables $\{ x_i \}_{i=1,...,N}$ 
as follows:
\begin{eqnarray}
z = \sum_{i=1}^{N} i x_i.
\end{eqnarray}
under the following constraints:
\begin{eqnarray}
\sum_{i=1}^{N} x_i = 1. \label{one-hot_constraint}
\end{eqnarray}
The integer variable $z$ ranges from 1 to $N$.
For instance, when $N = 4$,
$\{1, 0, 0, 0\} = 1$, $\{0, 1, 0, 0\} = 2$, $\{0, 0, 1, 0\} = 3$, and $\{0, 0, 0, 1\} = 4$ (see also Fig.~\ref{fig:types}).
If $L$ integer variables are considered,
then $L \times N$-dimensional binary variables are prepared.
Thus, the FM can be prepared as follows:
\begin{eqnarray}
H_{\rm FM} = w_0 + \sum_{i=1}^{LN} w_i x_i + \sum_{i=1}^{LN} \sum_{j=i+1}^{LN} \langle \mathbf{v}_i, \mathbf{v}_j \rangle x_i x_j. \label{FM_integer}
\end{eqnarray}
In addition, we should consider the constraints given by Eq.~(\ref{one-hot_constraint}).
To prevent this constraint, the QUBO model solved by IMs is given by
\begin{eqnarray}
H_{\rm QUBO} = H_{\rm FM} + \alpha \sum_{l = 1}^{L} \left( \sum_{i = 1}^N x_{i + N (l - 1)} -1 \right)^2, \label{onehot_QUBO}
\end{eqnarray}
where $\alpha$ denotes a positive hyperparameter.
The second term becomes zero when all constraints are satisfied.
The first term decreases when the surrogate model defined by the FM has a smaller objective function value.
Thus, by solving $H_{\rm QUBO}$ using IMs, a better solution to the integer optimization problem is obtained.
For the one-hot encoding, Endo et al.~\cite{PhysRevResearch.7.013149} reported that when continuous values are considered, introducing function smoothing regularization during the training of the FM model leads to improved optimization performance.

{\it Domain-wall encoding}: Domain-wall encoding\cite{Chancellor_2019,doi:10.1098/rsta.2021.0410} resembles one-hot encoding,
and the integer value of $z$ can be expressed as follows:
\begin{eqnarray}
z = \sum_{i=1}^{N-1} i (x_{i} -2 x_{i} x_{i+1} + x_{i+1}) + N x_N + 1.
\end{eqnarray}
The integer variable $z$ ranges from 0 to $N$.
For instance, when $N = 3$,
$\{0, 0, 0\} = 1$, $\{1, 0, 0\} = 2$, $\{1, 1, 0\} = 3$, and $\{1, 1, 1\} = 4$ (see also Fig.~\ref{fig:types}).
The integer variable can be represented with one bit less than in the case of binary encoding.
If $L$ integer variables are considered,
then $L \times N$-dimensional binary variables are prepared.
In addition, we should consider the constraint, and the QUBO model is defined as
\begin{eqnarray}
H_{\rm QUBO} =&& H_{\rm FM} \notag \\
&&+ \alpha \sum_{l=1}^L \left( \sum_{i=2}^{N} x_{i + N(l-1)} - \sum_{i=1}^{N-1} x_{i+N(l-1)} x_{i+1+N(l-1)} \right) \notag \\
\end{eqnarray}
where $\alpha$ denotes a positive hyperparameter.
The second term becomes zero when the all constraints are satisfied.

\subsubsection{Binary variational autoencoder}

A bVAE with binary variables as the latent space is useful as an encoding method for variables other than binary and integer variables, such as graphs, networks, and strings to binary variables.
The structure of bVAE is shown in Fig.~\ref{fig:types}, and was constructed using two neural networks: an encoder and a decoder.
The bVAE learns to make the decoded variables equal to the input using numerous inputs without objective functions.
The key point is that the values of the objective functions are not required to train the bVAE.
Using the encoder part,
transforming inputs into binary variables is possible.
If the dimension of the latent space is $N$, the input variable can be expressed as $\{ x_i \}_{i=1,...,N}$ where $x_i =$ 0 or 1.
Using a dataset in which the objective functions are already known, the FM is trained as 
\begin{eqnarray}
H_{\rm FM} = w_0 + \sum_{i=1}^{N} w_i x_i + \sum_{i=1}^{N} \sum_{j=i+1}^{N} \langle \mathbf{v}_i, \mathbf{v}_j \rangle x_i x_j.
\end{eqnarray}
Using the IMs, the lower energy state of $H_{\rm FM}$ is obtained.
Using the decoder of the trained bVAE, the input for BBF can be generated from the binary variables obtained by IMs in the latent space.
Thus, by applying the FMQA algorithm in the latent space, the BBO can be performed.

\subsection{Ising machine}

The IMs connected to the FMQA algorithm include the D-Wave quantum annealer, the Fixstars Amplify Annealing Engine (Amplify AE), and the Fujitsu digital annealer (DA). 
The following sections describe the characteristics of the machines.

\subsubsection{D-Wave quantum annealer}

The D-Wave One was discovered in 2011 as the first commercial quantum computer\cite{Johnson:2011aa}. 
D-Wave One has 128 qubits, which can directly realize quantum annealing\cite{PhysRevE.58.5355,Santoro_2006} computations in the hardware and solve QUBO. 
A superconducting circuit was used in the D-Wave system in which the right and left rotations of the superconducting current acted as binary variables, representing 0 or 1, respectively. 
The superconducting current is indeterminate until it is observed owing to quantum superposition, and the quantum annealing computation is performed by gradually reducing the quantum effects. 
The dilution refrigerator produces a temperature of 12 mK, which is close to absolute 0 K, and the shielding layer blocks external signals such as external magnetic fields and vibrations. 
The D-Wave system operates at less than 25 kW of power, mostly for the cooling system and the front-end server. 

The D-Wave 2000Q, a machine with 2048 qubits arranged on a coarsely connected graph known as a chimera graph, was launched in 2017. 
Although this is a coarsely connected graph, a fully connected graph can be created by making several bits act as a single bit in a process known as embedding. 
The latest machine is the D-Wave Advantage\cite{D-Wave}, which has more than 5,000 qubits and can solve large problems. 
In addition, the advantage system and hybrid solver service can also be combined to run up to 1,000,000 binary variable problems for coarsely connected graphs, and up to 20,000 binary variable problems for fully connected graphs. In recent years, state-of-the-art techniques such as fast annealing computations\cite{King:2022aa,King:2023aa} have been implemented.

\subsubsection{Fixstars Amplify Annealing Engine}

The Amplify AE is a GPU-based IM developed and operated by Fixstars Amplify Corporation in Japan \cite{fixstarsamplify}. Amplify AE have been available to the public since 2021, and as of today (October 2024), they have processed more than 45 million user requests. The IM can solve QUBO using up to 131,072 (fully connected graphs) and 262,144 (sparse graphs) binary variables. Thus, several practical combinatorial optimization problems, including FMQA, are solvable in terms of the decision variable dimensions.

The power of the Amplify AE can be exploited most effectively when the users formulate their combinatorial optimization problems using the Amplify SDK development environment. The Amplify SDK realizes fast and easy formulation of optimization problems. It supports more than ten different solvers, including gate-based quantum computers with QAOA, quantum annealing machines, and IMs of other types, including the Amplify AE. The SDK also supports software-based optimization solvers, such as Gurobi. The SDK bridges the user's formulation and the selected optimization solver, regardless of the decision variables, polynomial order, constraints, and hardware specifications. For this, the Amplify SDK performs conversions such as variable conversion, order reduction, penalization of constraints, and graph embedding. Thus, FMQA is no longer limited to binary-variable BBO. It can also solve problems with integer and real variables and explicit constraints without additional implementation of variable encoding or constraint penalization in the resulting QUBO equation.

\subsubsection{Digital annealer}

The DA developed by Fujitsu Limited, is a machine that searches for the optimal solution to the Ising model through simulation in a digital circuit.
DA is implemented using complementary metal-oxide-semiconductor (CMOS) technology, and several hundred thousand ($\sim 100,000$) bits have already been realized with all couplings\cite{url-ditigal,9045100,10.3389/fphy.2019.00048}.
Specifically, the hardware minimizes the evaluation function while updating the bit states using a stochastic search based on the Markov chain Monte Carlo (MCMC) method. 
The basic operation is based on the simulated annealing (SA) method\cite{doi:10.1126/science.220.4598.671,Cerny:1985aa}.
An acceleration technique using digital circuit technology is applied.
Typical innovations for quickly obtaining the optimal solution in DA are as follows:

{\it Parallel bit manipulation:} A Monte Carlo simulation using the Metropolis method is effective for determining the optimal solution to the Ising model. With this method, the change in the total energy of the system is calculated when a bit is flipped, and the algorithm determines whether the bit should be flipped. DA attempts to flip all bits in parallel, which improves the speed of the optimization search.

{\it Rejection-free search:} All the spin states are generated from the transition probabilities calculated from the energy change of the flip operation for each bit. This improves the search speed because the state is always updated in a single operation without rejection.

{\it Replica exchange method:} A replica exchange method is used to reduce the risk of becoming trapped in a local minimum and not reaching a global minimum\cite{PhysRevLett.57.2607,doi:10.1143/JPSJ.65.1604}. This technique overcomes trapping at the local minimum by simultaneously running Monte Carlo calculations for different temperatures (replicas) and implementing an exchange of replicas between each temperature.

\section{Implementation and Tools}
\label{sec:software}

Several implementations of the FMQA algorithm are available as open source under an MIT license.
These points are discussed in this section.

\subsubsection{Original FMQA package}

The original implementation of the FMQA algorithm is available at \url{https://github.com/tsudalab/fmqa}.
To install the package, obtain fmqa-master.zip from the GitHub page and execute the following commands:
\begin{verbatim}
$ python setup.py install
\end{verbatim}

To train the FM, \verb|xs| and \verb|ys| are prepared as the training data.
The former is a matrix containing bit sequences of the training data. 
The latter is a list of the values of the objective function.
The training was performed using the following commands:
\begin{lstlisting}[language=python]
model = fmqa.FMBQM.from_data(xs, ys)
\end{lstlisting}
The training was performed using Adam\cite{1412.6980}.
The next candidate was obtained by solving this model using an IM. 
For instance, when solving by SA, the dimod package provided by D-Wave can be used, and the following is an example to obtain the next candidate.
\begin{lstlisting}[language=python]
import dimod
sa_sampler = dimod.samplers.SimulatedAnnealingSampler()
res = sa_sampler.sample(model)
\end{lstlisting}
\verb|res| contains the bit sequence as the next candidate.

The coefficients of the trained FM are obtained using the following commands:
\begin{lstlisting}[language=python]
model.linear[k] for k in model.linear
model.quadratic[(k, l)] for (k, l) in model.quadratic
\end{lstlisting}
The QUBO model was constructed using these coefficients. 
The QUBO model can be solved using various IMs, and the FMQA algorithm can be applied.

\subsubsection{FMQA implementation using the Amplify SDK and PyTorch}

Using the aforementioned Amplify SDK, one can implement FMQA using Python in a straightforward manner. Several such examples are available at \url{https://amplify.fixstars.com/en/demo#blackbox} for a range of BBO problems such as optimization of material design, chemical plant operation conditions, airfoil design and city traffic signal control. These example programs utilize PyTorch\cite{1912.01703} to train the FM models on par with the original FMQA package. The programs convert the trained FM model using a PyTorch class to QUBO formulation, which is then optimized using an IM such as Amplify AE. Additionally, this implementation allows the use of ``hard'' constraints such as one-hot and domain-wall, as well as other equality and inequality constraints. Therefore, non-binary variables are implemented in a straightforward manner, and such examples are included in the above example programs. In addition, the user can consider additional constraints of variables in FMQA using this functionality of the Amplify SDK, regardless of the target optimization solvers.

\subsubsection{Python library: Amplify-BBOpt}

Fixstars Amplify has been developing a Python library known as Amplify-BBOpt\cite{fixstarsamplifybbopt}, which facilitates QA-based BBO, such as FMQA. Amplify-BBOpt leverages Fixstars Amplify’s features explained in the previous section ``Fixstars Amplify Annealing Engine''. In addition, the above aforementioned encoding from non-binary to binary variables is included. Thus, the user can exploit the fast optimization of various IMs while considering non-binary decision variables without additional encoder/decoder implementation. 
The user can install the library as follows:

\begin{verbatim}
$ pip install amplify_bbopt
\end{verbatim}

A typical usage of the library for a black-box function comprising of ``fluid dynamics simulation'' of a flow around an airfoil is shown below. 
The optimization results and the problem detail are introduced in the Sec. 3 as ``Wing shape optimization''.
The simulator, \texttt{wing\_simulator}, takes ``width,'' ``height,'' and ``attack\ angle'' of an airfoil and computes and returns the drag and lift forces. The return value from the black-box function is the corresponding negative lift-to-drag ratio.

\begin{lstlisting}[language=python]
from amplify_bbopt import RealVariable, blackbox

@blackbox
def my_blackbox_func(
    wing_width: float = RealVariable(bounds=(1, 20), nbins=100),
    wing_height: float = RealVariable(bounds=(1, 5), nbins=20),
    wing_angle: float = RealVariable(bounds=(0, 45), nbins=20),
) -> float:
    lift, drag = wing_simulator(wing_width, wing_height, wing_angle)
    return -lift / drag  # value to minimize
\end{lstlisting}

The initial training data samples are generated based on the randomly sampled inputs. In the BBO context, the training dataset does not need to be large. Here, we start with only ten samples.

\begin{lstlisting}[language=python]
from amplify_bbopt import DatasetGenerator

num_init_data = 10
data = DatasetGenerator(objective=my_blackbox_func).generate(num_samples=num_init_data)
\end{lstlisting}

Before setting up the optimizer, the solver client must be specified for use during the annealing process. Similar to Amplify SDK, Amplify-BBOpt supports various commercially available solvers; howeber, here, we use Amplify AE.

\begin{lstlisting}[language=python]
from datetime import timedelta
from amplify import FixstarsClient

client = FixstarsClient()
client.parameters.timeout = timedelta(milliseconds=2000)  # annealing timeout for 2 seconds
# client.token = "xxxxxxxxxxx"  # Enter your Amplify AE API token.
\end{lstlisting}

Finally, we instantiate an optimizer known as \texttt{FMQAOptimizer} and start the optimization for five iterations.

\begin{lstlisting}[language=python]
from amplify_bbopt import FMQAOptimizer

optimizer = FMQAOptimizer(data=data, client=client, objective=my_blackbox_func)
optimizer.optimize(num_cycles=5)
\end{lstlisting}

As QA-based BBO methods, such FMQA are relatively new and are still under maturity. Therefore, Amplify-BBOpt plans to implement features along with the research and development of relevant studies.

%%%%%%%%%%%%%%%
% Applications
%%%%%%%%%%%%%%%

\section{Application Examples} \label{sec:applications}

%%%%%%%%%%%%%%%
%%%%%%%%%%%%%%%
%%%%%%%%%%%%%%%
\subsection{Binary optimization problems without constraint}

If all the inputs, that is, the binary variables $\mathbf{x}$, to the FM are the same as the inputs to the BBF,
the FM is minimized by IMs, and the optimal solution to the FM can be used as the input for the BBF.
In this simple case, there is no need to add constraints to solve the FM in FMQA algorithm.
This type of problem has been applied to the design of metamaterials\cite{PhysRevResearch.2.013319,Kim2024}, layered materials\cite{Kim:2022aa,Kim:2023aa,KIM2024101847,arXiv:2408.05799}, and nanoparticles\cite{https://doi.org/10.1002/adpr.202300226}.
In addition, it can be applied to the lossy compression of matrices\cite{Kadowaki:2022aa} and feature selection to obtain an ML model with better prediction accuracy\cite{doi:10.1080/14686996.2024.2388016}.
In the following sections, we present some applications of metamaterial design, which is the first application of FMQA, layered material design, and feature selection, as representative examples.

%%%%%%%%%%%%%%%
\subsubsection{Metamaterial design} 

Metamaterials exhibit characteristic properties when artificially and intricately combined with multiple materials.
Because the material properties are signifiantly dependent on the elements, compounds, geometry, and their arrangements,
BBO is effective when the structure of the metamaterial is the input and the properties obtained from simulations or experiments are the output.
However, 
if the types of elements and compounds increase, and a complex arrangement is considered,
a combinatorial explosion will occur in the candidate materials.
Thus, FMQA algorithm is useful for designing metamaterials.

\begin{figure*}
  \begin{center}
    \includegraphics[scale=0.6]{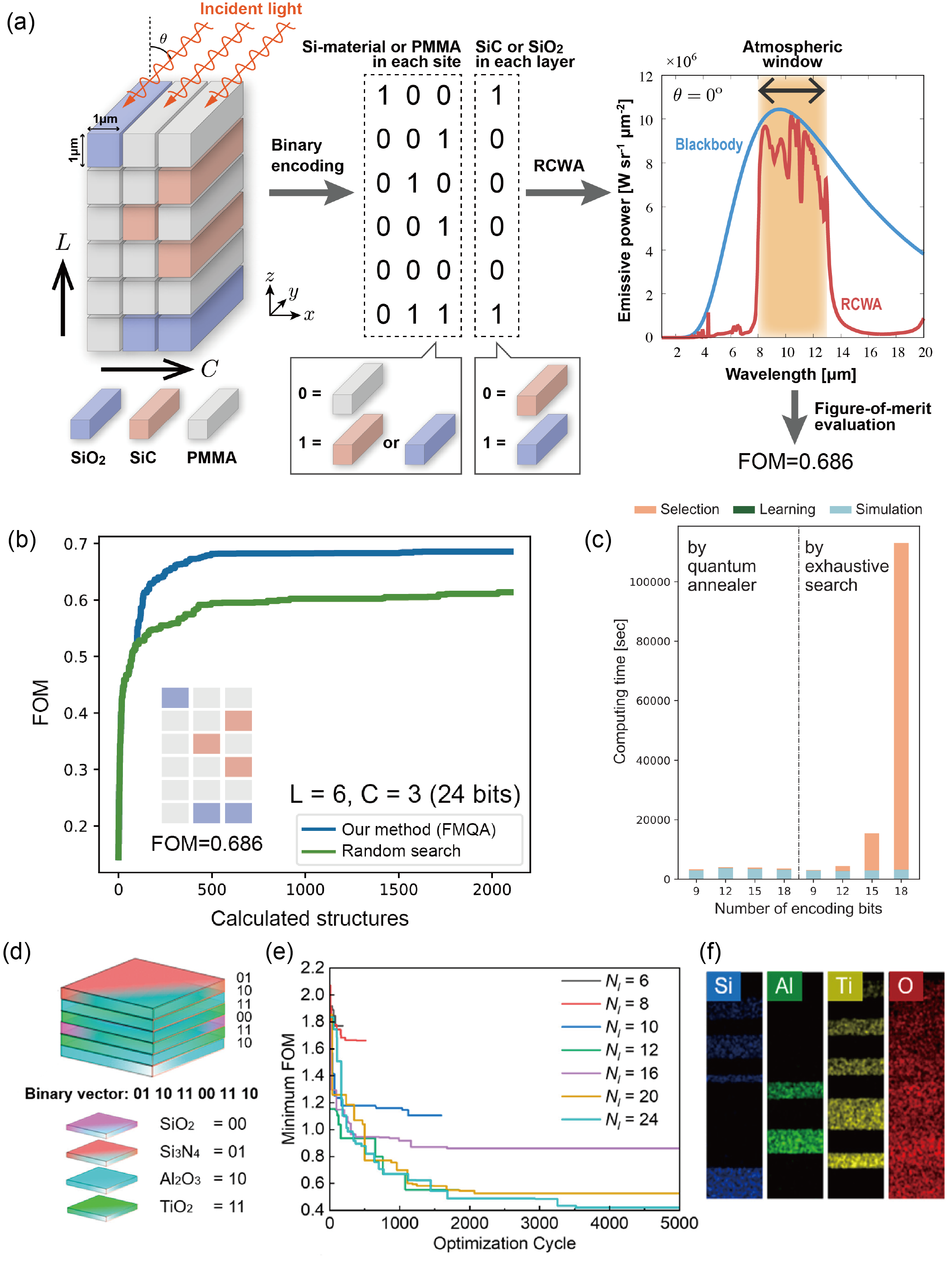}
  \end{center}
  \caption{
(a) Optimization target of metamaterial structure constructed by three types rod-like materials (SiO$_2$, SiC, and PMMA) to obtain better radiative cooling property.
This structure can be translated to binary variables, and emissive power was calculated by RCWA calculations.
The FOM can evaluate the shape of emissive power, and when all finite emissive power is located within the atmospheric window, FOM is maximized.
(b) FOM results depending on the number of calculated structures. The results by FMQA and RS were compared.
(c) Computation time of BBO with and without using the D-Wave 2000Q depending on the system size (number of encoding bits).
(d) Structure of layered materials expressed by binary variables.
(e) FOM results depending on the number of calculated structures depending on the number of layers $N_{\rm l}$ in the layered materials optimization.
(f) Energy-dispersive X-ray spectroscopy map for the fabricated layered materials.
(For (a), (b), and (c) reprinted from ref. \cite{PhysRevResearch.2.013319} under a Creative Commons Attribution 4.0 International license.
For (d), (e), and (f), adapted with permission from ref. \cite{Kim:2022aa}. Copyright 2022 American Chemical Society.)
  } 
  \label{fig:app_bin_cooling}
\end{figure*}

As an application of FMQA,
a metamaterial design for radiative cooling was reported by Kitai et al\cite{PhysRevResearch.2.013319}.
The arrangements of three types of rod-like compounds (SiO$_2$, SiC, and PMMA) were optimized to obtain better radiative cooling properties, as shown in Fig.~\ref{fig:app_bin_cooling} (a).
In this problem setting, three different materials are present, and it is impossible to represent them as binary variables.
However, by considering the constraint that each layer can contain only SiO$_2$ or SiC, it is possible to represent the arrangement of the rods as binary variables.
In other words, as shown in Fig.~\ref{fig:app_bin_cooling} (a), a binary variable representing Si compounds or PMMA for each position and another binary variable representing SiC = 0 and SiO$_2$ = 1 in each layer are prepared.
Let $C$ denote the number of compounds in the horizontal direction and $L$ denote the number of compounds in the vertical direction.
The arrangement can then be expressed as $(C+1) \times L$ bits.
When learning the FM, we flattened these bits into one dimension and used them as inputs.

The emissive power was evaluated using rigorous coupled-wave analysis (RCWA) calculations.
The difference between the calculated properties and atmospheric window was defined as the figure of merit (FOM).
This FOM was used as the output of BBF,
and a structure with a large value of FOM is required.
To maximize the properties in FMQA,
the output of BBF is defined as $y = - {\rm FOM}$ and the FMQA algorithm was applied.
In Fig.~\ref{fig:app_bin_cooling} (b), the results of the optimization processes are shown when $L=6$ and $C=3$, that is, 24 bits problem was considered.
As an IM, the D-Wave 2000Q was used,
and more than 2000 BBO cycles were performed.
For comparison, random search results are also shown.
If the FMQA algorithm is used, structures with large FOM values are found even if the number of calculated structures is smaller than that obtained by random sampling (RS).
In this study,
optimizations for larger systems with a maximum bit size of 60 have also been attempted. 
The metamaterials suitable for radiative cooling have also been developed.
In addition, the computation times of BBO with and without D-Wave 2000Q are summarized in Fig.~\ref{fig:app_bin_cooling} (c).
Without using IMs, the computation time required to select candidates increases rapidly depending on the problem size, whereas the search time remains almost constant as the problem size is increased when IM is used.
The results indicate that complex metamaterial structures can be optimized within a short time using the FMQA algorithm.

%%%%%%%%%%%%%%%
\subsubsection{Layered material design}

Materials exploration has been performed using FMQA, which focuses on layered materials\cite{Kim:2022aa}.
Layered materials were designed by layering multiple types of films (see Fig.~\ref{fig:app_bin_cooling} (d)).
Layered materials for high-performance transparent radiative cooling were designed by combining four materials: SiO$_2$, Si$_3$N$_4$, Al$_2$O$_3$, and TiO$_2$.
Each material can be represented by two bits,
that is,
the materials are defined as SiO$_2$ = 00, Si$_3$N$_4$ = 01, Al$_2$O$_3$ = 10, and TiO$_2$ = 11.
If the number of layers is $N_l$, then the one layered material can be represented by $2N_l$ bits.
Optimization calculations were performed up to the case $N_l = 24$; therefore, a 48-bit problem was treated at most.
When the materials to be used as layered materials and their arrangements are known, the wavelength-dependent optical properties are calculated using the transfer matrix method (TMM).
The thickness of each layer is fixed at 50 nm.
The difference between the calculated and desired optical properties was defined as FOM. 
This FOM was used as the output of BBF.
Using FMQA, the structures of layered materials with a small FOM were explored.
The results of the optimization processes are shown in Fig.~\ref{fig:app_bin_cooling} (e),
when $N_l=6, 8, 10, 12, 16, 20, 24$.
For $N_l=16, 20, 24$,
a total of 5000 iterations of FMQA optimization were performed.
D-Wave Advantage 4.1 was used as the IM.
For the $N_l = 24$ case,
the best structure with the smallest FOM was determined, and the layered material was fabricated using a thin-film deposition process.
In the 24 layers in the best structure, some of the neighboring layers are made of the same materials,
and a material with ten layers was proposed,
that is,
TiO$_2$ (100 nm) / SiO$_2$ (150 nm) / TiO$_2$ (100 nm) / SiO$_2$ (150 nm) / TiO$_2$ (100 nm) /  SiO$_2$ (50 nm) / Al$_2$O$_3$ (100 nm) / TiO$_2$ (200 nm) / Al$_2$O$_3$ (150 nm) / TiO$_2$ (100 nm).
In this material, Si$_3$N$_4$ was not used.
This material was fabricated, and its energy-dispersive X-ray spectroscopy map is shown in Fig.~\ref{fig:app_bin_cooling} (f),
resulting in a 10 layered material being correctly obtained.
It was confirmed using transmitted irradiance measurements that this material had the lowest FOM among the materials reported in the literature\cite{Kim:2022aa}.

The same technique is used to design a wide-angle spectral filter for energy saving windows.
In this case, the solar incident angle dependence of the optical properties was evaluated using the TMM,
and the averaged FOM for the angle dependence was used as the output of BBF.
The same four materials were used in addition to the high-performance transparent radiative cooling design, 
and the structural design using FMQA up to $N_l = 20$ was performed.
When $N_l = 20$, 8000 optimization cycles were performed,
and a world record material among other materials in the literature was also obtained\cite{KIM2024101847}.

%%%%%%%%%%%%%%%
\subsubsection{Polymer design} 

The polymer was designed using the FMQA algorithm\cite{Huang2024}. A total of 32 polymer motifs were prepared in advance and represented as 5-bit binary variables ($2^{15} = 32$), that is, the concept is similar to that in the case where the number of materials in a layered material is large. Three polymer motifs were selected to create triblock polymers (see Fig.~\ref{fig:app_bin_others} (a)) to optimize structures with high thermal conductivity (TC). Thus, the triblock polymer could be represented in 15 bits. The TC was obtained from a trained deep neural network (DNN) using Morgan fingerprints\cite{doi:10.1021/c160017a018} with frequency as a feature. A total of 1144 polymer datasets were used to train the DNN. In the QUBO, there is no need to consider the additional constraints, and only the FM model is included to predict the TC. A solver provided by D-Wave, dimod\cite{dimod}, was used to solve QUBO. The iterative dependence of the largest TC is shown in the Fig.~\ref{fig:app_bin_others} (b), and the TC gradually increases with the number of generations. However, FMQA did not exhibit better optimization performance than Bayesian optimization or the genetic algorithm (GA). Optimization performance can be improved by using IMs and improved encoding methods.

%%%%%%%%%%%%%%%
\subsubsection{Analog circuit design} 

An analog circuit was designed using FMQA algorithm\cite{HIDA2024}. The objective was to optimize the conductive structure of the filter used in the power amplifier, as shown in Fig.~\ref{fig:app_bin_others} (c). In a conductive structure with input and output ports, optimization is performed by targeting a specific frequency response. Two-dimensional (2D) square blocks are used to define the conductive structure. For each block, binary variables are defined such that if the conductor is present, it is 1; if the conductor is absent, it is 0. A $22 \times 22$ arrangement of blocks is considered with input and output ports on both sides. Given this structure, the finite element method (FEM) was used to simulate the frequency response of the conductive structure. The FOM is defined as the value that decreases when the desired frequency is reached, and optimization is performed to reduce it. In the QUBO for FMQA, there is no need to consider additional constraints, and only the FM model is considered. In this study, a DA was used as an Ising model solver. In addition, an optimization method that alternates between FMQA and GA is proposed to sample various structures. The iterative dependence of the FOM is illustrated in Fig.~\ref{fig:app_bin_others} (d). The results are compared with those of particle swarm optimization (PSO). The method combining FMQA and GA obtained the desired structure with fewer computations than PSO.

%%%%%%%%%%%%%%%
\subsubsection{Feature selection}

Feature selection is essential for improving the prediction accuracy of ML.
It has been reported that the FMQA algorithm can also be applied to feature selection \cite{doi:10.1080/14686996.2024.2388016}.
The least absolute shrinkage and selection operator (LASSO) method is a well-known feature selection method used to avoid overfitting in linear regression\cite{tibshirani_regression_1996}.
For a more precise implementation, the exhaustive search method\cite{Nagata:2015aa,doi:10.7566/JPSJ.87.044802} is useful for determining better combinations of features with the highest prediction accuracy.
However, the total number of combinations becomes $2^N -  1$ when the number of features is $N$.
Thus, as the number of features is increased, a combinatorial explosion occurs.
This feature selection can be considered as a BBO with binary variables.
In other words, the input of BBF is a combination of features, and the prediction accuracy of the ML models is used as the output.
To solve this problem using BBO, $N$-dimensional binary variables $\{ x_1, ..., x_N \}$ with $x_i = $ 0 or 1 are prepared.
If $x_i = 0$, the $i$th feature is not used, and if $x_i = 1$, the $i$th feature is used.
The feature selection was performed using by the FMQA algorithm.

\begin{figure*}
  \begin{center}
    \includegraphics[scale=1.0]{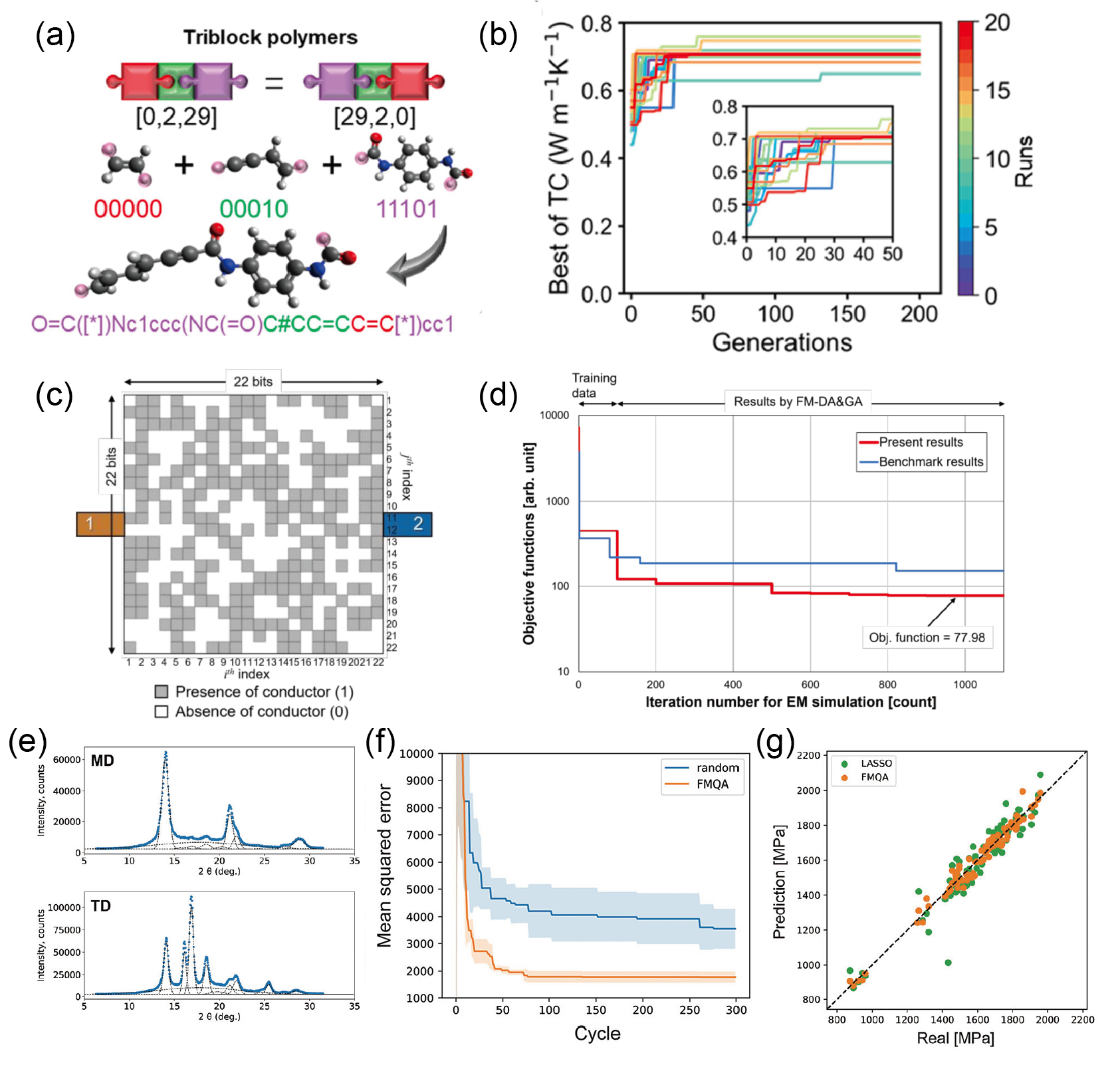}
  \end{center}
  \caption{
(a) Triblock polymers defined by 15 bits.
(b) Cycle dependence of the best TC for 20 independent runs.
(c) Example of conductive structures where 1 is input and 2 is output ports, respectively.
(d) Cycle dependence of the objective function defined by frequency response.
Present results were obtained by the combination method of FMQA and GA, and benchmark results were by PSO.
(e) Examples of peak deconvolution results for XRD measurements in the machine direction (MD; flow direction of the resin) and transverse direction (TD; width direction of the resin) of the homo polypropylene materials.
(f) Cycle dependence of the mean squared error between the predicted tensile modulus and the real one by FMQA algorithm and RS for the homo-polypropylene materials.
(g) Prediction results of the optimal model for the tensile modulus.
For comparison, the results by LASSO are also plotted.
(For (a) and (b), reprinted from ref. \cite{Huang2024} under a Creative Commons Attribution (CC BY) license.
For (c) and (d), reprinted from ref. \cite{HIDA2024} under a Creative Commons Attribution-NonCommercial-NoDerivs license.
For (e), (f), and (g), reprinted from ref. \cite{doi:10.1080/14686996.2024.2388016} under a Creative Commons Attribution license.)
  } 
  \label{fig:app_bin_others}
\end{figure*}

In Ref.~\cite{doi:10.1080/14686996.2024.2388016}, feature selection using the FMQA algorithm was performed to prepare the ML model with a high prediction accuracy for the mechanical properties of polymer materials.
A total of 75 samples of homo polypropylene materials were collected.
The target features were the parameters extracted from X-ray diffraction (XRD) patterns using Bayesian spectral deconvolution\cite{NAGATA201282,doi:10.1080/27660400.2022.2159753}.
Examples of peak deconvolution for XRD measurements in the machine direction (MD; flow direction of the resin) and transverse direction (TD; width direction of the resin) are shown in Fig.~\ref{fig:app_bin_others} (e).
Five feature types were generated for each peak, with 24 peaks for each sample. 
Thus, the number of dimensions of the prepared features was 120, and it was difficult to exhaustively search for $2^{120} - 1$ candidates.
In this study, the mean squared error between the predicted and real mechanical properties was used as the prediction accuracy, which is the output of the BBF.
A linear regression model and cross-validation were used to measure prediction accuracy.
Fig.~\ref{fig:app_bin_others} (f) shows the cycle dependence of the mean squared error by FMQA algorithm and RS when the tensile modulus is a materials property.
Even if the number of cycles is small, the FMQA algorithm can identify a better combination of features that exhibits a small mean squared error.
The comparison plots between the real tensile modulus and the prediction results obtained using the optimal model are summarized in Fig.~\ref{fig:app_bin_others} (g).
FMQA can successfully select features with a higher prediction accuracy than LASSO.
Therefore, this study shows that the FMQA algorithm can be applied as a feature selection method.

%%%%%%%%%%%%%%%
%%%%%%%%%%%%%%%
%%%%%%%%%%%%%%%
\subsection{Binary optimization problems with constraint}

Here, we introduce some applications of optimization problems that require constraints in addition to the FM model.
This is the case when not all the inputs to the FM are valid, and the solution must be selected with an appropriate constraint.
By minimizing the QUBO, which adds constraint terms to the FM, only valid solutions satisfying the constraint can be selected.
In the following section, we demonstrate the assignment of constraint terms to specific problem examples.

%%%%%%%%%%%%%%%
\subsubsection{Magnetic tunnel junction structure design}

\begin{figure*}
  \begin{center}
    \includegraphics[scale=0.8]{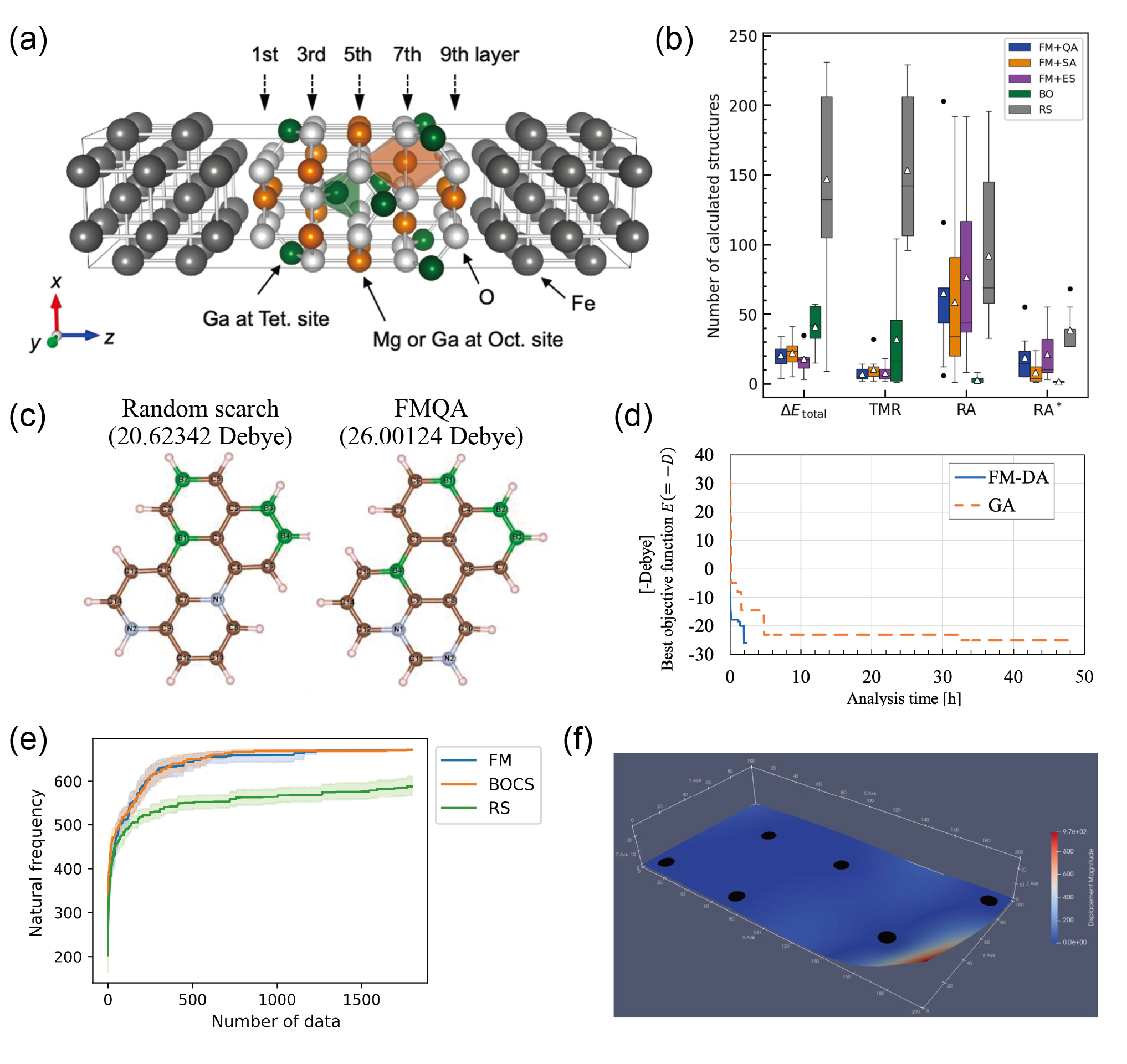}
  \end{center}
  \caption{
(a) Structure of MgGa$_2$O$_4$. The disordered site is the orange site.
(b) Optimization results by FMQA in MTJ structure design depending on the algorithms for each physical property.
FM + QA, FM + SA, and FM + ES are BBOs using FM with D-Wave quantum annealer, SA, and ES, respectively.
BO and RS are the Bayesian optimization and random sampling.
(c) Optimization results of atomistic configuration in perylene by RS and FMQA when 6-site substitution. 
The polarizability values are also shown.
The green, gray, and brown atoms are B, N, and C atoms, respectively.
(d) Objective function values against the computation time by FMQA and GA for the molecular structure optimization.
(e) Optimization results of the structural design for resonance avoidance by FMQA, BOCS, and RS.
(f) Result of frequency analysis obtained at the optimal mounting hole position.
(For (a) and (b), reprinted from ref. \cite{PhysRevApplied.20.024044} under a Creative Commons Attribution 4.0 International license.
For (c) and (d), reprinted from ref. \cite{202420244237} with permission.
For (e) and (f), reprinted from ref. \cite{Matsumori:2022aa} under a Creative Commons Attribution 4.0 International license.)
  } 
  \label{fig:app_bin_constraint}
\end{figure*}

Magnetic tunnel junctions (MTJs) have garnered considerable attention in the field of spintronics,
because they are expected to be applicable in non-volatile memory devices.
The MTJ structure comprises two layers of magnetic material with a very thin insulating layer between them.
An optimization study combining FMQA and first-principles calculations was conducted to understand the suitable structure of the insulating layer~\cite{PhysRevApplied.20.024044}.
The target material was MgGa$_2$O$_4$\cite{10.1063/1.4977946}, which has an inverse spinel structure as shown in Fig.~\ref{fig:app_bin_constraint} (a).
The optimization target is the arrangement of atoms to improve the physical properties when Mg or Ge ions are placed at the disordering sites.
A structure with 10 disordered sites was considered: 
Mg ions were placed in five of them, and Ge ions were placed in the remaining five.
In other words, there are $_{10}$C$_5 = 252$ possible configurations.
By providing a binary variable $x_i = 0$ or $1$ for each site, $x_i = 0$ when Mg ions are placed and $x_i = 1$ when Ge ions are placed.
Considering the case where there are $N$ disordered sites in total, the arrangement of the ions can be expressed by the binary variables: $\mathbf{x} = \{ x_1, x_2,..., x_N \}$.
By training the FM defined in Eq.~(\ref{eq:FM}) with known $y (\mathbf{x})$ as the target physical property, a surrogate model for predicting this property was trained.
However, there is no guarantee that the solution minimizing the FM will have five ions each.
In other words, when the FM is solved by the IM, the solution is selected from the $2^N$ possible structures expressed by $\mathbf{x}$ instead of from the $_{10}$C$_5$ configurations.
Therefore, to ensure that the obtained solution is correct for the number of ions,
that is, the numbers of $x_i = 1$ and $x_i = 0$ are the same,
a QUBO with the following constraint was prepared:
\begin{eqnarray}
H_{\rm QUBO} = H_{\rm FM} + \alpha \left( \sum_{i=1}^N x_i - \frac{N}{2} \right)^2,
\end{eqnarray}
where $\alpha$ denotes a positive hyperparameter.
The second term is a constraint that is minimized when $\sum_{i=1}^N x_i = N / 2$.
For this Hamiltonian, the optimal solution is one in which the output value of the FM is sufficiently small and $\sum_{i=1}^N x_i = N/2$.
Parameter $\alpha$ determines the constraint strength.
If it is excessively small, a solution that does not satisfy the constraint is obtained, whereas if it is excessively large, a solution with better properties is not obtained.
Therefore, these parameters must be set appropriately, 
and used to optimize the MTJ structure,
where $\alpha = 1$ is assumed.

In this study, three types of MTJ properties were considered.
The target properties were the total energy difference ($\Delta E_{\rm total}$), tunneling magnetoresistance (TMR), and the resistance area product, which were calculated using first-principles calculations. 
Small values for these parameters are required for spintronics devices.
Therefore, to train the FM, the properties multiplied by a negative sign were used as objective functions.
The optimization results obtained by FMQA are shown in Fig.~\ref{fig:app_bin_constraint} (b), 
which shows the number of cycles required to determine the best MTJ structure.
The results were compared with those obtained using the D-Wave quantum annealer, SA, and random sampling.
In addition, the results were obtained by the strategy in which the minimization of the FM was performed by an exhaustive search (ES) without using IMs and Bayesian optimization, 
because the size of the problem was small.
The FMQA algorithm is more efficient than random search, and the optimization of $\Delta E_{\rm total}$ and TMR requires fewer computations than the Bayesian optimization strategy.
This research shows that the FMQA algorithm is also a powerful tool for optimizing the arrangement of ions with the desired physical properties,
if a constraint is required.

%%%%%%%%%%%%%%%
\subsubsection{Atomistic configuration optimization in molecules}

The FMQA algorithm was used to optimize the atomistic configuration of carbon materials\cite{202420244237}.
Carbon materials are used such as catalysts in fuel cells and electrode materials in Li ion batteries, 
and their industrial applications are highly active.
This study focused on perylene (C$_{20}$H$_{12}$),
and substitution of carbon atoms was considered.
Substitution was performed by replacing N or B atoms with C atoms.
For each carbon position, two binary variables are used to determine the atomistic configuration.
In other words, if the first bit is 0, no substitution is performed, and if it is 1, substitution is performed for the N or B atoms.
If the second bit is 0, the C atom is substituted with a B atom; if the second bit is 1, the C atom is substituted with a N atom.
Thus, for each C atom position, `00' and `01' mean C atom, `10' is B atom, and `11' is C atom.
In perylene, $N=20$ carbons are present; 
thus, the substituted structure is represented by 40 binary variables ($\{ x_i \}_{i= 1,.... , 2N}$).
The polarizability calculated using the molecular orbital method was used as the objective function,
and maximization was desirable.
Therefore, when training the FM, polarizability multiplied by a negative sign was used as the objective function.
In addition to the MTJ structure design,
better structures were optimized for polarizability when the number of substituted atoms was fixed.
To ensure that the number of substituted atoms was $N_{\rm c}$,
the following QUBO model was developed:
\begin{eqnarray}
H_{\rm QUBO} = H_{\rm FM} + \alpha \left( \sum_{i=1}^{N} x_{2i -1} - N_{\rm c} \right)^2,
\end{eqnarray}
where $\alpha$ denotes a positive hyperparameter.
This is because the first bit of each carbon position expresses substitution.

The results for the case of 6-site substitution ($N_{\rm c} = 6$) are shown in Fig.~\ref{fig:app_bin_constraint} (c).
The number of cycles was 96, and the optimized structure and its polarizability values were compared using the FMQA algorithm and RS.
A DA was used as an IM.
The structure obtained by FMQA had a larger polarizability value than that obtained by RS.
In addition, Fig.~\ref{fig:app_bin_constraint} (d) shows the obtained objective function values against the computation time using the FMQA and GA.
The results demonstrate that the FMQA algorithm can find a better molecular structure with large polarizability within a short computation time.
It has been reported that FMQA exhibits high performance in solving molecular replacement problems.

%%%%%%%%%%%%%%%
\subsubsection{Structural design for resonance avoidance}

Printed circuit boards are used in the power control units of hybrid vehicles and require a proper structural design to avoid the resonance.
For instance, if the location of the mounting holes can be optimized to increase the natural frequency, low-frequency resonance can be avoided.
To solve the optimization problem of increasing the natural frequency, the FMQA algorithm was adopted\cite{Matsumori:2022aa}.
In this study, the candidate positions of mounting holes were determined on a 2D substrate.
If $N$ candidate positions exist,
then $N$-dimensional binary variables $\mathbf{x} = \{ x_1, x_2,..., x_N \}$ were used to determine the arrangement of the mounting holes.
If $x_i =1$, a mounting hole is formed at that position; when $x_i =0$, no mounting hole is formed at that position.
After the number of mounting holes is determined,
the natural frequency is calculated using the FEM. 
To obtain a higher natural frequency, the FM is trained using the natural frequency multiplied by the negative sign as the objective function.
When the number of mounting holes is fixed,
additional constraints are required.
For instance, when the number of mounting holes to be used is $N_{\rm c}$, the QUBO model constructed by the FM and constraint is as follows:
\begin{eqnarray}
H_{\rm QUBO} = H_{\rm FM} + \alpha \left( \sum_{i=1}^N x_i - N_{c} \right)^2,
\end{eqnarray}
where $\alpha$ denotes a positive hyperparameter.

Fig.~\ref{fig:app_bin_constraint} (e) shows the results of the optimization for problems with $N=20$ and $N_{\rm c} = 6$.
Here, the results of FMQA, BOCS, and RS are compared,
and the D-Wave Hybrid Solver was used as the IM.
The FMQA algorithm can determine a better mounting hole location with a higher natural frequency and fewer calculations than RS.
The results of the frequency analysis obtained at the optimal mounting hole position are shown in Fig.~\ref{fig:app_bin_constraint} (f).

%%%%%%%%%%%%%%%
%%%%%%%%%%%%%%%
%%%%%%%%%%%%%%%
\subsection{Integer variable problems}

If the integer variable problems are solved using the FMQA algorithm,
then encoding from integers to binary variables must be used.
Some applications were reported\cite{seki2022blackboxoptimizationintegervariableproblems,fixwing,fixtraffic,Lin2025},
and the following representative examples, 
the structural optimization of photonic crystal lasers\cite{Inoue:22},
wing shape optimization\cite{fixwing},
signal control optimization\cite{fixtraffic},
vehicle body structures optimization\cite{kondo2024} are introduced in this review.

%%%%%%%%%%%%%%%
\subsubsection{Structural optimization of photonic crystal lasers}

The demand for semiconductor lasers\cite{Samuel:2007aa} with high output power and beam quality is rapidly increasing for various applications, including optical detection in smart mobility and high-precision laser processing in smart manufacturing.
The FMQA algorithm is used to design the next generation of semiconductor lasers, known as photonic crystal lasers, which are characterized by their ability to operate at high power and high beam quality\cite{Inoue:22}.
In this study, 11 continuous variables were considered to determine the structure of photonic lasers.
Continuous variables cannot be handled by FMQA as is.
Thus, the upper and lower bounds of each variable are determined, and the discrete variables are defined by dividing them into 16 equal distances between the upper and lower bounds.
Using binary encoding, the 16 values are represented by four binary variables,
and 11 discrete parameters can be represented by 44 binary variables.
To simultaneously consider the three properties of output power ($P$), polarization ratio ($\eta$), and divergence angle ($\theta_x$ and $\theta_y$), the quantity $Q$ defined below is used as the objective function:
\begin{eqnarray}
Q = \frac{P \times \eta}{\theta_x \theta_y}.
\end{eqnarray}
The value of $Q$ increases as $P$ and $\eta$ increase, and $\theta_x$ and $\theta_y$ decrease.
Because photonic crystal lasers with large $Q$ values are desirable, 
the properties multiplied by a negative sign were used as objective functions to train the FM.
In this case, all inputs to the FM are valid, 
that is, all states defined by the 44 binary variables can be converted into 11-dimensional continuous variables for photonic crystal lasers.

\begin{figure*}
  \begin{center}
    \includegraphics[scale=0.8]{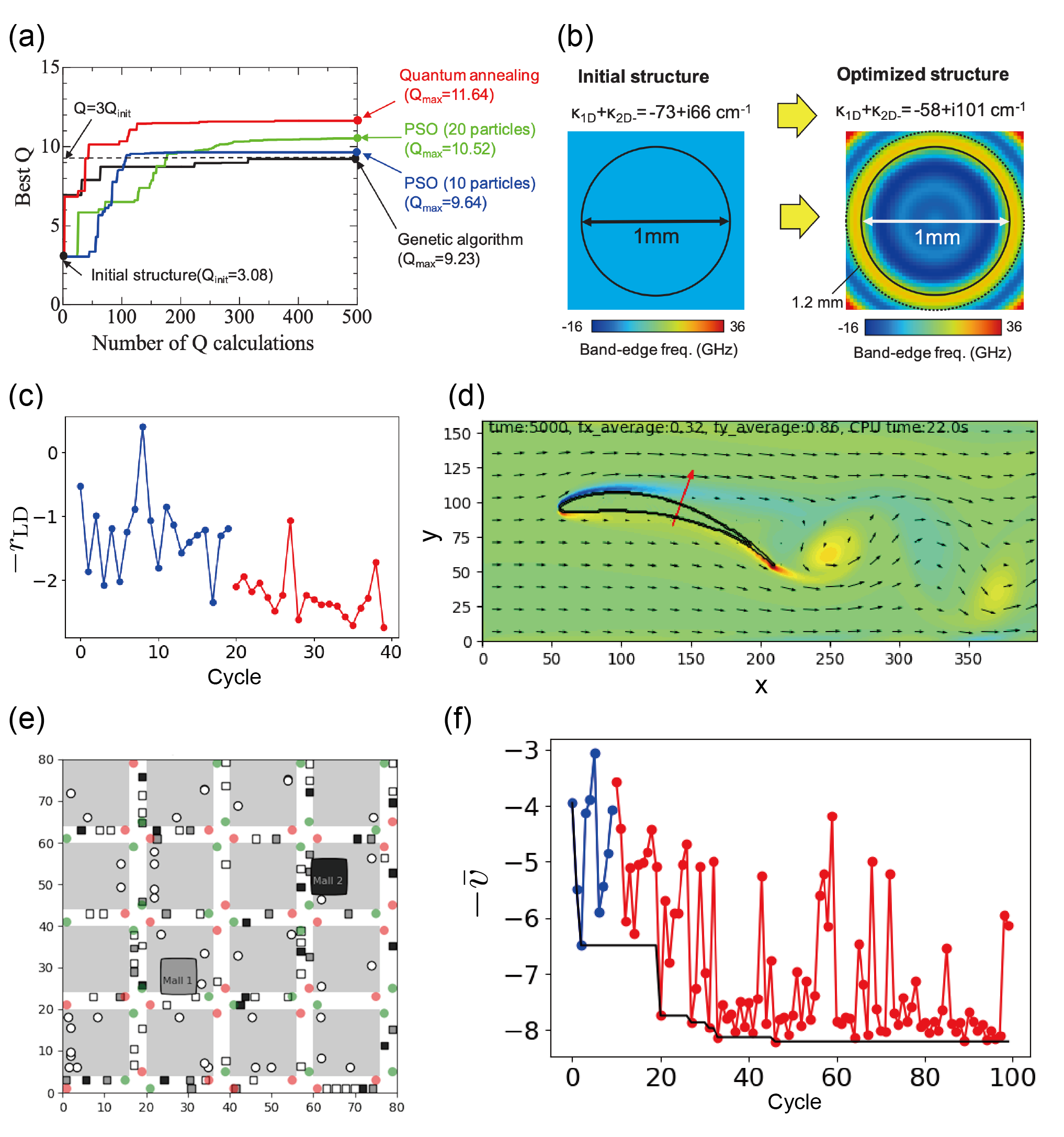}
  \end{center}
  \caption{
(a) Cycle dependence of the $Q$ value in the structural optimization of photonic crystal lasers.
The results by FMQA, PSO, and GAs are compared.
(b) Band edge frequency difference between initial structure and the optimized structure obtained by FMQA.
(c) Cycle dependence of $- r_{\rm LD}$ for the wing shape optimization by FMQA.
(d) Optimized wing shape by FMQA.
(e) Optimization target city for signal control optimization problem.
Squares represent the vehicles, whereas white circles represent the homes.
Red and green circles represent the signals, and two malls are present.
(f) Cycle dependence of the average of vehicle speeds $\bar{v}$ by FMQA.
(For (a) and (b), reprinted from ref. \cite{Inoue:22} under a Creative Commons license.
For (c) and (d), reprinted from ref. \cite{fixwing} with permission.
For (e) and (f), reprinted from ref. \cite{fixtraffic} with permission.)
}
  \label{fig:app_integer}
\end{figure*}

Fig.~\ref{fig:app_integer} (a) shows the cycle dependence of $Q$.
The IM used was a D-Wave Advantage quantum annealer.
It is shown that the FMQA algorithm can be used to find parameters with larger $Q$ values and fewer computations than other optimization methods.
Other methods include PSO\cite{488968,Wang:2018aa} and GAs\cite{Whitley:1994aa}.
The band edge frequency of the structure determined using the FMQA algorithm is shown in Fig.~\ref{fig:app_integer} (b).
For comparison, the results for the initial structure are also presented. 
The band edge frequency in the structure determined by the FMQA algorithm is larger than that of the initial structure.
This study demonstrated the effectiveness of FMQA for optimization problems in which continuous values are discretized using binary encoding.

%%%%%%%%%%%%%%%
\subsubsection{Wing shape optimization}

The wing shape of an aircraft is optimized to maximize the lift and minimize the drag.
These wing forces can be evaluated using actual experiments and computational fluid dynamics simulations.
For this optimization problem using the fluid dynamics simulations,
the FMQA algorithm was applied\cite{fixwing}.

The Joukowski transform was used to determine the shape of the wing, which is known as Joukowski airfoil.
Three continuous parameters ($\xi$, $\eta$, $\alpha$) are used to determine shape.
Thus, the task of the optimization problem is to find a better set of continuous parameters that yields the maximum lift and minimum drag forces.
To evaluate these forces, a 2D fluid analysis method based on the lattice Boltzmann method\cite{Suzuki_Minami_Inamuro_2015} was used.
Here, $F_{\rm L}$ and $F_{\rm D}$ denote the lift and drag forces, respectively.
Using these forces, 
the goal is to maximize the lift--drag ratio defined by $r_{\rm LD} = F_{\rm L} / F_{\rm D}$.
When training the FM, $- r_{\rm LD}$ was used as the objective function.
When the FMQA algorithm is applied, 
continuous variables must be discretized, and $\xi$ is divided into 20 equal parts between 1.0 and 9.55, $\eta$ is divided into 40 equal parts between 0 and 9.75, and $\alpha$ is divided into 40 equal parts between 0 and 39.
In this study, one-hot encoding was used, and a binary variable was prepared for each part.
Thus, 20 bits were used for $\xi$, whereas 40 bits were used for $\eta$ and $\alpha$.
In other words, 100 bits were used to determine the shape of the wing as $\{ x_i \}_{i=1,..., 100}$.
To implement one-hot encoding,
a QUBO model with the following constraint was prepared:
\begin{eqnarray}
H_{\rm QUBO} =&& H_{\rm FM} + \alpha \left[ \left( \sum_{i=1}^{20} x_i - 1 \right)^2 \right. \notag \\
&&\ \ \ \left.+ \left( \sum_{i=21}^{60} x_i - 1 \right)^2 + \left( \sum_{i=41}^{100} x_i - 1 \right)^2 \right],
\end{eqnarray}
where $\alpha$ denotes positive hyperparameter.
Fixstars Amplify AE is used to solve this QUBO model.
The values of $- r_{\rm LD}$ in the optimization process are shown in Fig.~\ref{fig:app_integer} (c).
In this process, the first 20 cycles were performed randomly and the remaining 20 cycles were performed using the FMQA algorithm.
Fig.~\ref{fig:app_integer} (d) shows the optimized wing shape after 40 cycles, 
which improved the lift--drag ratio by 17\% compared with that obtained after 20 cycles by RS.

%%%%%%%%%%%%%%%
\subsubsection{Signal control optimization}

Signal control using the FMQA algorithm was considered as an example of an application to a social science problem\cite{fixtraffic}.
In this application, the control of traffic signals was optimized to maximize the average speed of all vehicles traveling in a city.
The objective function was the average vehicle speed, which was evaluated using a multi-agent simulation (MAS).
Based on a simulation model known as the optimal speed model\cite{Bando:1994aa,PhysRevE.51.1035},
the driver of each vehicle accelerated and decelerated according to the optimal speed determined from the distance between the vehicles. 
The city, as an optimization target, is shown in Fig.~\ref{fig:app_integer} (e).
The city has 4 $\times$ 4 sections divided by roads and two shopping malls.
There are 16 intersections in the city. 
Traffic signals were present at each intersection.
The traffic signal can be controlled by three parameters: the length of the red light $L_{\rm r}$, length of the green light $L_{\rm g}$, and the start time of the red light $L_{\rm s}$, in seconds.
For instance, if the control parameters for a traffic signal at an intersection are $L_{\rm r} = 15.0$, $L_{\rm g} = 12.0$, and $L_{\rm s} = 5.0$, the north--south signal at the intersection will be red 5.0 s after the simulation start time, red for 15.0 s, and green for 12.0 s.
When using the FMQA algorithm, continuous variables must be expressed as binary variables.
One-hot encoding was used.
The lower and upper limits of the three parameters were set to 1 s and 20 s, respectively, and the discrete values were divided into 20 equal parts, that is, every second.
Thus, when the one-hot encoding was considered, 20 bits were prepared as binary variables for each continuous parameter.
The control parameters for each intersection can be expressed using 60 binary variables,
and 960 binary variables ($\{ x_i \}_{i=1,...,960}$) are required for all the control signals in the city.

In MAS, a problem setting in which the maximum number of vehicles is 20 and 50\% of these vehicles have destinations in one of the two shopping malls was considered.
The destinations of the remaining 50\% of vehicles were randomly selected from the city.
The homes for each vehicle were randomly located in the city, and a simple simulation was performed in which the vehicles continued to travel back and forth between the shopping mall and their homes or between their destinations and their homes.
When the control parameters of the signals are determined,
the average speed of vehicles, $\bar{v}$, was evaluated using MAS.
Thus, the control parameters that maximize $\bar{v}$ are searched for using the FMQA algorithm.
When training the FM, $-\bar{v}$ is used as the objective function.
To implement one-hot encoding,
we prepared a QUBO model with the following constraints:
\begin{eqnarray}
H_{\rm QUBO} = H_{\rm FM} + \alpha \sum_{l =1}^{16} \sum_{j=1}^3 \left( \sum_{i=1 + 20 (j-1) + 60 (l-1)}^{20 +  20 (j-1) + 60 (l-1)} x_i - 1 \right)^2
\end{eqnarray}
In this study, Fixstars Amplify AE was used as the IM.
The simulation was performed using 10 random parameters as the initial data.
BBO with the FMQA algorithm was executed for 90 cycles.
The $-  \bar{v}$ values for the optimization cycles are shown in Fig.~\ref{fig:app_integer} (f). 
This indicates that the use of FMQA improved the search for better control parameters for traffic signals.
The MAS simulation results for the optimized parameters are available at \url{https://amplify.fixstars.com/en/demo/fmqa_4_traffic}.

\subsubsection{Design optimization of vehicle body structures for multiple vehicles}

The benchmark problem for the optimization of vehicle body structures for multiple vehicles was based on Mazda's production models proposed in Ref.~\citep{kohira2018} and is available at \url{https://ladse.eng.isas.jaxa.jp/benchmark/}. The benchmark comprises three problem classes; however, the most difficult one is the multi-objective optimization of vehicle body structures for multiple vehicles, which is based on real-world problem settings and complexity. The problem involves:
\begin{itemize}
    \item A black-box objective function, which is the total weight of the three vehicle models (to be minimized),
    \item An objective function that can be expressed in a quadratic manner, representing the number of common parts across the three models (to be maximized) and 
    \item A total of 18 constraints among which there are 14 black-box constraint functions per vehicle (total of 42 black-box constraints). These black-box constraints are related to crashworthiness, eigenvalues, body rigidity, and manufacturing constraints. The values of these black-box functions are required to be zero or positive values such that the constraints are met.
\end{itemize}
A total of 222 real variables were used to control the objectives and constraints. Various groups have attempted to solve this benchmark problem based on various GAs including elitist non-dominated sorting genetic algorithm II (NSGA-II) \cite{Ootomo2018, Oyama2018} and Bayesian optimization algorithms \cite{Daulton2022}. All these experiments consider the number of cycles (number of evaluations of black-box functions) of 10,000 or 30,000, which is excessively large for the typical real-world design process standards.

Kondo et al.\cite{kondo2024} have tackled this problem with only less than 1,000 evaluations of black-box functions by adding the following two improvements to the original FMQA algorithm:
\begin{itemize}
    \item Weights to the objectives and constraints are determined based on the distances between the objective (constraint) values and corresponding target values. The target values are zero for the constraints, whereas for the objective functions, the target values are the average values in the dataset minus their standard deviation. Thus, the weights for the objectives change over cycles.
    \item Multiple searches per cycle. In each optimization cycle, the following three FMs are constructed. The first FM is constructed using a relatively large $K$ (see Eq.~\eqref{eq:FM}) to make the resulting model over-fit. The second FM is based on a relatively small $K$, expected to be an ``under-fitted'' model. The third model is constructed by linearly combining the first and the second FM models. Then, annealing is performed for each of these three FM models to obtain three input vectors. 
\end{itemize}
With these additions, Kondo et al. have performed three independent evaluations of black-box functions per cycle, multiplied by the number of cycles, which was 300, resulting in a total of 900 function evaluations. The Pareto solutions obtained exhibited improved solutions by the NSGA-II with 30,000 function evaluations.

%%%%%%%%%%%%%%%
%%%%%%%%%%%%%%%
%%%%%%%%%%%%%%%
\subsection{Others}

Using the bVAE, various optimization problems can be solved even if they are not binary or integer optimization problems\cite{Wilson:21}.
Applications in thermal emitter design\cite{10.1063/5.0060481},
molecular structure design\cite{D3DD00047H},
and peptide design\cite{Tucs:2023aa} are introduced.

%%%%%%%%%%%%%%%
\subsubsection{Thermal emitter design}

\begin{figure*}
  \begin{center}
    \includegraphics[scale=0.8]{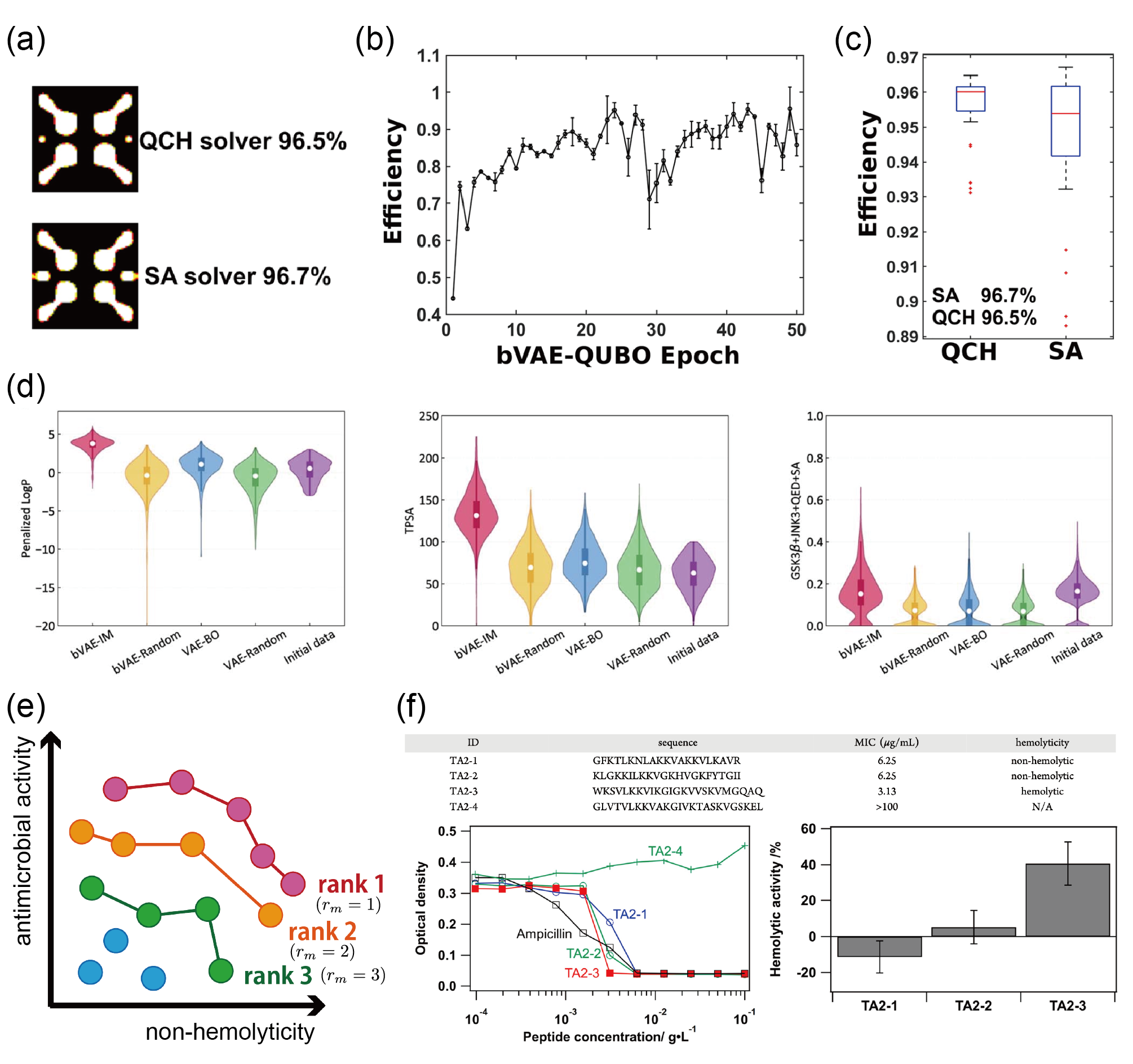}
  \end{center}
  \caption{
(a) Target topological shape of the thermal radiator for the thermal emitter design.
(b) Cycle dependence on efficiencies by FMQA.
(c) Results of the top 100 thermal emitter designs by D-Wave quantum annealer with hybrid solvers (QCH) and SA.
(d) Optimization results in molecular structure design.
Optimization results by combining bVAE and FMQA (bVAE + IM), and randomly generated in the latent space of bVAE (bVAE-Random) are shown. 
For comparison, the results using Bayesian optimization (VAE-BO) and RS (VAE-Random) using a non-binary VAE are shown.
The distribution of initial data is shown.
(e) Ranking of Pareto solutions for multiobjective optimizations.
(f) Peptide optimization results that were experimentally synthesized and evaluated.
Sequences, optical density, and hemolytic activity for four peptides are shown.
(For (a), (b), and (c), reprinted from ref. \cite{10.1063/5.0060481} with permission.
For (d), reprinted from ref. \cite{D3DD00047H} under a Creative Commons Attribution 3.0 Unported license.
For (e) and (f), reprinted from ref. \cite{Tucs:2023aa} under a Creative Commons Attribution-NonCommercial-NoDerivatives 4.0 International license.)
  } 
  \label{fig:app_bVAE}
\end{figure*}

The first study that used the bVAE and FMQA algorithms was the topology structure optimization of a thermal radiator by Wilson et al.\cite{10.1063/5.0060481}.
In this study, a thermal radiator was optimized for use in a thermophotovoltaics (TPV) engines.
The TPV engine generates electricity through radiative heat transfer between a heater and an array of photovoltaic cells.
The design of an efficient thermal radiator determines the efficiency of the TPV engine. 
Thus, an efficient design is essential.
The target topological shape of the thermal radiator is 2D, as shown in Fig.~\ref{fig:app_bVAE} (a).
The white colored area is TiN, whereas the black area is filled with air.
This topological shape is treated as a 64 $\times$ 64 binarized figure, 
and 5,000 training data points were available.
When the topological shape is converted into a binary variables,
4,096 bits are needed.
To reduce the dimensions of the explanatory variables,
the bVAE is useful.

Using the bVAE, a 500-dimensional binarized latent variable space was trained.
Thus, each topological shape could be expressed by 500-dimensional binary variables.
The objective function in the optimization problem was defined by the efficiency of the thermal radiator.
This efficiency was determined as the product of the in-band and out-of-band efficiencies,
and higher values are desirable for the TPV engines.
When training the FM, efficiency multiplied by the negative sign was used as the objective function.
In this case, all states defined by the 500-dimensional binary variables can be decoded to the topological shape of the thermal radiator using the trained bVAE.
Thus, there is no need to consider constraints to solve the QUBO model.
SA and a D-Wave Advantage hybrid sampler were used as IMs.
Fig.~\ref{fig:app_bVAE} (b) shows the efficiency as a function of the iterations using a D-Wave Advantage hybrid sampler.
It was shown that a structure with better efficiency could be obtained using the FMQA algorithm.
The results of the top 100 thermal emitter designs for both IMs are shown in Fig.~\ref{fig:app_bVAE} (c).
The optimal structures are shown in Fig.~\ref{fig:app_bVAE} (a).
This study reported that better topological structures could be effectively identified by combining bVAE and the FMQA algorithm.

%%%%%%%%%%%%%%%
\subsubsection{Molecular structure design}

Molecular structures can be represented by strings, known as SMILES (simplified molecular input line entry system)\cite{Weininger:1988aa}.
For instance, benzene is represented as C1=CC=CC=C1 and acetic acid is represented as CC(=O)O.
The length of the SMILES varies depending on the molecule.
To convert SMILES to binary variables,
junction tree variational autoencoder (JTVAE) is useful for treating molecular structures as graphs\cite{pmlr-v80-jin18a}.
JTVAE was modified to the bJTVAE using Gumbel softmax reparameterization\cite{jang2017categoricalreparameterizationgumbelsoftmax} to obtain the latent space with binary variables.
A total of 250,000 molecules were used to train the bJTVAE.
The dimension of the latent space was 300\cite{D3DD00047H}.

The three types of molecular properties (penalized LogP, TPSA, GSK3$\beta$ + JNK3 + QED + SA) were optimized to verify the performance of the combined bJTVAE and the FMQA algorithm\cite{D3DD00047H}.
These properties were calculated from the molecular structures,
and the larger values were required.
Thus, to train the FM, the negative sign was multiplied by these properties.
Fig.~\ref{fig:app_bVAE} (d) compares the results of the optimizations combining bVAE and FMQA and randomly generated in the latent space of bVAE.
For comparison, it also shows the results using a non-binary VAE.
Bayesian optimization and RS were performed using a non-binary VAE.
In all cases, the combined search for bJTVAE and FMQA generated several molecules with excellent properties.
For this verification, Fixstars Amplify AE was used as the IM.
In addition, 
this study reported that increasing the number of binary variables in the latent space is expected to improve the accuracy of bVAE, and IM with several binary variables would be effective for molecular structure optimization.

%%%%%%%%%%%%%%%
\subsubsection{Peptide design}

Peptide design was performed using bVAE and the FMQA algorithm\cite{Tucs:2023aa}.
A peptide is a combination of amino acids, and the 20 amino acids can be represented by strings such as `A,' `C,' and `D.' 
Thus, the peptide sequence was represented by a string of letters.
To convert a sequence into binary variables, bVAE is useful for text modeling\cite{10.1007/978-3-030-34518-1_10}.
The training data for the sequences included 19,530 antimicrobial and 5,583 non-antimicrobial peptides.
The latent space has 64 dimensions; thus, each peptide sequence can be represented by 64-dimensional binary variables.

In this study, peptide sequence optimization was performed to simultaneously increase antimicrobial and non-hemolytic properties.
These are generally known to have trade-offs, and designing peptides that simultaneously achieve both is an important issue in the development of therapeutic peptides.
To perform this optimization, it is necessary to set up a single objective function that handles multi-objective optimization.
Multi-objective optimization is performed to find the maximum number of Pareto solutions when two or more objective functions with a trade-off are present.
The Pareto solution is the optimal solution after the balance of the objective function is determined.
Depending on this balance, a different solution becomes a Pareto solution.
This set of Pareto solutions is known as the Pareto front and is a trade-off curve.
To treat the multi-objective optimization problem with a single objective function, a ranking variable $r_m$ is introduced into the currently available data, which indicates the distance between each data point and the Pareto front.
If the solution is on the Pareto front, $r_m = 1$ is assigned.
After removing all the solutions with $r_m = 1$, a new Pareto front is defined in the remaining samples, and $r_m = 2$ is assigned to the sample on the new Pareto front.
This procedure is repeated until a ranking is obtained for all samples (see Fig.~\ref{fig:app_bVAE} (e)).
To train the FM, $y_m=-1/r_m$ is used as the objective function; thus, a small value of $y_m$ is desired.
In this way, multi-objective optimization can be treated as single-objective optimization and solved using the FMQA algorithm.
The BBO method using FMQA that can handle this multi-objective optimization problem is known as multi-objective optimization by quantum annealing (MOQA).
MOQA was used to design 200,000 peptides.
The D-Wave Advantage System 4.1 was used as IM.
Four peptides were synthesized and evaluated (Fig.~\ref{fig:app_bVAE} (f)) from peptides designed using MOQA.
Two of these peptides were experimentally determined to be both antimicrobial and non-hemolytic.
This study showed that peptides were successfully designed by multi-objective optimization using the FMQA algorithm.

%%%%%%%%%%%%%%%
% Perspective
%%%%%%%%%%%%%%%

\section{Closing Statements and Outlook}
\label{sec:outlook}

In this review,
we provide an overview of the BBO algorithm using IMs, known as FMQA, and its applications.
In this algorithm, the FM, which can be expressed as a QUBO model,
is used as a surrogate model.
The FM can then be solved efficiently using the IMs, and a fast BBO is realized.
In particular, using IMs, the FMQA algorithm has been explored in a large input space.
To make the FMQA algorithm easy to run, a Python package will be developed and released; thus, R \& D using FMQA will become increasingly popular in both academia and industry.
We strongly believe that BBO using FMQA is a key technique for IMs.
Finally, outlook of the FMQA algorithm is briefly described.

{\it Using a quantum computer instead of IM}: 
Gao et al. designed a highly efficient organic light-emitting diode emitter using the FMQA algorithm\cite{doi:10.34133/icomputing.0037}.
This study used the QAOA\cite{QAOA} to solve the QUBO.
QAOA can be treated using quantum computers.
Although the small-sized problems with six bits were attempted, the QAOA calculations were performed using a state vector simulator and the IBM Quantum System One\cite{IBMQ}.
In the future, when the number of qubits in quantum computers increases, large-scale BBO problems can be handled using FMQA and quantum computers instead of IMs.

{\it Optimization of the experimental results}:
The applications introduced in this review are all optimization results when the objective functions are obtained through simulations.
Using the results obtained from real experiments as the objective function, they can contribute directly to the basic science, such as materials development and drug discovery.
However, although FMQA can handle a large search space, it requires several BBF readings,
which is time-consuming.
Therefore, when real experiments are conducted in optimization cycles, FMQA algorithm is expected to be more powerful in combination with automated experimental systems such as self-driving laboratories.
Recently, self-driving laboratories have been developed for various fields\cite{Schneider:2018aa,Burger:2020aa,doi:10.1126/sciadv.aaz8867,doi:10.1080/27660400.2023.2232297,Tom2024,Yoshikawa2025}.
The linkage between these factors and FMQA may lead to the development of innovative materials and drug discovery strategies.

{\it Applications to larger-scale problems}:
Up to 1,000,000 binary variables can be used in the current IMs.
In contrast, the application problems discussed in this review can all be regarded as small scale problems.
Therefore, IMs can potentially be applied to larger-scale problems using the FMQA algorithm.
To address the lager-scale problem, room for further innovation exists in the FMQA algorithm.
One is the development of effective batch processing methods, in which multiple suggestions of inputs for BBF are effectively generated.
For instance, batch versions of Bayesian optimization have been widely developed in the field of informatics\cite{pmlr-v51-gonzalez16a, NIPS2010_e702e51d, NEURIPS2023_727a5a5c}.
However, no definitive method has been developed yet, and this is still the subject of research.
Similarly, effective batch techniques for the FMQA algorithm will definitely be needed in the future.
In particular, IM-specific batch techniques, such as using uncertainty in QA and sampling methods using IM, may be effective and will be an important technological developments.
The second is the development of a surrogate model suitable for BBO using IMs, which is a more advanced version of the FM.
Although not directly solvable using the current IM, the introduction of higher order binary optimization (HUBO)\cite{Glos:2022aa,Domino:2022aa}, which has a higher order than QUBO,
is useful for ensuring the prediction accuracy.
Furthermore, model fitting to the graph structure of IMs can be expected to improve the performance of IMs as well as the FMQA.

\begin{acknowledgments}
The authors thank Taiga Hayashi, Ryo Ogawa, Tetsuro Abe, Mayumi Nakano, Tokiya Fukuda, Masashi Yamashita, Hyakka Nakada, Shuta Kikuchi, Kotaro Terada, Yosuke Mukasa, Kotaro Tanahashi, Tadashi Kadowaki, Tengfei Luo, and Junichiro Shiomi for the discussions about FMQA algorithm.

This work was partially supported by Japan Science and Technology Agency (JST) PRESTO (JPMJPR24T8), JST CREST (JPMJCR21O2), the Japan Society for the Promotion of Science (JSPS) KAKENHI (Grant Numbers JP23H05447, 25K07172), the Council for Science, Technology, and Innovation (CSTI) through the Cross-ministerial Strategic Innovation Promotion Program (SIP), ``Promoting the application of advanced quantum technology platforms to social issues'' (Funding agency: QST), JST (Grant Number JPMJPF2221). 
In addition, this paper is partially based on results obtained from a project, JPNP23003, commissioned by the New Energy and Industrial Technology Development Organization (NEDO).
S.T. wishes to express their gratitude to the World Premier International Research Center Initiative (WPI), MEXT, Japan, for their support of the Human Biology-Microbiome-Quantum Research Center (Bio2Q).

\end{acknowledgments}

\section*{AUTHOR DECLARATIONS}

\subsection*{Conflict of Interest}
The authors have no conflicts to disclose.

\subsection*{Author Contributions}
{\bf Ryo Tamura}: Conceptualization (lead); Writing – original draft (lead); Supervision (equal). 
{\bf Yuya Seki}: Writing – review \& editing (supporting).
{\bf Yuki Minamoto}: Writing – original draft (supporting); Writing – review \& editing (supporting). 
{\bf Koki Kitai}: Writing – review \& editing (supporting).
{\bf Yoshiki Matsuda}: Supervision (supporting); Writing – review \& editing (supporting). 
{\bf Shu Tanaka}: Conceptualization (supporting); Supervision (equal); Writing – review \& editing (supporting). 
{\bf Koji Tsuda}: Conceptualization (supporting); Supervision (equal); Writing – review \& editing (supporting).

\section*{Data Availability Statement}

Data sharing is not applicable to this article as no new data were
created or analyzed in this study.

\section*{REFERENCES}
%\bibliography{aipsamp}% Produces the bibliography via BibTeX.
%\bibliographystyle{aipnum4-1}
%\bibliographystyle{apsrev4-1}
\bibliography{FM_review}

\begin{thebibliography}{140}%
\makeatletter
\providecommand \@ifxundefined [1]{%
 \@ifx{#1\undefined}
}%
\providecommand \@ifnum [1]{%
 \ifnum #1\expandafter \@firstoftwo
 \else \expandafter \@secondoftwo
 \fi
}%
\providecommand \@ifx [1]{%
 \ifx #1\expandafter \@firstoftwo
 \else \expandafter \@secondoftwo
 \fi
}%
\providecommand \natexlab [1]{#1}%
\providecommand \enquote  [1]{``#1''}%
\providecommand \bibnamefont  [1]{#1}%
\providecommand \bibfnamefont [1]{#1}%
\providecommand \citenamefont [1]{#1}%
\providecommand \href@noop [0]{\@secondoftwo}%
\providecommand \href [0]{\begingroup \@sanitize@url \@href}%
\providecommand \@href[1]{\@@startlink{#1}\@@href}%
\providecommand \@@href[1]{\endgroup#1\@@endlink}%
\providecommand \@sanitize@url [0]{\catcode `\\12\catcode `\$12\catcode `\&12\catcode `\#12\catcode `\^12\catcode `\_12\catcode `\%12\relax}%
\providecommand \@@startlink[1]{}%
\providecommand \@@endlink[0]{}%
\providecommand \url  [0]{\begingroup\@sanitize@url \@url }%
\providecommand \@url [1]{\endgroup\@href {#1}{\urlprefix }}%
\providecommand \urlprefix  [0]{URL }%
\providecommand \Eprint [0]{\href }%
\providecommand \doibase [0]{http://dx.doi.org/}%
\providecommand \selectlanguage [0]{\@gobble}%
\providecommand \bibinfo  [0]{\@secondoftwo}%
\providecommand \bibfield  [0]{\@secondoftwo}%
\providecommand \translation [1]{[#1]}%
\providecommand \BibitemOpen [0]{}%
\providecommand \bibitemStop [0]{}%
\providecommand \bibitemNoStop [0]{.\EOS\space}%
\providecommand \EOS [0]{\spacefactor3000\relax}%
\providecommand \BibitemShut  [1]{\csname bibitem#1\endcsname}%
\let\auto@bib@innerbib\@empty
%</preamble>
\bibitem [{\citenamefont {Golovin}\ \emph {et~al.}(2017)\citenamefont {Golovin}, \citenamefont {Solnik}, \citenamefont {Moitra}, \citenamefont {Kochanski}, \citenamefont {Karro},\ and\ \citenamefont {Sculley}}]{10.1145/3097983.3098043}%
  \BibitemOpen
  \bibfield  {author} {\bibinfo {author} {\bibfnamefont {D.}~\bibnamefont {Golovin}}, \bibinfo {author} {\bibfnamefont {B.}~\bibnamefont {Solnik}}, \bibinfo {author} {\bibfnamefont {S.}~\bibnamefont {Moitra}}, \bibinfo {author} {\bibfnamefont {G.}~\bibnamefont {Kochanski}}, \bibinfo {author} {\bibfnamefont {J.}~\bibnamefont {Karro}}, \ and\ \bibinfo {author} {\bibfnamefont {D.}~\bibnamefont {Sculley}},\ }\bibfield  {title} {\enquote {\bibinfo {title} {Google vizier: A service for black-box optimization},}\ }in\ \href {\doibase 10.1145/3097983.3098043} {\emph {\bibinfo {booktitle} {Proceedings of the 23rd ACM SIGKDD International Conference on Knowledge Discovery and Data Mining}}},\ \bibinfo {series and number} {KDD '17}\ (\bibinfo  {publisher} {Association for Computing Machinery},\ \bibinfo {address} {New York, NY, USA},\ \bibinfo {year} {2017})\ pp.\ \bibinfo {pages} {1487--1495}\BibitemShut {NoStop}%
\bibitem [{\citenamefont {Alarie}\ \emph {et~al.}(2021)\citenamefont {Alarie}, \citenamefont {Audet}, \citenamefont {Gheribi}, \citenamefont {Kokkolaras},\ and\ \citenamefont {{Le Digabel}}}]{ALARIE2021100011}%
  \BibitemOpen
  \bibfield  {author} {\bibinfo {author} {\bibfnamefont {S.}~\bibnamefont {Alarie}}, \bibinfo {author} {\bibfnamefont {C.}~\bibnamefont {Audet}}, \bibinfo {author} {\bibfnamefont {A.~E.}\ \bibnamefont {Gheribi}}, \bibinfo {author} {\bibfnamefont {M.}~\bibnamefont {Kokkolaras}}, \ and\ \bibinfo {author} {\bibfnamefont {S.}~\bibnamefont {{Le Digabel}}},\ }\bibfield  {title} {\enquote {\bibinfo {title} {Two decades of blackbox optimization applications},}\ }\href {\doibase https://doi.org/10.1016/j.ejco.2021.100011} {\bibfield  {journal} {\bibinfo  {journal} {EURO Journal on Computational Optimization}\ }\textbf {\bibinfo {volume} {9}},\ \bibinfo {pages} {100011} (\bibinfo {year} {2021})}\BibitemShut {NoStop}%
\bibitem [{\citenamefont {Terayama}\ \emph {et~al.}(2021)\citenamefont {Terayama}, \citenamefont {Sumita}, \citenamefont {Tamura},\ and\ \citenamefont {Tsuda}}]{Terayama:2021aa}%
  \BibitemOpen
  \bibfield  {author} {\bibinfo {author} {\bibfnamefont {K.}~\bibnamefont {Terayama}}, \bibinfo {author} {\bibfnamefont {M.}~\bibnamefont {Sumita}}, \bibinfo {author} {\bibfnamefont {R.}~\bibnamefont {Tamura}}, \ and\ \bibinfo {author} {\bibfnamefont {K.}~\bibnamefont {Tsuda}},\ }\bibfield  {title} {\enquote {\bibinfo {title} {Black-box optimization for automated discovery},}\ }\href {\doibase 10.1021/acs.accounts.0c00713} {\bibfield  {journal} {\bibinfo  {journal} {Accounts of Chemical Research}\ }\textbf {\bibinfo {volume} {54}},\ \bibinfo {pages} {1334--1346} (\bibinfo {year} {2021})}\BibitemShut {NoStop}%
\bibitem [{\citenamefont {Kumagai}\ and\ \citenamefont {Yasuda}(2023)}]{Kumagai2023}%
  \BibitemOpen
  \bibfield  {author} {\bibinfo {author} {\bibfnamefont {W.}~\bibnamefont {Kumagai}}\ and\ \bibinfo {author} {\bibfnamefont {K.}~\bibnamefont {Yasuda}},\ }\enquote {\bibinfo {title} {Black-box optimization and its applications},}\ in\ \href {\doibase 10.1007/978-981-19-7776-3_6} {\emph {\bibinfo {booktitle} {Innovative Systems Approach for Facilitating Smarter World}}},\ \bibinfo {editor} {edited by\ \bibinfo {editor} {\bibfnamefont {T.}~\bibnamefont {Kaihara}}, \bibinfo {editor} {\bibfnamefont {H.}~\bibnamefont {Kita}}, \bibinfo {editor} {\bibfnamefont {S.}~\bibnamefont {Takahashi}}, \ and\ \bibinfo {editor} {\bibfnamefont {M.}~\bibnamefont {Funabashi}}}\ (\bibinfo  {publisher} {Springer Nature Singapore},\ \bibinfo {address} {Singapore},\ \bibinfo {year} {2023})\ pp.\ \bibinfo {pages} {81--100}\BibitemShut {NoStop}%
\bibitem [{\citenamefont {Belevitch}(1962)}]{Belevitch1962SummaryOT}%
  \BibitemOpen
  \bibfield  {author} {\bibinfo {author} {\bibfnamefont {V.}~\bibnamefont {Belevitch}},\ }\bibfield  {title} {\enquote {\bibinfo {title} {Summary of the history of circuit theory},}\ }\href {https://api.semanticscholar.org/CorpusID:51666316} {\bibfield  {journal} {\bibinfo  {journal} {Proceedings of the IRE}\ }\textbf {\bibinfo {volume} {50}},\ \bibinfo {pages} {848--855} (\bibinfo {year} {1962})}\BibitemShut {NoStop}%
\bibitem [{\citenamefont {Wiener}(1961)}]{10.7551/mitpress/11810.001.0001}%
  \BibitemOpen
  \bibfield  {author} {\bibinfo {author} {\bibfnamefont {N.}~\bibnamefont {Wiener}},\ }\href {\doibase 10.7551/mitpress/11810.001.0001} {\emph {\bibinfo {title} {{Cybernetics or Control and Communication in the Animal and the Machine}}}}\ (\bibinfo  {publisher} {The MIT Press},\ \bibinfo {year} {1961})\BibitemShut {NoStop}%
\bibitem [{\citenamefont {Frazier}(2018)}]{frazier2018tutorialbayesianoptimization}%
  \BibitemOpen
  \bibfield  {author} {\bibinfo {author} {\bibfnamefont {P.~I.}\ \bibnamefont {Frazier}},\ }\href {https://arxiv.org/abs/1807.02811} {\enquote {\bibinfo {title} {A tutorial on {Bayesian} optimization},}\ } (\bibinfo {year} {2018}),\ \Eprint {http://arxiv.org/abs/1807.02811} {arXiv:1807.02811 [stat.ML]} \BibitemShut {NoStop}%
\bibitem [{\citenamefont {Snoek}, \citenamefont {Larochelle},\ and\ \citenamefont {Adams}(2012)}]{NIPS2012_05311655}%
  \BibitemOpen
  \bibfield  {author} {\bibinfo {author} {\bibfnamefont {J.}~\bibnamefont {Snoek}}, \bibinfo {author} {\bibfnamefont {H.}~\bibnamefont {Larochelle}}, \ and\ \bibinfo {author} {\bibfnamefont {R.~P.}\ \bibnamefont {Adams}},\ }\bibfield  {title} {\enquote {\bibinfo {title} {Practical {Bayesian} optimization of machine learning algorithms},}\ }in\ \href {https://proceedings.neurips.cc/paper_files/paper/2012/file/05311655a15b75fab86956663e1819cd-Paper.pdf} {\emph {\bibinfo {booktitle} {Advances in Neural Information Processing Systems}}},\ Vol.~\bibinfo {volume} {25},\ \bibinfo {editor} {edited by\ \bibinfo {editor} {\bibfnamefont {F.}~\bibnamefont {Pereira}}, \bibinfo {editor} {\bibfnamefont {C.}~\bibnamefont {Burges}}, \bibinfo {editor} {\bibfnamefont {L.}~\bibnamefont {Bottou}}, \ and\ \bibinfo {editor} {\bibfnamefont {K.}~\bibnamefont {Weinberger}}}\ (\bibinfo  {publisher} {Curran Associates, Inc.},\ \bibinfo {year} {2012})\BibitemShut {NoStop}%
\bibitem [{\citenamefont {Shahriari}\ \emph {et~al.}(2016)\citenamefont {Shahriari}, \citenamefont {Swersky}, \citenamefont {Wang}, \citenamefont {Adams},\ and\ \citenamefont {de~Freitas}}]{7352306}%
  \BibitemOpen
  \bibfield  {author} {\bibinfo {author} {\bibfnamefont {B.}~\bibnamefont {Shahriari}}, \bibinfo {author} {\bibfnamefont {K.}~\bibnamefont {Swersky}}, \bibinfo {author} {\bibfnamefont {Z.}~\bibnamefont {Wang}}, \bibinfo {author} {\bibfnamefont {R.~P.}\ \bibnamefont {Adams}}, \ and\ \bibinfo {author} {\bibfnamefont {N.}~\bibnamefont {de~Freitas}},\ }\bibfield  {title} {\enquote {\bibinfo {title} {Taking the human out of the loop: A review of {Bayesian} optimization},}\ }\href {\doibase 10.1109/JPROC.2015.2494218} {\bibfield  {journal} {\bibinfo  {journal} {Proceedings of the IEEE}\ }\textbf {\bibinfo {volume} {104}},\ \bibinfo {pages} {148--175} (\bibinfo {year} {2016})}\BibitemShut {NoStop}%
\bibitem [{\citenamefont {Wang}\ \emph {et~al.}(2023)\citenamefont {Wang}, \citenamefont {Jin}, \citenamefont {Schmitt},\ and\ \citenamefont {Olhofer}}]{10.1145/3582078}%
  \BibitemOpen
  \bibfield  {author} {\bibinfo {author} {\bibfnamefont {X.}~\bibnamefont {Wang}}, \bibinfo {author} {\bibfnamefont {Y.}~\bibnamefont {Jin}}, \bibinfo {author} {\bibfnamefont {S.}~\bibnamefont {Schmitt}}, \ and\ \bibinfo {author} {\bibfnamefont {M.}~\bibnamefont {Olhofer}},\ }\bibfield  {title} {\enquote {\bibinfo {title} {Recent advances in {Bayesian} optimization},}\ }\href {https://doi.org/10.1145/3582078} {\bibfield  {journal} {\bibinfo  {journal} {ACM Comput. Surv.}\ }\textbf {\bibinfo {volume} {55}} (\bibinfo {year} {2023})}\BibitemShut {NoStop}%
\bibitem [{\citenamefont {Seko}\ \emph {et~al.}(2015)\citenamefont {Seko}, \citenamefont {Togo}, \citenamefont {Hayashi}, \citenamefont {Tsuda}, \citenamefont {Chaput},\ and\ \citenamefont {Tanaka}}]{PhysRevLett.115.205901}%
  \BibitemOpen
  \bibfield  {author} {\bibinfo {author} {\bibfnamefont {A.}~\bibnamefont {Seko}}, \bibinfo {author} {\bibfnamefont {A.}~\bibnamefont {Togo}}, \bibinfo {author} {\bibfnamefont {H.}~\bibnamefont {Hayashi}}, \bibinfo {author} {\bibfnamefont {K.}~\bibnamefont {Tsuda}}, \bibinfo {author} {\bibfnamefont {L.}~\bibnamefont {Chaput}}, \ and\ \bibinfo {author} {\bibfnamefont {I.}~\bibnamefont {Tanaka}},\ }\bibfield  {title} {\enquote {\bibinfo {title} {Prediction of low-thermal-conductivity compounds with first-principles anharmonic lattice-dynamics calculations and {Bayesian} optimization},}\ }\href {\doibase 10.1103/PhysRevLett.115.205901} {\bibfield  {journal} {\bibinfo  {journal} {Physical Review Letters}\ }\textbf {\bibinfo {volume} {115}},\ \bibinfo {pages} {205901} (\bibinfo {year} {2015})}\BibitemShut {NoStop}%
\bibitem [{\citenamefont {Ju}\ \emph {et~al.}(2017)\citenamefont {Ju}, \citenamefont {Shiga}, \citenamefont {Feng}, \citenamefont {Hou}, \citenamefont {Tsuda},\ and\ \citenamefont {Shiomi}}]{PhysRevX.7.021024}%
  \BibitemOpen
  \bibfield  {author} {\bibinfo {author} {\bibfnamefont {S.}~\bibnamefont {Ju}}, \bibinfo {author} {\bibfnamefont {T.}~\bibnamefont {Shiga}}, \bibinfo {author} {\bibfnamefont {L.}~\bibnamefont {Feng}}, \bibinfo {author} {\bibfnamefont {Z.}~\bibnamefont {Hou}}, \bibinfo {author} {\bibfnamefont {K.}~\bibnamefont {Tsuda}}, \ and\ \bibinfo {author} {\bibfnamefont {J.}~\bibnamefont {Shiomi}},\ }\bibfield  {title} {\enquote {\bibinfo {title} {Designing nanostructures for phonon transport via {Bayesian} optimization},}\ }\href {\doibase 10.1103/PhysRevX.7.021024} {\bibfield  {journal} {\bibinfo  {journal} {Physical Review X}\ }\textbf {\bibinfo {volume} {7}},\ \bibinfo {pages} {021024} (\bibinfo {year} {2017})}\BibitemShut {NoStop}%
\bibitem [{\citenamefont {Jalas}\ \emph {et~al.}(2021)\citenamefont {Jalas}, \citenamefont {Kirchen}, \citenamefont {Messner}, \citenamefont {Winkler}, \citenamefont {H\"ubner}, \citenamefont {Dirkwinkel}, \citenamefont {Schnepp}, \citenamefont {Lehe},\ and\ \citenamefont {Maier}}]{PhysRevLett.126.104801}%
  \BibitemOpen
  \bibfield  {author} {\bibinfo {author} {\bibfnamefont {S.}~\bibnamefont {Jalas}}, \bibinfo {author} {\bibfnamefont {M.}~\bibnamefont {Kirchen}}, \bibinfo {author} {\bibfnamefont {P.}~\bibnamefont {Messner}}, \bibinfo {author} {\bibfnamefont {P.}~\bibnamefont {Winkler}}, \bibinfo {author} {\bibfnamefont {L.}~\bibnamefont {H\"ubner}}, \bibinfo {author} {\bibfnamefont {J.}~\bibnamefont {Dirkwinkel}}, \bibinfo {author} {\bibfnamefont {M.}~\bibnamefont {Schnepp}}, \bibinfo {author} {\bibfnamefont {R.}~\bibnamefont {Lehe}}, \ and\ \bibinfo {author} {\bibfnamefont {A.~R.}\ \bibnamefont {Maier}},\ }\bibfield  {title} {\enquote {\bibinfo {title} {Bayesian optimization of a laser-plasma accelerator},}\ }\href {\doibase 10.1103/PhysRevLett.126.104801} {\bibfield  {journal} {\bibinfo  {journal} {Physical Review Letters}\ }\textbf {\bibinfo {volume} {126}},\ \bibinfo {pages} {104801} (\bibinfo {year} {2021})}\BibitemShut {NoStop}%
\bibitem [{\citenamefont {Yu}\ \emph {et~al.}(2020)\citenamefont {Yu}, \citenamefont {Yang}, \citenamefont {Wu},\ and\ \citenamefont {Marom}}]{Yu:2020aa}%
  \BibitemOpen
  \bibfield  {author} {\bibinfo {author} {\bibfnamefont {M.}~\bibnamefont {Yu}}, \bibinfo {author} {\bibfnamefont {S.}~\bibnamefont {Yang}}, \bibinfo {author} {\bibfnamefont {C.}~\bibnamefont {Wu}}, \ and\ \bibinfo {author} {\bibfnamefont {N.}~\bibnamefont {Marom}},\ }\bibfield  {title} {\enquote {\bibinfo {title} {Machine learning the hubbard {U} parameter in {DFT+U} using {Bayesian} optimization},}\ }\href {\doibase 10.1038/s41524-020-00446-9} {\bibfield  {journal} {\bibinfo  {journal} {npj Computational Materials}\ }\textbf {\bibinfo {volume} {6}},\ \bibinfo {pages} {180} (\bibinfo {year} {2020})}\BibitemShut {NoStop}%
\bibitem [{\citenamefont {H{\"a}se}\ \emph {et~al.}(2018)\citenamefont {H{\"a}se}, \citenamefont {Roch}, \citenamefont {Kreisbeck},\ and\ \citenamefont {Aspuru-Guzik}}]{Hase:2018aa}%
  \BibitemOpen
  \bibfield  {author} {\bibinfo {author} {\bibfnamefont {F.}~\bibnamefont {H{\"a}se}}, \bibinfo {author} {\bibfnamefont {L.~M.}\ \bibnamefont {Roch}}, \bibinfo {author} {\bibfnamefont {C.}~\bibnamefont {Kreisbeck}}, \ and\ \bibinfo {author} {\bibfnamefont {A.}~\bibnamefont {Aspuru-Guzik}},\ }\bibfield  {title} {\enquote {\bibinfo {title} {Phoenics: A {Bayesian} optimizer for chemistry},}\ }\href {\doibase 10.1021/acscentsci.8b00307} {\bibfield  {journal} {\bibinfo  {journal} {ACS Central Science}\ }\textbf {\bibinfo {volume} {4}},\ \bibinfo {pages} {1134--1145} (\bibinfo {year} {2018})}\BibitemShut {NoStop}%
\bibitem [{\citenamefont {Homma}\ \emph {et~al.}(2020)\citenamefont {Homma}, \citenamefont {Liu}, \citenamefont {Sumita}, \citenamefont {Tamura}, \citenamefont {Fushimi}, \citenamefont {Iwata}, \citenamefont {Tsuda},\ and\ \citenamefont {Kaneta}}]{Homma:2020aa}%
  \BibitemOpen
  \bibfield  {author} {\bibinfo {author} {\bibfnamefont {K.}~\bibnamefont {Homma}}, \bibinfo {author} {\bibfnamefont {Y.}~\bibnamefont {Liu}}, \bibinfo {author} {\bibfnamefont {M.}~\bibnamefont {Sumita}}, \bibinfo {author} {\bibfnamefont {R.}~\bibnamefont {Tamura}}, \bibinfo {author} {\bibfnamefont {N.}~\bibnamefont {Fushimi}}, \bibinfo {author} {\bibfnamefont {J.}~\bibnamefont {Iwata}}, \bibinfo {author} {\bibfnamefont {K.}~\bibnamefont {Tsuda}}, \ and\ \bibinfo {author} {\bibfnamefont {C.}~\bibnamefont {Kaneta}},\ }\bibfield  {title} {\enquote {\bibinfo {title} {Optimization of a heterogeneous ternary {Li$_3$PO$_4$}--{Li$_3$BO$_3$}--{Li$_2$SO$_4$} mixture for {Li}-ion conductivity by machine learning},}\ }\href {\doibase 10.1021/acs.jpcc.9b11654} {\bibfield  {journal} {\bibinfo  {journal} {The Journal of Physical Chemistry C}\ }\textbf {\bibinfo {volume} {124}},\ \bibinfo {pages} {12865--12870} (\bibinfo {year} {2020})}\BibitemShut {NoStop}%
\bibitem [{\citenamefont {Fang}\ \emph {et~al.}(2021)\citenamefont {Fang}, \citenamefont {Makkonen}, \citenamefont {Todorovi{\'c}}, \citenamefont {Rinke},\ and\ \citenamefont {Chen}}]{Fang:2021aa}%
  \BibitemOpen
  \bibfield  {author} {\bibinfo {author} {\bibfnamefont {L.}~\bibnamefont {Fang}}, \bibinfo {author} {\bibfnamefont {E.}~\bibnamefont {Makkonen}}, \bibinfo {author} {\bibfnamefont {M.}~\bibnamefont {Todorovi{\'c}}}, \bibinfo {author} {\bibfnamefont {P.}~\bibnamefont {Rinke}}, \ and\ \bibinfo {author} {\bibfnamefont {X.}~\bibnamefont {Chen}},\ }\bibfield  {title} {\enquote {\bibinfo {title} {Efficient amino acid conformer search with {Bayesian} optimization},}\ }\href {\doibase 10.1021/acs.jctc.0c00648} {\bibfield  {journal} {\bibinfo  {journal} {Journal of Chemical Theory and Computation}\ }\textbf {\bibinfo {volume} {17}},\ \bibinfo {pages} {1955--1966} (\bibinfo {year} {2021})}\BibitemShut {NoStop}%
\bibitem [{\citenamefont {Wang}\ and\ \citenamefont {Dowling}(2022)}]{WANG2022100728}%
  \BibitemOpen
  \bibfield  {author} {\bibinfo {author} {\bibfnamefont {K.}~\bibnamefont {Wang}}\ and\ \bibinfo {author} {\bibfnamefont {A.~W.}\ \bibnamefont {Dowling}},\ }\bibfield  {title} {\enquote {\bibinfo {title} {Bayesian optimization for chemical products and functional materials},}\ }\href {\doibase https://doi.org/10.1016/j.coche.2021.100728} {\bibfield  {journal} {\bibinfo  {journal} {Current Opinion in Chemical Engineering}\ }\textbf {\bibinfo {volume} {36}},\ \bibinfo {pages} {100728} (\bibinfo {year} {2022})}\BibitemShut {NoStop}%
\bibitem [{\citenamefont {Wakabayashi}\ \emph {et~al.}(2019)\citenamefont {Wakabayashi}, \citenamefont {Otsuka}, \citenamefont {Krockenberger}, \citenamefont {Sawada}, \citenamefont {Taniyasu},\ and\ \citenamefont {Yamamoto}}]{10.1063/1.5123019}%
  \BibitemOpen
  \bibfield  {author} {\bibinfo {author} {\bibfnamefont {Y.~K.}\ \bibnamefont {Wakabayashi}}, \bibinfo {author} {\bibfnamefont {T.}~\bibnamefont {Otsuka}}, \bibinfo {author} {\bibfnamefont {Y.}~\bibnamefont {Krockenberger}}, \bibinfo {author} {\bibfnamefont {H.}~\bibnamefont {Sawada}}, \bibinfo {author} {\bibfnamefont {Y.}~\bibnamefont {Taniyasu}}, \ and\ \bibinfo {author} {\bibfnamefont {H.}~\bibnamefont {Yamamoto}},\ }\bibfield  {title} {\enquote {\bibinfo {title} {{Machine-learning-assisted thin-film growth: Bayesian optimization in molecular beam epitaxy of SrRuO$_3$ thin films}},}\ }\href {\doibase 10.1063/1.5123019} {\bibfield  {journal} {\bibinfo  {journal} {APL Materials}\ }\textbf {\bibinfo {volume} {7}},\ \bibinfo {pages} {101114} (\bibinfo {year} {2019})}\BibitemShut {NoStop}%
\bibitem [{\citenamefont {Zhang}, \citenamefont {Apley},\ and\ \citenamefont {Chen}(2020)}]{Zhang:2020aa}%
  \BibitemOpen
  \bibfield  {author} {\bibinfo {author} {\bibfnamefont {Y.}~\bibnamefont {Zhang}}, \bibinfo {author} {\bibfnamefont {D.~W.}\ \bibnamefont {Apley}}, \ and\ \bibinfo {author} {\bibfnamefont {W.}~\bibnamefont {Chen}},\ }\bibfield  {title} {\enquote {\bibinfo {title} {Bayesian optimization for materials design with mixed quantitative and qualitative variables},}\ }\href {\doibase 10.1038/s41598-020-60652-9} {\bibfield  {journal} {\bibinfo  {journal} {Scientific Reports}\ }\textbf {\bibinfo {volume} {10}},\ \bibinfo {pages} {4924} (\bibinfo {year} {2020})}\BibitemShut {NoStop}%
\bibitem [{\citenamefont {Tamura}\ \emph {et~al.}(2021)\citenamefont {Tamura}, \citenamefont {Osada}, \citenamefont {Minagawa}, \citenamefont {Kohata}, \citenamefont {Hirosawa}, \citenamefont {Tsuda},\ and\ \citenamefont {Kawagishi}}]{TAMURA2021109290}%
  \BibitemOpen
  \bibfield  {author} {\bibinfo {author} {\bibfnamefont {R.}~\bibnamefont {Tamura}}, \bibinfo {author} {\bibfnamefont {T.}~\bibnamefont {Osada}}, \bibinfo {author} {\bibfnamefont {K.}~\bibnamefont {Minagawa}}, \bibinfo {author} {\bibfnamefont {T.}~\bibnamefont {Kohata}}, \bibinfo {author} {\bibfnamefont {M.}~\bibnamefont {Hirosawa}}, \bibinfo {author} {\bibfnamefont {K.}~\bibnamefont {Tsuda}}, \ and\ \bibinfo {author} {\bibfnamefont {K.}~\bibnamefont {Kawagishi}},\ }\bibfield  {title} {\enquote {\bibinfo {title} {Machine learning-driven optimization in powder manufacturing of {Ni}-{Co} based superalloy},}\ }\href {\doibase https://doi.org/10.1016/j.matdes.2020.109290} {\bibfield  {journal} {\bibinfo  {journal} {Materials and Design}\ }\textbf {\bibinfo {volume} {198}},\ \bibinfo {pages} {109290} (\bibinfo {year} {2021})}\BibitemShut {NoStop}%
\bibitem [{\citenamefont {Liang}\ \emph {et~al.}(2021)\citenamefont {Liang}, \citenamefont {Gongora}, \citenamefont {Ren}, \citenamefont {Tiihonen}, \citenamefont {Liu}, \citenamefont {Sun}, \citenamefont {Deneault}, \citenamefont {Bash}, \citenamefont {Mekki-Berrada}, \citenamefont {Khan}, \citenamefont {Hippalgaonkar}, \citenamefont {Maruyama}, \citenamefont {Brown}, \citenamefont {Fisher~III},\ and\ \citenamefont {Buonassisi}}]{Liang:2021aa}%
  \BibitemOpen
  \bibfield  {author} {\bibinfo {author} {\bibfnamefont {Q.}~\bibnamefont {Liang}}, \bibinfo {author} {\bibfnamefont {A.~E.}\ \bibnamefont {Gongora}}, \bibinfo {author} {\bibfnamefont {Z.}~\bibnamefont {Ren}}, \bibinfo {author} {\bibfnamefont {A.}~\bibnamefont {Tiihonen}}, \bibinfo {author} {\bibfnamefont {Z.}~\bibnamefont {Liu}}, \bibinfo {author} {\bibfnamefont {S.}~\bibnamefont {Sun}}, \bibinfo {author} {\bibfnamefont {J.~R.}\ \bibnamefont {Deneault}}, \bibinfo {author} {\bibfnamefont {D.}~\bibnamefont {Bash}}, \bibinfo {author} {\bibfnamefont {F.}~\bibnamefont {Mekki-Berrada}}, \bibinfo {author} {\bibfnamefont {S.~A.}\ \bibnamefont {Khan}}, \bibinfo {author} {\bibfnamefont {K.}~\bibnamefont {Hippalgaonkar}}, \bibinfo {author} {\bibfnamefont {B.}~\bibnamefont {Maruyama}}, \bibinfo {author} {\bibfnamefont {K.~A.}\ \bibnamefont {Brown}}, \bibinfo {author} {\bibfnamefont {J.}~\bibnamefont {Fisher~III}}, \ and\ \bibinfo {author} {\bibfnamefont {T.}~\bibnamefont {Buonassisi}},\ }\bibfield  {title} {\enquote
  {\bibinfo {title} {Benchmarking the performance of {Bayesian} optimization across multiple experimental materials science domains},}\ }\href {\doibase 10.1038/s41524-021-00656-9} {\bibfield  {journal} {\bibinfo  {journal} {npj Computational Materials}\ }\textbf {\bibinfo {volume} {7}},\ \bibinfo {pages} {188} (\bibinfo {year} {2021})}\BibitemShut {NoStop}%
\bibitem [{\citenamefont {Lucas}(2014)}]{10.3389/fphy.2014.00005}%
  \BibitemOpen
  \bibfield  {author} {\bibinfo {author} {\bibfnamefont {A.}~\bibnamefont {Lucas}},\ }\bibfield  {title} {\enquote {\bibinfo {title} {Ising formulations of many {NP} problems},}\ }\href {https://www.frontiersin.org/journals/physics/articles/10.3389/fphy.2014.00005} {\bibfield  {journal} {\bibinfo  {journal} {Frontiers in Physics}\ }\textbf {\bibinfo {volume} {2}} (\bibinfo {year} {2014})}\BibitemShut {NoStop}%
\bibitem [{\citenamefont {Cook}\ \emph {et~al.}(1998)\citenamefont {Cook}, \citenamefont {Cunningham}, \citenamefont {Pulleyblank},\ and\ \citenamefont {Schrijver}}]{Combinatorial_book}%
  \BibitemOpen
  \bibfield  {author} {\bibinfo {author} {\bibfnamefont {W.~J.}\ \bibnamefont {Cook}}, \bibinfo {author} {\bibfnamefont {W.~H.}\ \bibnamefont {Cunningham}}, \bibinfo {author} {\bibfnamefont {W.~R.}\ \bibnamefont {Pulleyblank}}, \ and\ \bibinfo {author} {\bibfnamefont {A.}~\bibnamefont {Schrijver}},\ }\href@noop {} {\emph {\bibinfo {title} {Combinatorial Optimization}}}\ (\bibinfo  {publisher} {John Wiley \& Sons, Inc.},\ \bibinfo {year} {1998})\BibitemShut {NoStop}%
\bibitem [{\citenamefont {Mohseni}, \citenamefont {McMahon},\ and\ \citenamefont {Byrnes}(2022)}]{Mohseni:2022aa}%
  \BibitemOpen
  \bibfield  {author} {\bibinfo {author} {\bibfnamefont {N.}~\bibnamefont {Mohseni}}, \bibinfo {author} {\bibfnamefont {P.~L.}\ \bibnamefont {McMahon}}, \ and\ \bibinfo {author} {\bibfnamefont {T.}~\bibnamefont {Byrnes}},\ }\bibfield  {title} {\enquote {\bibinfo {title} {Ising machines as hardware solvers of combinatorial optimization problems},}\ }\href {\doibase 10.1038/s42254-022-00440-8} {\bibfield  {journal} {\bibinfo  {journal} {Nature Reviews Physics}\ }\textbf {\bibinfo {volume} {4}},\ \bibinfo {pages} {363--379} (\bibinfo {year} {2022})}\BibitemShut {NoStop}%
\bibitem [{\citenamefont {Tanaka}, \citenamefont {Tamura},\ and\ \citenamefont {Chakrabarti}(2017)}]{10.5555/3159044}%
  \BibitemOpen
  \bibfield  {author} {\bibinfo {author} {\bibfnamefont {S.}~\bibnamefont {Tanaka}}, \bibinfo {author} {\bibfnamefont {R.}~\bibnamefont {Tamura}}, \ and\ \bibinfo {author} {\bibfnamefont {B.~K.}\ \bibnamefont {Chakrabarti}},\ }\href@noop {} {\emph {\bibinfo {title} {Quantum Spin Glasses, Annealing and Computation}}},\ \bibinfo {edition} {1st}\ ed.\ (\bibinfo  {publisher} {Cambridge University Press},\ \bibinfo {address} {USA},\ \bibinfo {year} {2017})\BibitemShut {NoStop}%
\bibitem [{\citenamefont {Tanahashi}\ \emph {et~al.}(2019)\citenamefont {Tanahashi}, \citenamefont {Takayanagi}, \citenamefont {Motohashi},\ and\ \citenamefont {Tanaka}}]{doi:10.7566/JPSJ.88.061010}%
  \BibitemOpen
  \bibfield  {author} {\bibinfo {author} {\bibfnamefont {K.}~\bibnamefont {Tanahashi}}, \bibinfo {author} {\bibfnamefont {S.}~\bibnamefont {Takayanagi}}, \bibinfo {author} {\bibfnamefont {T.}~\bibnamefont {Motohashi}}, \ and\ \bibinfo {author} {\bibfnamefont {S.}~\bibnamefont {Tanaka}},\ }\bibfield  {title} {\enquote {\bibinfo {title} {Application of {Ising} machines and a software development for {Ising} machines},}\ }\href {\doibase 10.7566/JPSJ.88.061010} {\bibfield  {journal} {\bibinfo  {journal} {Journal of the Physical Society of Japan}\ }\textbf {\bibinfo {volume} {88}},\ \bibinfo {pages} {061010} (\bibinfo {year} {2019})}\BibitemShut {NoStop}%
\bibitem [{\citenamefont {Yarkoni}\ \emph {et~al.}(2022)\citenamefont {Yarkoni}, \citenamefont {Raponi}, \citenamefont {B{\"a}ck},\ and\ \citenamefont {Schmitt}}]{Yarkoni_2022}%
  \BibitemOpen
  \bibfield  {author} {\bibinfo {author} {\bibfnamefont {S.}~\bibnamefont {Yarkoni}}, \bibinfo {author} {\bibfnamefont {E.}~\bibnamefont {Raponi}}, \bibinfo {author} {\bibfnamefont {T.}~\bibnamefont {B{\"a}ck}}, \ and\ \bibinfo {author} {\bibfnamefont {S.}~\bibnamefont {Schmitt}},\ }\bibfield  {title} {\enquote {\bibinfo {title} {Quantum annealing for industry applications: {Introduction} and review},}\ }\href {\doibase 10.1088/1361-6633/ac8c54} {\bibfield  {journal} {\bibinfo  {journal} {Reports on Progress in Physics}\ }\textbf {\bibinfo {volume} {85}},\ \bibinfo {pages} {104001} (\bibinfo {year} {2022})}\BibitemShut {NoStop}%
\bibitem [{\citenamefont {Wilson}(1975)}]{RevModPhys.47.773}%
  \BibitemOpen
  \bibfield  {author} {\bibinfo {author} {\bibfnamefont {K.~G.}\ \bibnamefont {Wilson}},\ }\bibfield  {title} {\enquote {\bibinfo {title} {The renormalization group: Critical phenomena and the kondo problem},}\ }\href {\doibase 10.1103/RevModPhys.47.773} {\bibfield  {journal} {\bibinfo  {journal} {Rev. Mod. Phys.}\ }\textbf {\bibinfo {volume} {47}},\ \bibinfo {pages} {773--840} (\bibinfo {year} {1975})}\BibitemShut {NoStop}%
\bibitem [{\citenamefont {Pelissetto}\ and\ \citenamefont {Vicari}(2002)}]{PELISSETTO2002549}%
  \BibitemOpen
  \bibfield  {author} {\bibinfo {author} {\bibfnamefont {A.}~\bibnamefont {Pelissetto}}\ and\ \bibinfo {author} {\bibfnamefont {E.}~\bibnamefont {Vicari}},\ }\bibfield  {title} {\enquote {\bibinfo {title} {Critical phenomena and renormalization-group theory},}\ }\href {\doibase https://doi.org/10.1016/S0370-1573(02)00219-3} {\bibfield  {journal} {\bibinfo  {journal} {Physics Reports}\ }\textbf {\bibinfo {volume} {368}},\ \bibinfo {pages} {549--727} (\bibinfo {year} {2002})}\BibitemShut {NoStop}%
\bibitem [{\citenamefont {Kochenberger}\ \emph {et~al.}(2014)\citenamefont {Kochenberger}, \citenamefont {Hao}, \citenamefont {Glover}, \citenamefont {Lewis}, \citenamefont {L{\"u}}, \citenamefont {Wang},\ and\ \citenamefont {Wang}}]{Kochenberger:2014aa}%
  \BibitemOpen
  \bibfield  {author} {\bibinfo {author} {\bibfnamefont {G.}~\bibnamefont {Kochenberger}}, \bibinfo {author} {\bibfnamefont {J.-K.}\ \bibnamefont {Hao}}, \bibinfo {author} {\bibfnamefont {F.}~\bibnamefont {Glover}}, \bibinfo {author} {\bibfnamefont {M.}~\bibnamefont {Lewis}}, \bibinfo {author} {\bibfnamefont {Z.}~\bibnamefont {L{\"u}}}, \bibinfo {author} {\bibfnamefont {H.}~\bibnamefont {Wang}}, \ and\ \bibinfo {author} {\bibfnamefont {Y.}~\bibnamefont {Wang}},\ }\bibfield  {title} {\enquote {\bibinfo {title} {The unconstrained binary quadratic programming problem: a survey},}\ }\href {\doibase 10.1007/s10878-014-9734-0} {\bibfield  {journal} {\bibinfo  {journal} {Journal of Combinatorial Optimization}\ }\textbf {\bibinfo {volume} {28}},\ \bibinfo {pages} {58--81} (\bibinfo {year} {2014})}\BibitemShut {NoStop}%
\bibitem [{\citenamefont {Yamaoka}\ \emph {et~al.}(2016)\citenamefont {Yamaoka}, \citenamefont {Yoshimura}, \citenamefont {Hayashi}, \citenamefont {Okuyama}, \citenamefont {Aoki},\ and\ \citenamefont {Mizuno}}]{7350099}%
  \BibitemOpen
  \bibfield  {author} {\bibinfo {author} {\bibfnamefont {M.}~\bibnamefont {Yamaoka}}, \bibinfo {author} {\bibfnamefont {C.}~\bibnamefont {Yoshimura}}, \bibinfo {author} {\bibfnamefont {M.}~\bibnamefont {Hayashi}}, \bibinfo {author} {\bibfnamefont {T.}~\bibnamefont {Okuyama}}, \bibinfo {author} {\bibfnamefont {H.}~\bibnamefont {Aoki}}, \ and\ \bibinfo {author} {\bibfnamefont {H.}~\bibnamefont {Mizuno}},\ }\bibfield  {title} {\enquote {\bibinfo {title} {A 20k-spin {Ising} chip to solve combinatorial optimization problems with {CMOS} annealing},}\ }\href {\doibase 10.1109/JSSC.2015.2498601} {\bibfield  {journal} {\bibinfo  {journal} {IEEE Journal of Solid-State Circuits}\ }\textbf {\bibinfo {volume} {51}},\ \bibinfo {pages} {303--309} (\bibinfo {year} {2016})}\BibitemShut {NoStop}%
\bibitem [{\citenamefont {Matsubara}\ \emph {et~al.}(2018)\citenamefont {Matsubara}, \citenamefont {Tamura}, \citenamefont {Takatsu}, \citenamefont {Yoo}, \citenamefont {Vatankhahghadim}, \citenamefont {Yamasaki}, \citenamefont {Miyazawa}, \citenamefont {Tsukamoto}, \citenamefont {Watanabe}, \citenamefont {Takemoto},\ and\ \citenamefont {Sheikholeslami}}]{10.1007/978-3-319-61566-0_39}%
  \BibitemOpen
  \bibfield  {author} {\bibinfo {author} {\bibfnamefont {S.}~\bibnamefont {Matsubara}}, \bibinfo {author} {\bibfnamefont {H.}~\bibnamefont {Tamura}}, \bibinfo {author} {\bibfnamefont {M.}~\bibnamefont {Takatsu}}, \bibinfo {author} {\bibfnamefont {D.}~\bibnamefont {Yoo}}, \bibinfo {author} {\bibfnamefont {B.}~\bibnamefont {Vatankhahghadim}}, \bibinfo {author} {\bibfnamefont {H.}~\bibnamefont {Yamasaki}}, \bibinfo {author} {\bibfnamefont {T.}~\bibnamefont {Miyazawa}}, \bibinfo {author} {\bibfnamefont {S.}~\bibnamefont {Tsukamoto}}, \bibinfo {author} {\bibfnamefont {Y.}~\bibnamefont {Watanabe}}, \bibinfo {author} {\bibfnamefont {K.}~\bibnamefont {Takemoto}}, \ and\ \bibinfo {author} {\bibfnamefont {A.}~\bibnamefont {Sheikholeslami}},\ }\bibfield  {title} {\enquote {\bibinfo {title} {Ising-model optimizer with parallel-trial bit-sieve engine},}\ }in\ \href@noop {} {\emph {\bibinfo {booktitle} {Complex, Intelligent, and Software Intensive Systems}}},\ \bibinfo {editor} {edited by\ \bibinfo {editor} {\bibfnamefont
  {L.}~\bibnamefont {Barolli}}\ and\ \bibinfo {editor} {\bibfnamefont {O.}~\bibnamefont {Terzo}}}\ (\bibinfo  {publisher} {Springer International Publishing},\ \bibinfo {address} {Cham},\ \bibinfo {year} {2018})\ pp.\ \bibinfo {pages} {432--438}\BibitemShut {NoStop}%
\bibitem [{\citenamefont {Goto}(2016)}]{Goto:2016aa}%
  \BibitemOpen
  \bibfield  {author} {\bibinfo {author} {\bibfnamefont {H.}~\bibnamefont {Goto}},\ }\bibfield  {title} {\enquote {\bibinfo {title} {Bifurcation-based adiabatic quantum computation with a nonlinear oscillator network},}\ }\href {\doibase 10.1038/srep21686} {\bibfield  {journal} {\bibinfo  {journal} {Scientific Reports}\ }\textbf {\bibinfo {volume} {6}},\ \bibinfo {pages} {21686} (\bibinfo {year} {2016})}\BibitemShut {NoStop}%
\bibitem [{\citenamefont {Goto}, \citenamefont {Tatsumura},\ and\ \citenamefont {Dixon}(2019)}]{doi:10.1126/sciadv.aav2372}%
  \BibitemOpen
  \bibfield  {author} {\bibinfo {author} {\bibfnamefont {H.}~\bibnamefont {Goto}}, \bibinfo {author} {\bibfnamefont {K.}~\bibnamefont {Tatsumura}}, \ and\ \bibinfo {author} {\bibfnamefont {A.~R.}\ \bibnamefont {Dixon}},\ }\bibfield  {title} {\enquote {\bibinfo {title} {Combinatorial optimization by simulating adiabatic bifurcations in nonlinear hamiltonian systems},}\ }\href {\doibase 10.1126/sciadv.aav2372} {\bibfield  {journal} {\bibinfo  {journal} {Science Advances}\ }\textbf {\bibinfo {volume} {5}},\ \bibinfo {pages} {eaav2372} (\bibinfo {year} {2019})}\BibitemShut {NoStop}%
\bibitem [{\citenamefont {Johnson}\ \emph {et~al.}(2011)\citenamefont {Johnson}, \citenamefont {Amin}, \citenamefont {Gildert}, \citenamefont {Lanting}, \citenamefont {Hamze}, \citenamefont {Dickson}, \citenamefont {Harris}, \citenamefont {Berkley}, \citenamefont {Johansson}, \citenamefont {Bunyk}, \citenamefont {Chapple}, \citenamefont {Enderud}, \citenamefont {Hilton}, \citenamefont {Karimi}, \citenamefont {Ladizinsky}, \citenamefont {Ladizinsky}, \citenamefont {Oh}, \citenamefont {Perminov}, \citenamefont {Rich}, \citenamefont {Thom}, \citenamefont {Tolkacheva}, \citenamefont {Truncik}, \citenamefont {Uchaikin}, \citenamefont {Wang}, \citenamefont {Wilson},\ and\ \citenamefont {Rose}}]{Johnson:2011aa}%
  \BibitemOpen
  \bibfield  {author} {\bibinfo {author} {\bibfnamefont {M.~W.}\ \bibnamefont {Johnson}}, \bibinfo {author} {\bibfnamefont {M.~H.~S.}\ \bibnamefont {Amin}}, \bibinfo {author} {\bibfnamefont {S.}~\bibnamefont {Gildert}}, \bibinfo {author} {\bibfnamefont {T.}~\bibnamefont {Lanting}}, \bibinfo {author} {\bibfnamefont {F.}~\bibnamefont {Hamze}}, \bibinfo {author} {\bibfnamefont {N.}~\bibnamefont {Dickson}}, \bibinfo {author} {\bibfnamefont {R.}~\bibnamefont {Harris}}, \bibinfo {author} {\bibfnamefont {A.~J.}\ \bibnamefont {Berkley}}, \bibinfo {author} {\bibfnamefont {J.}~\bibnamefont {Johansson}}, \bibinfo {author} {\bibfnamefont {P.}~\bibnamefont {Bunyk}}, \bibinfo {author} {\bibfnamefont {E.~M.}\ \bibnamefont {Chapple}}, \bibinfo {author} {\bibfnamefont {C.}~\bibnamefont {Enderud}}, \bibinfo {author} {\bibfnamefont {J.~P.}\ \bibnamefont {Hilton}}, \bibinfo {author} {\bibfnamefont {K.}~\bibnamefont {Karimi}}, \bibinfo {author} {\bibfnamefont {E.}~\bibnamefont {Ladizinsky}}, \bibinfo {author} {\bibfnamefont
  {N.}~\bibnamefont {Ladizinsky}}, \bibinfo {author} {\bibfnamefont {T.}~\bibnamefont {Oh}}, \bibinfo {author} {\bibfnamefont {I.}~\bibnamefont {Perminov}}, \bibinfo {author} {\bibfnamefont {C.}~\bibnamefont {Rich}}, \bibinfo {author} {\bibfnamefont {M.~C.}\ \bibnamefont {Thom}}, \bibinfo {author} {\bibfnamefont {E.}~\bibnamefont {Tolkacheva}}, \bibinfo {author} {\bibfnamefont {C.~J.~S.}\ \bibnamefont {Truncik}}, \bibinfo {author} {\bibfnamefont {S.}~\bibnamefont {Uchaikin}}, \bibinfo {author} {\bibfnamefont {J.}~\bibnamefont {Wang}}, \bibinfo {author} {\bibfnamefont {B.}~\bibnamefont {Wilson}}, \ and\ \bibinfo {author} {\bibfnamefont {G.}~\bibnamefont {Rose}},\ }\bibfield  {title} {\enquote {\bibinfo {title} {Quantum annealing with manufactured spins},}\ }\href {\doibase 10.1038/nature10012} {\bibfield  {journal} {\bibinfo  {journal} {Nature}\ }\textbf {\bibinfo {volume} {473}},\ \bibinfo {pages} {194--198} (\bibinfo {year} {2011})}\BibitemShut {NoStop}%
\bibitem [{\citenamefont {Easttom}(2024)}]{Easttom2024}%
  \BibitemOpen
  \bibfield  {author} {\bibinfo {author} {\bibfnamefont {C.}~\bibnamefont {Easttom}},\ }\enquote {\bibinfo {title} {Quantum annealing},}\ in\ \href {\doibase 10.1007/978-3-031-66477-9_9} {\emph {\bibinfo {booktitle} {Hardware for Quantum Computing}}}\ (\bibinfo  {publisher} {Springer Nature Switzerland},\ \bibinfo {address} {Cham},\ \bibinfo {year} {2024})\ pp.\ \bibinfo {pages} {101--112}\BibitemShut {NoStop}%
\bibitem [{\citenamefont {Inagaki}\ \emph {et~al.}(2016)\citenamefont {Inagaki}, \citenamefont {Haribara}, \citenamefont {Igarashi}, \citenamefont {Sonobe}, \citenamefont {Tamate}, \citenamefont {Honjo}, \citenamefont {Marandi}, \citenamefont {McMahon}, \citenamefont {Umeki}, \citenamefont {Enbutsu}, \citenamefont {Tadanaga}, \citenamefont {Takenouchi}, \citenamefont {Aihara}, \citenamefont {Kawarabayashi}, \citenamefont {Inoue}, \citenamefont {Utsunomiya},\ and\ \citenamefont {Takesue}}]{doi:10.1126/science.aah4243}%
  \BibitemOpen
  \bibfield  {author} {\bibinfo {author} {\bibfnamefont {T.}~\bibnamefont {Inagaki}}, \bibinfo {author} {\bibfnamefont {Y.}~\bibnamefont {Haribara}}, \bibinfo {author} {\bibfnamefont {K.}~\bibnamefont {Igarashi}}, \bibinfo {author} {\bibfnamefont {T.}~\bibnamefont {Sonobe}}, \bibinfo {author} {\bibfnamefont {S.}~\bibnamefont {Tamate}}, \bibinfo {author} {\bibfnamefont {T.}~\bibnamefont {Honjo}}, \bibinfo {author} {\bibfnamefont {A.}~\bibnamefont {Marandi}}, \bibinfo {author} {\bibfnamefont {P.~L.}\ \bibnamefont {McMahon}}, \bibinfo {author} {\bibfnamefont {T.}~\bibnamefont {Umeki}}, \bibinfo {author} {\bibfnamefont {K.}~\bibnamefont {Enbutsu}}, \bibinfo {author} {\bibfnamefont {O.}~\bibnamefont {Tadanaga}}, \bibinfo {author} {\bibfnamefont {H.}~\bibnamefont {Takenouchi}}, \bibinfo {author} {\bibfnamefont {K.}~\bibnamefont {Aihara}}, \bibinfo {author} {\bibfnamefont {K.}~\bibnamefont {Kawarabayashi}}, \bibinfo {author} {\bibfnamefont {K.}~\bibnamefont {Inoue}}, \bibinfo {author} {\bibfnamefont
  {S.}~\bibnamefont {Utsunomiya}}, \ and\ \bibinfo {author} {\bibfnamefont {H.}~\bibnamefont {Takesue}},\ }\bibfield  {title} {\enquote {\bibinfo {title} {A coherent {Ising} machine for 2000-node optimization problems},}\ }\href {\doibase 10.1126/science.aah4243} {\bibfield  {journal} {\bibinfo  {journal} {Science}\ }\textbf {\bibinfo {volume} {354}},\ \bibinfo {pages} {603--606} (\bibinfo {year} {2016})}\BibitemShut {NoStop}%
\bibitem [{\citenamefont {Honjo}\ \emph {et~al.}(2021)\citenamefont {Honjo}, \citenamefont {Sonobe}, \citenamefont {Inaba}, \citenamefont {Inagaki}, \citenamefont {Ikuta}, \citenamefont {Yamada}, \citenamefont {Kazama}, \citenamefont {Enbutsu}, \citenamefont {Umeki}, \citenamefont {Kasahara}, \citenamefont {Kawarabayashi},\ and\ \citenamefont {Takesue}}]{doi:10.1126/sciadv.abh0952}%
  \BibitemOpen
  \bibfield  {author} {\bibinfo {author} {\bibfnamefont {T.}~\bibnamefont {Honjo}}, \bibinfo {author} {\bibfnamefont {T.}~\bibnamefont {Sonobe}}, \bibinfo {author} {\bibfnamefont {K.}~\bibnamefont {Inaba}}, \bibinfo {author} {\bibfnamefont {T.}~\bibnamefont {Inagaki}}, \bibinfo {author} {\bibfnamefont {T.}~\bibnamefont {Ikuta}}, \bibinfo {author} {\bibfnamefont {Y.}~\bibnamefont {Yamada}}, \bibinfo {author} {\bibfnamefont {T.}~\bibnamefont {Kazama}}, \bibinfo {author} {\bibfnamefont {K.}~\bibnamefont {Enbutsu}}, \bibinfo {author} {\bibfnamefont {T.}~\bibnamefont {Umeki}}, \bibinfo {author} {\bibfnamefont {R.}~\bibnamefont {Kasahara}}, \bibinfo {author} {\bibfnamefont {K.}~\bibnamefont {Kawarabayashi}}, \ and\ \bibinfo {author} {\bibfnamefont {H.}~\bibnamefont {Takesue}},\ }\bibfield  {title} {\enquote {\bibinfo {title} {100,000-spin coherent {Ising} machine},}\ }\href {\doibase 10.1126/sciadv.abh0952} {\bibfield  {journal} {\bibinfo  {journal} {Science Advances}\ }\textbf {\bibinfo {volume} {7}},\ \bibinfo
  {pages} {eabh0952} (\bibinfo {year} {2021})}\BibitemShut {NoStop}%
\bibitem [{\citenamefont {Farhi}, \citenamefont {Goldstone},\ and\ \citenamefont {Gutmann}(2017)}]{QAOA}%
  \BibitemOpen
  \bibfield  {author} {\bibinfo {author} {\bibfnamefont {E.}~\bibnamefont {Farhi}}, \bibinfo {author} {\bibfnamefont {J.}~\bibnamefont {Goldstone}}, \ and\ \bibinfo {author} {\bibfnamefont {S.}~\bibnamefont {Gutmann}},\ }\href {http://arxiv.org/abs/1411.4028} {\enquote {\bibinfo {title} {A {Quantum} {Approximate} {Optimization} {Algorithm}},}\ } (\bibinfo {year} {2017}),\ \bibinfo {note} {arXiv:1411.4028 [quant-ph]}\BibitemShut {NoStop}%
\bibitem [{\citenamefont {Kitai}\ \emph {et~al.}(2020)\citenamefont {Kitai}, \citenamefont {Guo}, \citenamefont {Ju}, \citenamefont {Tanaka}, \citenamefont {Tsuda}, \citenamefont {Shiomi},\ and\ \citenamefont {Tamura}}]{PhysRevResearch.2.013319}%
  \BibitemOpen
  \bibfield  {author} {\bibinfo {author} {\bibfnamefont {K.}~\bibnamefont {Kitai}}, \bibinfo {author} {\bibfnamefont {J.}~\bibnamefont {Guo}}, \bibinfo {author} {\bibfnamefont {S.}~\bibnamefont {Ju}}, \bibinfo {author} {\bibfnamefont {S.}~\bibnamefont {Tanaka}}, \bibinfo {author} {\bibfnamefont {K.}~\bibnamefont {Tsuda}}, \bibinfo {author} {\bibfnamefont {J.}~\bibnamefont {Shiomi}}, \ and\ \bibinfo {author} {\bibfnamefont {R.}~\bibnamefont {Tamura}},\ }\bibfield  {title} {\enquote {\bibinfo {title} {Designing metamaterials with quantum annealing and factorization machines},}\ }\href {\doibase 10.1103/PhysRevResearch.2.013319} {\bibfield  {journal} {\bibinfo  {journal} {Physical Review Research}\ }\textbf {\bibinfo {volume} {2}},\ \bibinfo {pages} {013319} (\bibinfo {year} {2020})}\BibitemShut {NoStop}%
\bibitem [{\citenamefont {Rendle}(2010)}]{5694074}%
  \BibitemOpen
  \bibfield  {author} {\bibinfo {author} {\bibfnamefont {S.}~\bibnamefont {Rendle}},\ }\bibfield  {title} {\enquote {\bibinfo {title} {Factorization machines},}\ }in\ \href {\doibase 10.1109/ICDM.2010.127} {\emph {\bibinfo {booktitle} {2010 IEEE International Conference on Data Mining}}}\ (\bibinfo {year} {2010})\ pp.\ \bibinfo {pages} {995--1000}\BibitemShut {NoStop}%
\bibitem [{\citenamefont {Baptista}\ and\ \citenamefont {Poloczek}(2018)}]{baptista2018bayesianoptimizationcombinatorialstructures}%
  \BibitemOpen
  \bibfield  {author} {\bibinfo {author} {\bibfnamefont {R.}~\bibnamefont {Baptista}}\ and\ \bibinfo {author} {\bibfnamefont {M.}~\bibnamefont {Poloczek}},\ }\href {https://arxiv.org/abs/1806.08838} {\enquote {\bibinfo {title} {Bayesian optimization of combinatorial structures},}\ } (\bibinfo {year} {2018}),\ \Eprint {http://arxiv.org/abs/1806.08838} {arXiv:1806.08838 [stat.ML]} \BibitemShut {NoStop}%
\bibitem [{\citenamefont {Koshikawa}\ \emph {et~al.}(2021)\citenamefont {Koshikawa}, \citenamefont {Ohzeki}, \citenamefont {Kadowaki},\ and\ \citenamefont {Tanaka}}]{doi:10.7566/JPSJ.90.064001}%
  \BibitemOpen
  \bibfield  {author} {\bibinfo {author} {\bibfnamefont {A.~S.}\ \bibnamefont {Koshikawa}}, \bibinfo {author} {\bibfnamefont {M.}~\bibnamefont {Ohzeki}}, \bibinfo {author} {\bibfnamefont {T.}~\bibnamefont {Kadowaki}}, \ and\ \bibinfo {author} {\bibfnamefont {K.}~\bibnamefont {Tanaka}},\ }\bibfield  {title} {\enquote {\bibinfo {title} {Benchmark test of black-box optimization using {D-Wave} quantum annealer},}\ }\href {\doibase 10.7566/JPSJ.90.064001} {\bibfield  {journal} {\bibinfo  {journal} {Journal of the Physical Society of Japan}\ }\textbf {\bibinfo {volume} {90}},\ \bibinfo {pages} {064001} (\bibinfo {year} {2021})}\BibitemShut {NoStop}%
\bibitem [{\citenamefont {Okada}\ \emph {et~al.}(2023)\citenamefont {Okada}, \citenamefont {Yoshida}, \citenamefont {Kidono}, \citenamefont {Matsumori}, \citenamefont {Takeno},\ and\ \citenamefont {Kadowaki}}]{10113307}%
  \BibitemOpen
  \bibfield  {author} {\bibinfo {author} {\bibfnamefont {A.}~\bibnamefont {Okada}}, \bibinfo {author} {\bibfnamefont {H.}~\bibnamefont {Yoshida}}, \bibinfo {author} {\bibfnamefont {K.}~\bibnamefont {Kidono}}, \bibinfo {author} {\bibfnamefont {T.}~\bibnamefont {Matsumori}}, \bibinfo {author} {\bibfnamefont {T.}~\bibnamefont {Takeno}}, \ and\ \bibinfo {author} {\bibfnamefont {T.}~\bibnamefont {Kadowaki}},\ }\bibfield  {title} {\enquote {\bibinfo {title} {Design optimization of noise filter using quantum annealer},}\ }\href {\doibase 10.1109/ACCESS.2023.3271969} {\bibfield  {journal} {\bibinfo  {journal} {IEEE Access}\ }\textbf {\bibinfo {volume} {11}},\ \bibinfo {pages} {44343--44349} (\bibinfo {year} {2023})}\BibitemShut {NoStop}%
\bibitem [{\citenamefont {Morita}, \citenamefont {Nishikawa},\ and\ \citenamefont {Ohzeki}(2023)}]{doi:10.7566/JPSJ.92.123801}%
  \BibitemOpen
  \bibfield  {author} {\bibinfo {author} {\bibfnamefont {K.}~\bibnamefont {Morita}}, \bibinfo {author} {\bibfnamefont {Y.}~\bibnamefont {Nishikawa}}, \ and\ \bibinfo {author} {\bibfnamefont {M.}~\bibnamefont {Ohzeki}},\ }\bibfield  {title} {\enquote {\bibinfo {title} {Random postprocessing for combinatorial {Bayesian} optimization},}\ }\href {\doibase 10.7566/JPSJ.92.123801} {\bibfield  {journal} {\bibinfo  {journal} {Journal of the Physical Society of Japan}\ }\textbf {\bibinfo {volume} {92}},\ \bibinfo {pages} {123801} (\bibinfo {year} {2023})}\BibitemShut {NoStop}%
\bibitem [{\citenamefont {Hatakeyama-Sato}\ \emph {et~al.}(2021)\citenamefont {Hatakeyama-Sato}, \citenamefont {Kashikawa}, \citenamefont {Kimura},\ and\ \citenamefont {Oyaizu}}]{https://doi.org/10.1002/aisy.202000209}%
  \BibitemOpen
  \bibfield  {author} {\bibinfo {author} {\bibfnamefont {K.}~\bibnamefont {Hatakeyama-Sato}}, \bibinfo {author} {\bibfnamefont {T.}~\bibnamefont {Kashikawa}}, \bibinfo {author} {\bibfnamefont {K.}~\bibnamefont {Kimura}}, \ and\ \bibinfo {author} {\bibfnamefont {K.}~\bibnamefont {Oyaizu}},\ }\bibfield  {title} {\enquote {\bibinfo {title} {Tackling the challenge of a huge materials science search space with quantum-inspired annealing},}\ }\href {\doibase https://doi.org/10.1002/aisy.202000209} {\bibfield  {journal} {\bibinfo  {journal} {Advanced Intelligent Systems}\ }\textbf {\bibinfo {volume} {3}},\ \bibinfo {pages} {2000209} (\bibinfo {year} {2021})}\BibitemShut {NoStop}%
\bibitem [{\citenamefont {Kim}\ \emph {et~al.}(2022)\citenamefont {Kim}, \citenamefont {Shang}, \citenamefont {Moon}, \citenamefont {Pastega}, \citenamefont {Lee},\ and\ \citenamefont {Luo}}]{Kim:2022aa}%
  \BibitemOpen
  \bibfield  {author} {\bibinfo {author} {\bibfnamefont {S.}~\bibnamefont {Kim}}, \bibinfo {author} {\bibfnamefont {W.}~\bibnamefont {Shang}}, \bibinfo {author} {\bibfnamefont {S.}~\bibnamefont {Moon}}, \bibinfo {author} {\bibfnamefont {T.}~\bibnamefont {Pastega}}, \bibinfo {author} {\bibfnamefont {E.}~\bibnamefont {Lee}}, \ and\ \bibinfo {author} {\bibfnamefont {T.}~\bibnamefont {Luo}},\ }\bibfield  {title} {\enquote {\bibinfo {title} {High-performance transparent radiative cooler designed by quantum computing},}\ }\href {\doibase 10.1021/acsenergylett.2c01969} {\bibfield  {journal} {\bibinfo  {journal} {ACS Energy Letters}\ }\textbf {\bibinfo {volume} {7}},\ \bibinfo {pages} {4134--4141} (\bibinfo {year} {2022})}\BibitemShut {NoStop}%
\bibitem [{\citenamefont {Nawa}\ \emph {et~al.}(2023)\citenamefont {Nawa}, \citenamefont {Suzuki}, \citenamefont {Masuda}, \citenamefont {Tanaka},\ and\ \citenamefont {Miura}}]{PhysRevApplied.20.024044}%
  \BibitemOpen
  \bibfield  {author} {\bibinfo {author} {\bibfnamefont {K.}~\bibnamefont {Nawa}}, \bibinfo {author} {\bibfnamefont {T.}~\bibnamefont {Suzuki}}, \bibinfo {author} {\bibfnamefont {K.}~\bibnamefont {Masuda}}, \bibinfo {author} {\bibfnamefont {S.}~\bibnamefont {Tanaka}}, \ and\ \bibinfo {author} {\bibfnamefont {Y.}~\bibnamefont {Miura}},\ }\bibfield  {title} {\enquote {\bibinfo {title} {Quantum annealing optimization method for the design of barrier materials in magnetic tunnel junctions},}\ }\href {\doibase 10.1103/PhysRevApplied.20.024044} {\bibfield  {journal} {\bibinfo  {journal} {Physical Review Applied}\ }\textbf {\bibinfo {volume} {20}},\ \bibinfo {pages} {024044} (\bibinfo {year} {2023})}\BibitemShut {NoStop}%
\bibitem [{\citenamefont {Luo}\ \emph {et~al.}(2024)\citenamefont {Luo}, \citenamefont {Xu}, \citenamefont {Shang}, \citenamefont {Kim},\ and\ \citenamefont {Lee}}]{QALO}%
  \BibitemOpen
  \bibfield  {author} {\bibinfo {author} {\bibfnamefont {T.}~\bibnamefont {Luo}}, \bibinfo {author} {\bibfnamefont {Z.}~\bibnamefont {Xu}}, \bibinfo {author} {\bibfnamefont {W.}~\bibnamefont {Shang}}, \bibinfo {author} {\bibfnamefont {S.}~\bibnamefont {Kim}}, \ and\ \bibinfo {author} {\bibfnamefont {E.}~\bibnamefont {Lee}},\ }\href {\doibase 10.21203/rs.3.rs-4518513/v1} {\enquote {\bibinfo {title} {{QALO}: {Quantum} annealing-assisted lattice optimization},}\ } (\bibinfo {year} {2024})\BibitemShut {NoStop}%
\bibitem [{\citenamefont {Couzini\'{e}}\ \emph {et~al.}(2025)\citenamefont {Couzini\'{e}}, \citenamefont {Seki}, \citenamefont {Nishiya}, \citenamefont {Nishi}, \citenamefont {Kosugi}, \citenamefont {Tanaka},\ and\ \citenamefont {Matsushita}}]{couzinie2024machinelearningsupportedannealing}%
  \BibitemOpen
  \bibfield  {author} {\bibinfo {author} {\bibfnamefont {Y.}~\bibnamefont {Couzini\'{e}}}, \bibinfo {author} {\bibfnamefont {Y.}~\bibnamefont {Seki}}, \bibinfo {author} {\bibfnamefont {Y.}~\bibnamefont {Nishiya}}, \bibinfo {author} {\bibfnamefont {H.}~\bibnamefont {Nishi}}, \bibinfo {author} {\bibfnamefont {T.}~\bibnamefont {Kosugi}}, \bibinfo {author} {\bibfnamefont {S.}~\bibnamefont {Tanaka}}, \ and\ \bibinfo {author} {\bibfnamefont {Y.-i.}\ \bibnamefont {Matsushita}},\ }\bibfield  {title} {\enquote {\bibinfo {title} {Machine learning supported annealing for prediction of grand canonical crystal structures},}\ }\href {\doibase 10.7566/JPSJ.94.044802} {\bibfield  {journal} {\bibinfo  {journal} {Journal of the Physical Society of Japan}\ }\textbf {\bibinfo {volume} {94}},\ \bibinfo {pages} {044802} (\bibinfo {year} {2025})}\BibitemShut {NoStop}%
\bibitem [{\citenamefont {Lin}\ \emph {et~al.}(2025{\natexlab{a}})\citenamefont {Lin}, \citenamefont {Tada}, \citenamefont {Koizumi}, \citenamefont {Sumita}, \citenamefont {Tsuda},\ and\ \citenamefont {Tamura}}]{Lin}%
  \BibitemOpen
  \bibfield  {author} {\bibinfo {author} {\bibfnamefont {J.}~\bibnamefont {Lin}}, \bibinfo {author} {\bibfnamefont {T.}~\bibnamefont {Tada}}, \bibinfo {author} {\bibfnamefont {A.}~\bibnamefont {Koizumi}}, \bibinfo {author} {\bibfnamefont {M.}~\bibnamefont {Sumita}}, \bibinfo {author} {\bibfnamefont {K.}~\bibnamefont {Tsuda}}, \ and\ \bibinfo {author} {\bibfnamefont {R.}~\bibnamefont {Tamura}},\ }\bibfield  {title} {\enquote {\bibinfo {title} {Determination of stable proton configurations by black-box optimization using an {Ising} machine},}\ }\href {\doibase 10.1021/acs.jpcc.4c07104} {\bibfield  {journal} {\bibinfo  {journal} {The Journal of Physical Chemistry C}\ }\textbf {\bibinfo {volume} {129}},\ \bibinfo {pages} {2332--2340} (\bibinfo {year} {2025}{\natexlab{a}})}\BibitemShut {NoStop}%
\bibitem [{\citenamefont {Tu{\v c}s}\ \emph {et~al.}(2023)\citenamefont {Tu{\v c}s}, \citenamefont {Berenger}, \citenamefont {Yumoto}, \citenamefont {Tamura}, \citenamefont {Uzawa},\ and\ \citenamefont {Tsuda}}]{Tucs:2023aa}%
  \BibitemOpen
  \bibfield  {author} {\bibinfo {author} {\bibfnamefont {A.}~\bibnamefont {Tu{\v c}s}}, \bibinfo {author} {\bibfnamefont {F.}~\bibnamefont {Berenger}}, \bibinfo {author} {\bibfnamefont {A.}~\bibnamefont {Yumoto}}, \bibinfo {author} {\bibfnamefont {R.}~\bibnamefont {Tamura}}, \bibinfo {author} {\bibfnamefont {T.}~\bibnamefont {Uzawa}}, \ and\ \bibinfo {author} {\bibfnamefont {K.}~\bibnamefont {Tsuda}},\ }\bibfield  {title} {\enquote {\bibinfo {title} {Quantum annealing designs nonhemolytic antimicrobial peptides in a discrete latent space},}\ }\href {\doibase 10.1021/acsmedchemlett.2c00487} {\bibfield  {journal} {\bibinfo  {journal} {ACS Medicinal Chemistry Letters}\ }\textbf {\bibinfo {volume} {14}},\ \bibinfo {pages} {577--582} (\bibinfo {year} {2023})}\BibitemShut {NoStop}%
\bibitem [{\citenamefont {Kadowaki}\ and\ \citenamefont {Ambai}(2022)}]{Kadowaki:2022aa}%
  \BibitemOpen
  \bibfield  {author} {\bibinfo {author} {\bibfnamefont {T.}~\bibnamefont {Kadowaki}}\ and\ \bibinfo {author} {\bibfnamefont {M.}~\bibnamefont {Ambai}},\ }\bibfield  {title} {\enquote {\bibinfo {title} {Lossy compression of matrices by black box optimisation of mixed integer nonlinear programming},}\ }\href {\doibase 10.1038/s41598-022-19763-8} {\bibfield  {journal} {\bibinfo  {journal} {Scientific Reports}\ }\textbf {\bibinfo {volume} {12}},\ \bibinfo {pages} {15482} (\bibinfo {year} {2022})}\BibitemShut {NoStop}%
\bibitem [{fix({\natexlab{a}})}]{fixtraffic}%
  \BibitemOpen
  \href@noop {} {\enquote {\bibinfo {title} {Black-box optimiztion of traffic light control},}\ } ({\natexlab{a}}),\ \bibinfo {note} {\url{https://amplify.fixstars.com/en/demo/fmqa_4_traffic}}\BibitemShut {NoStop}%
\bibitem [{\citenamefont {Mena}\ and\ \citenamefont {{\~{N}}anculef}(2019)}]{10.1007/978-3-030-33904-3_12}%
  \BibitemOpen
  \bibfield  {author} {\bibinfo {author} {\bibfnamefont {F.}~\bibnamefont {Mena}}\ and\ \bibinfo {author} {\bibfnamefont {R.}~\bibnamefont {{\~{N}}anculef}},\ }\bibfield  {title} {\enquote {\bibinfo {title} {A binary variational autoencoder for hashing},}\ }in\ \href@noop {} {\emph {\bibinfo {booktitle} {Progress in Pattern Recognition, Image Analysis, Computer Vision, and Applications}}},\ \bibinfo {editor} {edited by\ \bibinfo {editor} {\bibfnamefont {I.}~\bibnamefont {Nystr{\"o}m}}, \bibinfo {editor} {\bibfnamefont {Y.}~\bibnamefont {Hern{\'a}ndez~Heredia}}, \ and\ \bibinfo {editor} {\bibfnamefont {V.}~\bibnamefont {Mili{\'a}n~N{\'u}{\~{n}}ez}}}\ (\bibinfo  {publisher} {Springer International Publishing},\ \bibinfo {address} {Cham},\ \bibinfo {year} {2019})\ pp.\ \bibinfo {pages} {131--141}\BibitemShut {NoStop}%
\bibitem [{\citenamefont {Akiba}\ \emph {et~al.}(2019)\citenamefont {Akiba}, \citenamefont {Sano}, \citenamefont {Yanase}, \citenamefont {Ohta},\ and\ \citenamefont {Koyama}}]{optuna_2019}%
  \BibitemOpen
  \bibfield  {author} {\bibinfo {author} {\bibfnamefont {T.}~\bibnamefont {Akiba}}, \bibinfo {author} {\bibfnamefont {S.}~\bibnamefont {Sano}}, \bibinfo {author} {\bibfnamefont {T.}~\bibnamefont {Yanase}}, \bibinfo {author} {\bibfnamefont {T.}~\bibnamefont {Ohta}}, \ and\ \bibinfo {author} {\bibfnamefont {M.}~\bibnamefont {Koyama}},\ }\bibfield  {title} {\enquote {\bibinfo {title} {Optuna: A next-generation hyperparameter optimization framework},}\ }in\ \href@noop {} {\emph {\bibinfo {booktitle} {Proceedings of the 25th {ACM} {SIGKDD} International Conference on Knowledge Discovery and Data Mining}}}\ (\bibinfo {year} {2019})\BibitemShut {NoStop}%
\bibitem [{HPy()}]{HPyOpt}%
  \BibitemOpen
  \href@noop {} {\enquote {\bibinfo {title} {Gpyopt},}\ }\bibinfo {note} {\url{https://sheffieldml.github.io/GPyOpt/}}\BibitemShut {NoStop}%
\bibitem [{\citenamefont {Felton}, \citenamefont {Rittig},\ and\ \citenamefont {Lapkin}(2021)}]{https://doi.org/10.1002/cmtd.202000051}%
  \BibitemOpen
  \bibfield  {author} {\bibinfo {author} {\bibfnamefont {K.~C.}\ \bibnamefont {Felton}}, \bibinfo {author} {\bibfnamefont {J.~G.}\ \bibnamefont {Rittig}}, \ and\ \bibinfo {author} {\bibfnamefont {A.~A.}\ \bibnamefont {Lapkin}},\ }\bibfield  {title} {\enquote {\bibinfo {title} {Summit: Benchmarking machine learning methods for reaction optimisation},}\ }\href {\doibase https://doi.org/10.1002/cmtd.202000051} {\bibfield  {journal} {\bibinfo  {journal} {Chemistry--Methods}\ }\textbf {\bibinfo {volume} {1}},\ \bibinfo {pages} {116--122} (\bibinfo {year} {2021})}\BibitemShut {NoStop}%
\bibitem [{\citenamefont {Balandat}\ \emph {et~al.}(2020)\citenamefont {Balandat}, \citenamefont {Karrer}, \citenamefont {Jiang}, \citenamefont {Daulton}, \citenamefont {Letham}, \citenamefont {Wilson},\ and\ \citenamefont {Bakshy}}]{balandat2020botorchframeworkefficientmontecarlo}%
  \BibitemOpen
  \bibfield  {author} {\bibinfo {author} {\bibfnamefont {M.}~\bibnamefont {Balandat}}, \bibinfo {author} {\bibfnamefont {B.}~\bibnamefont {Karrer}}, \bibinfo {author} {\bibfnamefont {D.~R.}\ \bibnamefont {Jiang}}, \bibinfo {author} {\bibfnamefont {S.}~\bibnamefont {Daulton}}, \bibinfo {author} {\bibfnamefont {B.}~\bibnamefont {Letham}}, \bibinfo {author} {\bibfnamefont {A.~G.}\ \bibnamefont {Wilson}}, \ and\ \bibinfo {author} {\bibfnamefont {E.}~\bibnamefont {Bakshy}},\ }\href {https://arxiv.org/abs/1910.06403} {\enquote {\bibinfo {title} {{BoTorch}: A framework for efficient {Monte-Carlo Bayesian} optimization},}\ } (\bibinfo {year} {2020}),\ \Eprint {http://arxiv.org/abs/1910.06403} {arXiv:1910.06403 [cs.LG]} \BibitemShut {NoStop}%
\bibitem [{\citenamefont {Ueno}\ \emph {et~al.}(2016)\citenamefont {Ueno}, \citenamefont {Rhone}, \citenamefont {Hou}, \citenamefont {Mizoguchi},\ and\ \citenamefont {Tsuda}}]{UENO201618}%
  \BibitemOpen
  \bibfield  {author} {\bibinfo {author} {\bibfnamefont {T.}~\bibnamefont {Ueno}}, \bibinfo {author} {\bibfnamefont {T.~D.}\ \bibnamefont {Rhone}}, \bibinfo {author} {\bibfnamefont {Z.}~\bibnamefont {Hou}}, \bibinfo {author} {\bibfnamefont {T.}~\bibnamefont {Mizoguchi}}, \ and\ \bibinfo {author} {\bibfnamefont {K.}~\bibnamefont {Tsuda}},\ }\bibfield  {title} {\enquote {\bibinfo {title} {{COMBO}: An efficient bayesian optimization library for materials science},}\ }\href {\doibase https://doi.org/10.1016/j.md.2016.04.001} {\bibfield  {journal} {\bibinfo  {journal} {Materials Discovery}\ }\textbf {\bibinfo {volume} {4}},\ \bibinfo {pages} {18--21} (\bibinfo {year} {2016})}\BibitemShut {NoStop}%
\bibitem [{\citenamefont {Motoyama}\ \emph {et~al.}(2022)\citenamefont {Motoyama}, \citenamefont {Tamura}, \citenamefont {Yoshimi}, \citenamefont {Terayama}, \citenamefont {Ueno},\ and\ \citenamefont {Tsuda}}]{MOTOYAMA2022108405}%
  \BibitemOpen
  \bibfield  {author} {\bibinfo {author} {\bibfnamefont {Y.}~\bibnamefont {Motoyama}}, \bibinfo {author} {\bibfnamefont {R.}~\bibnamefont {Tamura}}, \bibinfo {author} {\bibfnamefont {K.}~\bibnamefont {Yoshimi}}, \bibinfo {author} {\bibfnamefont {K.}~\bibnamefont {Terayama}}, \bibinfo {author} {\bibfnamefont {T.}~\bibnamefont {Ueno}}, \ and\ \bibinfo {author} {\bibfnamefont {K.}~\bibnamefont {Tsuda}},\ }\bibfield  {title} {\enquote {\bibinfo {title} {Bayesian optimization package: {PHYSBO}},}\ }\href {\doibase https://doi.org/10.1016/j.cpc.2022.108405} {\bibfield  {journal} {\bibinfo  {journal} {Computer Physics Communications}\ }\textbf {\bibinfo {volume} {278}},\ \bibinfo {pages} {108405} (\bibinfo {year} {2022})}\BibitemShut {NoStop}%
\bibitem [{\citenamefont {Rendle}(2012)}]{rendle:tist2012}%
  \BibitemOpen
  \bibfield  {author} {\bibinfo {author} {\bibfnamefont {S.}~\bibnamefont {Rendle}},\ }\bibfield  {title} {\enquote {\bibinfo {title} {Factorization machines with {libFM}},}\ }\href@noop {} {\bibfield  {journal} {\bibinfo  {journal} {ACM Trans. Intell. Syst. Technol.}\ }\textbf {\bibinfo {volume} {3}},\ \bibinfo {pages} {57:1--57:22} (\bibinfo {year} {2012})}\BibitemShut {NoStop}%
\bibitem [{\citenamefont {Bayer}(2016)}]{JMLR:v17:15-355}%
  \BibitemOpen
  \bibfield  {author} {\bibinfo {author} {\bibfnamefont {I.}~\bibnamefont {Bayer}},\ }\bibfield  {title} {\enquote {\bibinfo {title} {fastfm: A library for factorization machines},}\ }\href {http://jmlr.org/papers/v17/15-355.html} {\bibfield  {journal} {\bibinfo  {journal} {Journal of Machine Learning Research}\ }\textbf {\bibinfo {volume} {17}},\ \bibinfo {pages} {1--5} (\bibinfo {year} {2016})}\BibitemShut {NoStop}%
\bibitem [{\citenamefont {Paszke}\ \emph {et~al.}(2019{\natexlab{a}})\citenamefont {Paszke}, \citenamefont {Gross}, \citenamefont {Massa}, \citenamefont {Lerer}, \citenamefont {Bradbury}, \citenamefont {Chanan}, \citenamefont {Killeen}, \citenamefont {Lin}, \citenamefont {Gimelshein}, \citenamefont {Antiga}, \citenamefont {Desmaison}, \citenamefont {Kopf}, \citenamefont {Yang}, \citenamefont {DeVito}, \citenamefont {Raison}, \citenamefont {Tejani}, \citenamefont {Chilamkurthy}, \citenamefont {Steiner}, \citenamefont {Fang}, \citenamefont {Bai},\ and\ \citenamefont {Chintala}}]{NEURIPS2019_9015}%
  \BibitemOpen
  \bibfield  {author} {\bibinfo {author} {\bibfnamefont {A.}~\bibnamefont {Paszke}}, \bibinfo {author} {\bibfnamefont {S.}~\bibnamefont {Gross}}, \bibinfo {author} {\bibfnamefont {F.}~\bibnamefont {Massa}}, \bibinfo {author} {\bibfnamefont {A.}~\bibnamefont {Lerer}}, \bibinfo {author} {\bibfnamefont {J.}~\bibnamefont {Bradbury}}, \bibinfo {author} {\bibfnamefont {G.}~\bibnamefont {Chanan}}, \bibinfo {author} {\bibfnamefont {T.}~\bibnamefont {Killeen}}, \bibinfo {author} {\bibfnamefont {Z.}~\bibnamefont {Lin}}, \bibinfo {author} {\bibfnamefont {N.}~\bibnamefont {Gimelshein}}, \bibinfo {author} {\bibfnamefont {L.}~\bibnamefont {Antiga}}, \bibinfo {author} {\bibfnamefont {A.}~\bibnamefont {Desmaison}}, \bibinfo {author} {\bibfnamefont {A.}~\bibnamefont {Kopf}}, \bibinfo {author} {\bibfnamefont {E.}~\bibnamefont {Yang}}, \bibinfo {author} {\bibfnamefont {Z.}~\bibnamefont {DeVito}}, \bibinfo {author} {\bibfnamefont {M.}~\bibnamefont {Raison}}, \bibinfo {author} {\bibfnamefont {A.}~\bibnamefont {Tejani}}, \bibinfo
  {author} {\bibfnamefont {S.}~\bibnamefont {Chilamkurthy}}, \bibinfo {author} {\bibfnamefont {B.}~\bibnamefont {Steiner}}, \bibinfo {author} {\bibfnamefont {L.}~\bibnamefont {Fang}}, \bibinfo {author} {\bibfnamefont {J.}~\bibnamefont {Bai}}, \ and\ \bibinfo {author} {\bibfnamefont {S.}~\bibnamefont {Chintala}},\ }\bibfield  {title} {\enquote {\bibinfo {title} {Pytorch: An imperative style, high-performance deep learning library},}\ }in\ \href {http://papers.neurips.cc/paper/9015-pytorch-an-imperative-style-high-performance-deep-learning-library.pdf} {\emph {\bibinfo {booktitle} {Advances in Neural Information Processing Systems 32}}}\ (\bibinfo  {publisher} {Curran Associates, Inc.},\ \bibinfo {year} {2019})\ pp.\ \bibinfo {pages} {8024--8035}\BibitemShut {NoStop}%
\bibitem [{FMQ()}]{FMQA_git}%
  \BibitemOpen
  \href@noop {} {\enquote {\bibinfo {title} {{GitHub}, fmqa},}\ }\bibinfo {note} {\url{https://github.com/tsudalab/fmqa}}\BibitemShut {NoStop}%
\bibitem [{\citenamefont {Seki}, \citenamefont {Tamura},\ and\ \citenamefont {Tanaka}(2022)}]{seki2022blackboxoptimizationintegervariableproblems}%
  \BibitemOpen
  \bibfield  {author} {\bibinfo {author} {\bibfnamefont {Y.}~\bibnamefont {Seki}}, \bibinfo {author} {\bibfnamefont {R.}~\bibnamefont {Tamura}}, \ and\ \bibinfo {author} {\bibfnamefont {S.}~\bibnamefont {Tanaka}},\ }\href {https://arxiv.org/abs/2209.01016} {\enquote {\bibinfo {title} {Black-box optimization for integer-variable problems using {Ising} machines and factorization machines},}\ } (\bibinfo {year} {2022}),\ \Eprint {http://arxiv.org/abs/2209.01016} {arXiv:2209.01016 [cs.LG]} \BibitemShut {NoStop}%
\bibitem [{\citenamefont {Agrell}\ \emph {et~al.}(2004)\citenamefont {Agrell}, \citenamefont {Lassing}, \citenamefont {Strom},\ and\ \citenamefont {Ottosson}}]{1362904}%
  \BibitemOpen
  \bibfield  {author} {\bibinfo {author} {\bibfnamefont {E.}~\bibnamefont {Agrell}}, \bibinfo {author} {\bibfnamefont {J.}~\bibnamefont {Lassing}}, \bibinfo {author} {\bibfnamefont {E.}~\bibnamefont {Strom}}, \ and\ \bibinfo {author} {\bibfnamefont {T.}~\bibnamefont {Ottosson}},\ }\bibfield  {title} {\enquote {\bibinfo {title} {On the optimality of the binary reflected gray code},}\ }\href {\doibase 10.1109/TIT.2004.838367} {\bibfield  {journal} {\bibinfo  {journal} {IEEE Transactions on Information Theory}\ }\textbf {\bibinfo {volume} {50}},\ \bibinfo {pages} {3170--3182} (\bibinfo {year} {2004})}\BibitemShut {NoStop}%
\bibitem [{\citenamefont {Endo}\ and\ \citenamefont {Takahashi}(2025)}]{PhysRevResearch.7.013149}%
  \BibitemOpen
  \bibfield  {author} {\bibinfo {author} {\bibfnamefont {K.}~\bibnamefont {Endo}}\ and\ \bibinfo {author} {\bibfnamefont {K.~Z.}\ \bibnamefont {Takahashi}},\ }\bibfield  {title} {\enquote {\bibinfo {title} {Function smoothing regularization for precision factorization machine annealing in continuous variable optimization problems},}\ }\href {\doibase 10.1103/PhysRevResearch.7.013149} {\bibfield  {journal} {\bibinfo  {journal} {Physical Review Research}\ }\textbf {\bibinfo {volume} {7}},\ \bibinfo {pages} {013149} (\bibinfo {year} {2025})}\BibitemShut {NoStop}%
\bibitem [{\citenamefont {Chancellor}(2019)}]{Chancellor_2019}%
  \BibitemOpen
  \bibfield  {author} {\bibinfo {author} {\bibfnamefont {N.}~\bibnamefont {Chancellor}},\ }\bibfield  {title} {\enquote {\bibinfo {title} {Domain wall encoding of discrete variables for quantum annealing and {QAOA}},}\ }\href {\doibase 10.1088/2058-9565/ab33c2} {\bibfield  {journal} {\bibinfo  {journal} {Quantum Science and Technology}\ }\textbf {\bibinfo {volume} {4}},\ \bibinfo {pages} {045004} (\bibinfo {year} {2019})}\BibitemShut {NoStop}%
\bibitem [{\citenamefont {Berwald}, \citenamefont {Chancellor},\ and\ \citenamefont {Dridi}(2023)}]{doi:10.1098/rsta.2021.0410}%
  \BibitemOpen
  \bibfield  {author} {\bibinfo {author} {\bibfnamefont {J.}~\bibnamefont {Berwald}}, \bibinfo {author} {\bibfnamefont {N.}~\bibnamefont {Chancellor}}, \ and\ \bibinfo {author} {\bibfnamefont {R.}~\bibnamefont {Dridi}},\ }\bibfield  {title} {\enquote {\bibinfo {title} {Understanding domain-wall encoding theoretically and experimentally},}\ }\href {\doibase 10.1098/rsta.2021.0410} {\bibfield  {journal} {\bibinfo  {journal} {Philosophical Transactions of the Royal Society A: Mathematical, Physical and Engineering Sciences}\ }\textbf {\bibinfo {volume} {381}},\ \bibinfo {pages} {20210410} (\bibinfo {year} {2023})}\BibitemShut {NoStop}%
\bibitem [{\citenamefont {Kadowaki}\ and\ \citenamefont {Nishimori}(1998)}]{PhysRevE.58.5355}%
  \BibitemOpen
  \bibfield  {author} {\bibinfo {author} {\bibfnamefont {T.}~\bibnamefont {Kadowaki}}\ and\ \bibinfo {author} {\bibfnamefont {H.}~\bibnamefont {Nishimori}},\ }\bibfield  {title} {\enquote {\bibinfo {title} {Quantum annealing in the transverse {Ising} model},}\ }\href {\doibase 10.1103/PhysRevE.58.5355} {\bibfield  {journal} {\bibinfo  {journal} {Physical Review E}\ }\textbf {\bibinfo {volume} {58}},\ \bibinfo {pages} {5355--5363} (\bibinfo {year} {1998})}\BibitemShut {NoStop}%
\bibitem [{\citenamefont {Santoro}\ and\ \citenamefont {Tosatti}(2006)}]{Santoro_2006}%
  \BibitemOpen
  \bibfield  {author} {\bibinfo {author} {\bibfnamefont {G.~E.}\ \bibnamefont {Santoro}}\ and\ \bibinfo {author} {\bibfnamefont {E.}~\bibnamefont {Tosatti}},\ }\bibfield  {title} {\enquote {\bibinfo {title} {Optimization using quantum mechanics: {Quantum} annealing through adiabatic evolution},}\ }\href {\doibase 10.1088/0305-4470/39/36/R01} {\bibfield  {journal} {\bibinfo  {journal} {Journal of Physics A: Mathematical and General}\ }\textbf {\bibinfo {volume} {39}},\ \bibinfo {pages} {R393} (\bibinfo {year} {2006})}\BibitemShut {NoStop}%
\bibitem [{D-W()}]{D-Wave}%
  \BibitemOpen
  \href@noop {} {\enquote {\bibinfo {title} {{D-Wave, Advantage2}},}\ }\bibinfo {note} {\url{https://www.dwavesys.com/solutions-and-products/systems/}}\BibitemShut {NoStop}%
\bibitem [{\citenamefont {King}\ \emph {et~al.}(2022)\citenamefont {King}, \citenamefont {Suzuki}, \citenamefont {Raymond}, \citenamefont {Zucca}, \citenamefont {Lanting}, \citenamefont {Altomare}, \citenamefont {Berkley}, \citenamefont {Ejtemaee}, \citenamefont {Hoskinson}, \citenamefont {Huang}, \citenamefont {Ladizinsky}, \citenamefont {MacDonald}, \citenamefont {Marsden}, \citenamefont {Oh}, \citenamefont {Poulin-Lamarre}, \citenamefont {Reis}, \citenamefont {Rich}, \citenamefont {Sato}, \citenamefont {Whittaker}, \citenamefont {Yao}, \citenamefont {Harris}, \citenamefont {Lidar}, \citenamefont {Nishimori},\ and\ \citenamefont {Amin}}]{King:2022aa}%
  \BibitemOpen
  \bibfield  {author} {\bibinfo {author} {\bibfnamefont {A.~D.}\ \bibnamefont {King}}, \bibinfo {author} {\bibfnamefont {S.}~\bibnamefont {Suzuki}}, \bibinfo {author} {\bibfnamefont {J.}~\bibnamefont {Raymond}}, \bibinfo {author} {\bibfnamefont {A.}~\bibnamefont {Zucca}}, \bibinfo {author} {\bibfnamefont {T.}~\bibnamefont {Lanting}}, \bibinfo {author} {\bibfnamefont {F.}~\bibnamefont {Altomare}}, \bibinfo {author} {\bibfnamefont {A.~J.}\ \bibnamefont {Berkley}}, \bibinfo {author} {\bibfnamefont {S.}~\bibnamefont {Ejtemaee}}, \bibinfo {author} {\bibfnamefont {E.}~\bibnamefont {Hoskinson}}, \bibinfo {author} {\bibfnamefont {S.}~\bibnamefont {Huang}}, \bibinfo {author} {\bibfnamefont {E.}~\bibnamefont {Ladizinsky}}, \bibinfo {author} {\bibfnamefont {A.~J.~R.}\ \bibnamefont {MacDonald}}, \bibinfo {author} {\bibfnamefont {G.}~\bibnamefont {Marsden}}, \bibinfo {author} {\bibfnamefont {T.}~\bibnamefont {Oh}}, \bibinfo {author} {\bibfnamefont {G.}~\bibnamefont {Poulin-Lamarre}}, \bibinfo {author} {\bibfnamefont
  {M.}~\bibnamefont {Reis}}, \bibinfo {author} {\bibfnamefont {C.}~\bibnamefont {Rich}}, \bibinfo {author} {\bibfnamefont {Y.}~\bibnamefont {Sato}}, \bibinfo {author} {\bibfnamefont {J.~D.}\ \bibnamefont {Whittaker}}, \bibinfo {author} {\bibfnamefont {J.}~\bibnamefont {Yao}}, \bibinfo {author} {\bibfnamefont {R.}~\bibnamefont {Harris}}, \bibinfo {author} {\bibfnamefont {D.~A.}\ \bibnamefont {Lidar}}, \bibinfo {author} {\bibfnamefont {H.}~\bibnamefont {Nishimori}}, \ and\ \bibinfo {author} {\bibfnamefont {M.~H.}\ \bibnamefont {Amin}},\ }\bibfield  {title} {\enquote {\bibinfo {title} {Coherent quantum annealing in a programmable 2,000 qubit {Ising} chain},}\ }\href {\doibase 10.1038/s41567-022-01741-6} {\bibfield  {journal} {\bibinfo  {journal} {Nature Physics}\ }\textbf {\bibinfo {volume} {18}},\ \bibinfo {pages} {1324--1328} (\bibinfo {year} {2022})}\BibitemShut {NoStop}%
\bibitem [{\citenamefont {King}\ \emph {et~al.}(2023)\citenamefont {King}, \citenamefont {Raymond}, \citenamefont {Lanting}, \citenamefont {Harris}, \citenamefont {Zucca}, \citenamefont {Altomare}, \citenamefont {Berkley}, \citenamefont {Boothby}, \citenamefont {Ejtemaee}, \citenamefont {Enderud}, \citenamefont {Hoskinson}, \citenamefont {Huang}, \citenamefont {Ladizinsky}, \citenamefont {MacDonald}, \citenamefont {Marsden}, \citenamefont {Molavi}, \citenamefont {Oh}, \citenamefont {Poulin-Lamarre}, \citenamefont {Reis}, \citenamefont {Rich}, \citenamefont {Sato}, \citenamefont {Tsai}, \citenamefont {Volkmann}, \citenamefont {Whittaker}, \citenamefont {Yao}, \citenamefont {Sandvik},\ and\ \citenamefont {Amin}}]{King:2023aa}%
  \BibitemOpen
  \bibfield  {author} {\bibinfo {author} {\bibfnamefont {A.~D.}\ \bibnamefont {King}}, \bibinfo {author} {\bibfnamefont {J.}~\bibnamefont {Raymond}}, \bibinfo {author} {\bibfnamefont {T.}~\bibnamefont {Lanting}}, \bibinfo {author} {\bibfnamefont {R.}~\bibnamefont {Harris}}, \bibinfo {author} {\bibfnamefont {A.}~\bibnamefont {Zucca}}, \bibinfo {author} {\bibfnamefont {F.}~\bibnamefont {Altomare}}, \bibinfo {author} {\bibfnamefont {A.~J.}\ \bibnamefont {Berkley}}, \bibinfo {author} {\bibfnamefont {K.}~\bibnamefont {Boothby}}, \bibinfo {author} {\bibfnamefont {S.}~\bibnamefont {Ejtemaee}}, \bibinfo {author} {\bibfnamefont {C.}~\bibnamefont {Enderud}}, \bibinfo {author} {\bibfnamefont {E.}~\bibnamefont {Hoskinson}}, \bibinfo {author} {\bibfnamefont {S.}~\bibnamefont {Huang}}, \bibinfo {author} {\bibfnamefont {E.}~\bibnamefont {Ladizinsky}}, \bibinfo {author} {\bibfnamefont {A.~J.~R.}\ \bibnamefont {MacDonald}}, \bibinfo {author} {\bibfnamefont {G.}~\bibnamefont {Marsden}}, \bibinfo {author} {\bibfnamefont
  {R.}~\bibnamefont {Molavi}}, \bibinfo {author} {\bibfnamefont {T.}~\bibnamefont {Oh}}, \bibinfo {author} {\bibfnamefont {G.}~\bibnamefont {Poulin-Lamarre}}, \bibinfo {author} {\bibfnamefont {M.}~\bibnamefont {Reis}}, \bibinfo {author} {\bibfnamefont {C.}~\bibnamefont {Rich}}, \bibinfo {author} {\bibfnamefont {Y.}~\bibnamefont {Sato}}, \bibinfo {author} {\bibfnamefont {N.}~\bibnamefont {Tsai}}, \bibinfo {author} {\bibfnamefont {M.}~\bibnamefont {Volkmann}}, \bibinfo {author} {\bibfnamefont {J.~D.}\ \bibnamefont {Whittaker}}, \bibinfo {author} {\bibfnamefont {J.}~\bibnamefont {Yao}}, \bibinfo {author} {\bibfnamefont {A.~W.}\ \bibnamefont {Sandvik}}, \ and\ \bibinfo {author} {\bibfnamefont {M.~H.}\ \bibnamefont {Amin}},\ }\bibfield  {title} {\enquote {\bibinfo {title} {Quantum critical dynamics in a 5,000-qubit programmable spin glass},}\ }\href {\doibase 10.1038/s41586-023-05867-2} {\bibfield  {journal} {\bibinfo  {journal} {Nature}\ }\textbf {\bibinfo {volume} {617}},\ \bibinfo {pages} {61--66} (\bibinfo
  {year} {2023})}\BibitemShut {NoStop}%
\bibitem [{fix({\natexlab{b}})}]{fixstarsamplify}%
  \BibitemOpen
  \href@noop {} {\enquote {\bibinfo {title} {{Fixstars Amplify}},}\ } ({\natexlab{b}}),\ \bibinfo {note} {\url{https://amplify.fixstars.com/en/}}\BibitemShut {NoStop}%
\bibitem [{url()}]{url-ditigal}%
  \BibitemOpen
  \href@noop {} {\enquote {\bibinfo {title} {Fujitsu digital annealer},}\ }\bibinfo {note} {\url{https://www.fujitsu.com/global/services/business-services/digital-annealer/}}\BibitemShut {NoStop}%
\bibitem [{\citenamefont {Matsubara}\ \emph {et~al.}(2020)\citenamefont {Matsubara}, \citenamefont {Takatsu}, \citenamefont {Miyazawa}, \citenamefont {Shibasaki}, \citenamefont {Watanabe}, \citenamefont {Takemoto},\ and\ \citenamefont {Tamura}}]{9045100}%
  \BibitemOpen
  \bibfield  {author} {\bibinfo {author} {\bibfnamefont {S.}~\bibnamefont {Matsubara}}, \bibinfo {author} {\bibfnamefont {M.}~\bibnamefont {Takatsu}}, \bibinfo {author} {\bibfnamefont {T.}~\bibnamefont {Miyazawa}}, \bibinfo {author} {\bibfnamefont {T.}~\bibnamefont {Shibasaki}}, \bibinfo {author} {\bibfnamefont {Y.}~\bibnamefont {Watanabe}}, \bibinfo {author} {\bibfnamefont {K.}~\bibnamefont {Takemoto}}, \ and\ \bibinfo {author} {\bibfnamefont {H.}~\bibnamefont {Tamura}},\ }\bibfield  {title} {\enquote {\bibinfo {title} {Digital annealer for high-speed solving of combinatorial optimization problems and its applications},}\ }in\ \href {\doibase 10.1109/ASP-DAC47756.2020.9045100} {\emph {\bibinfo {booktitle} {2020 25th Asia and South Pacific Design Automation Conference (ASP-DAC)}}}\ (\bibinfo {year} {2020})\ pp.\ \bibinfo {pages} {667--672}\BibitemShut {NoStop}%
\bibitem [{\citenamefont {Aramon}\ \emph {et~al.}(2019)\citenamefont {Aramon}, \citenamefont {Rosenberg}, \citenamefont {Valiante}, \citenamefont {Miyazawa}, \citenamefont {Tamura},\ and\ \citenamefont {Katzgraber}}]{10.3389/fphy.2019.00048}%
  \BibitemOpen
  \bibfield  {author} {\bibinfo {author} {\bibfnamefont {M.}~\bibnamefont {Aramon}}, \bibinfo {author} {\bibfnamefont {G.}~\bibnamefont {Rosenberg}}, \bibinfo {author} {\bibfnamefont {E.}~\bibnamefont {Valiante}}, \bibinfo {author} {\bibfnamefont {T.}~\bibnamefont {Miyazawa}}, \bibinfo {author} {\bibfnamefont {H.}~\bibnamefont {Tamura}}, \ and\ \bibinfo {author} {\bibfnamefont {H.~G.}\ \bibnamefont {Katzgraber}},\ }\bibfield  {title} {\enquote {\bibinfo {title} {Physics-inspired optimization for quadratic unconstrained problems using a digital annealer},}\ }\href {https://www.frontiersin.org/journals/physics/articles/10.3389/fphy.2019.00048} {\bibfield  {journal} {\bibinfo  {journal} {Frontiers in Physics}\ }\textbf {\bibinfo {volume} {7}} (\bibinfo {year} {2019})}\BibitemShut {NoStop}%
\bibitem [{\citenamefont {Kirkpatrick}, \citenamefont {Gelatt},\ and\ \citenamefont {Vecchi}(1983)}]{doi:10.1126/science.220.4598.671}%
  \BibitemOpen
  \bibfield  {author} {\bibinfo {author} {\bibfnamefont {S.}~\bibnamefont {Kirkpatrick}}, \bibinfo {author} {\bibfnamefont {C.~D.}\ \bibnamefont {Gelatt}}, \ and\ \bibinfo {author} {\bibfnamefont {M.~P.}\ \bibnamefont {Vecchi}},\ }\bibfield  {title} {\enquote {\bibinfo {title} {Optimization by simulated annealing},}\ }\href {\doibase 10.1126/science.220.4598.671} {\bibfield  {journal} {\bibinfo  {journal} {Science}\ }\textbf {\bibinfo {volume} {220}},\ \bibinfo {pages} {671--680} (\bibinfo {year} {1983})}\BibitemShut {NoStop}%
\bibitem [{\citenamefont {{\v C}ern{\'y}}(1985)}]{Cerny:1985aa}%
  \BibitemOpen
  \bibfield  {author} {\bibinfo {author} {\bibfnamefont {V.}~\bibnamefont {{\v C}ern{\'y}}},\ }\bibfield  {title} {\enquote {\bibinfo {title} {Thermodynamical approach to the traveling salesman problem: An efficient simulation algorithm},}\ }\href {\doibase 10.1007/BF00940812} {\bibfield  {journal} {\bibinfo  {journal} {Journal of Optimization Theory and Applications}\ }\textbf {\bibinfo {volume} {45}},\ \bibinfo {pages} {41--51} (\bibinfo {year} {1985})}\BibitemShut {NoStop}%
\bibitem [{\citenamefont {Swendsen}\ and\ \citenamefont {Wang}(1986)}]{PhysRevLett.57.2607}%
  \BibitemOpen
  \bibfield  {author} {\bibinfo {author} {\bibfnamefont {R.~H.}\ \bibnamefont {Swendsen}}\ and\ \bibinfo {author} {\bibfnamefont {J.-S.}\ \bibnamefont {Wang}},\ }\bibfield  {title} {\enquote {\bibinfo {title} {Replica {Monte} {Carlo} simulation of spin-glasses},}\ }\href {\doibase 10.1103/PhysRevLett.57.2607} {\bibfield  {journal} {\bibinfo  {journal} {Physical Review Letters}\ }\textbf {\bibinfo {volume} {57}},\ \bibinfo {pages} {2607--2609} (\bibinfo {year} {1986})}\BibitemShut {NoStop}%
\bibitem [{\citenamefont {Hukushima}\ and\ \citenamefont {Nemoto}(1996)}]{doi:10.1143/JPSJ.65.1604}%
  \BibitemOpen
  \bibfield  {author} {\bibinfo {author} {\bibfnamefont {K.}~\bibnamefont {Hukushima}}\ and\ \bibinfo {author} {\bibfnamefont {K.}~\bibnamefont {Nemoto}},\ }\bibfield  {title} {\enquote {\bibinfo {title} {Exchange {Monte} {Carlo} method and application to spin glass simulations},}\ }\href {\doibase 10.1143/JPSJ.65.1604} {\bibfield  {journal} {\bibinfo  {journal} {Journal of the Physical Society of Japan}\ }\textbf {\bibinfo {volume} {65}},\ \bibinfo {pages} {1604--1608} (\bibinfo {year} {1996})}\BibitemShut {NoStop}%
\bibitem [{\citenamefont {Kingma}\ and\ \citenamefont {Ba}(2014)}]{1412.6980}%
  \BibitemOpen
  \bibfield  {author} {\bibinfo {author} {\bibfnamefont {D.~P.}\ \bibnamefont {Kingma}}\ and\ \bibinfo {author} {\bibfnamefont {J.}~\bibnamefont {Ba}},\ }\href@noop {} {\enquote {\bibinfo {title} {Adam: A method for stochastic optimization},}\ } (\bibinfo {year} {2014}),\ \Eprint {http://arxiv.org/abs/arXiv:1412.6980} {arXiv:1412.6980} \BibitemShut {NoStop}%
\bibitem [{\citenamefont {Paszke}\ \emph {et~al.}(2019{\natexlab{b}})\citenamefont {Paszke}, \citenamefont {Gross}, \citenamefont {Massa}, \citenamefont {Lerer}, \citenamefont {Bradbury}, \citenamefont {Chanan}, \citenamefont {Killeen}, \citenamefont {Lin}, \citenamefont {Gimelshein}, \citenamefont {Antiga}, \citenamefont {Desmaison}, \citenamefont {K\"{o}pf}, \citenamefont {Yang}, \citenamefont {DeVito}, \citenamefont {Raison}, \citenamefont {Tejani}, \citenamefont {Chilamkurthy}, \citenamefont {Steiner}, \citenamefont {Fang}, \citenamefont {Bai},\ and\ \citenamefont {Chintala}}]{1912.01703}%
  \BibitemOpen
  \bibfield  {author} {\bibinfo {author} {\bibfnamefont {A.}~\bibnamefont {Paszke}}, \bibinfo {author} {\bibfnamefont {S.}~\bibnamefont {Gross}}, \bibinfo {author} {\bibfnamefont {F.}~\bibnamefont {Massa}}, \bibinfo {author} {\bibfnamefont {A.}~\bibnamefont {Lerer}}, \bibinfo {author} {\bibfnamefont {J.}~\bibnamefont {Bradbury}}, \bibinfo {author} {\bibfnamefont {G.}~\bibnamefont {Chanan}}, \bibinfo {author} {\bibfnamefont {T.}~\bibnamefont {Killeen}}, \bibinfo {author} {\bibfnamefont {Z.}~\bibnamefont {Lin}}, \bibinfo {author} {\bibfnamefont {N.}~\bibnamefont {Gimelshein}}, \bibinfo {author} {\bibfnamefont {L.}~\bibnamefont {Antiga}}, \bibinfo {author} {\bibfnamefont {A.}~\bibnamefont {Desmaison}}, \bibinfo {author} {\bibfnamefont {A.}~\bibnamefont {K\"{o}pf}}, \bibinfo {author} {\bibfnamefont {E.}~\bibnamefont {Yang}}, \bibinfo {author} {\bibfnamefont {Z.}~\bibnamefont {DeVito}}, \bibinfo {author} {\bibfnamefont {M.}~\bibnamefont {Raison}}, \bibinfo {author} {\bibfnamefont {A.}~\bibnamefont {Tejani}},
  \bibinfo {author} {\bibfnamefont {S.}~\bibnamefont {Chilamkurthy}}, \bibinfo {author} {\bibfnamefont {B.}~\bibnamefont {Steiner}}, \bibinfo {author} {\bibfnamefont {L.}~\bibnamefont {Fang}}, \bibinfo {author} {\bibfnamefont {J.}~\bibnamefont {Bai}}, \ and\ \bibinfo {author} {\bibfnamefont {S.}~\bibnamefont {Chintala}},\ }\enquote {\bibinfo {title} {Pytorch: an imperative style, high-performance deep learning library},}\ in\ \href@noop {} {\emph {\bibinfo {booktitle} {Proceedings of the 33rd International Conference on Neural Information Processing Systems}}}\ (\bibinfo  {publisher} {Curran Associates Inc.},\ \bibinfo {address} {Red Hook, NY, USA},\ \bibinfo {year} {2019})\BibitemShut {NoStop}%
\bibitem [{fix({\natexlab{c}})}]{fixstarsamplifybbopt}%
  \BibitemOpen
  \href@noop {} {\enquote {\bibinfo {title} {{Amplify-BBOpt}},}\ } ({\natexlab{c}}),\ \bibinfo {note} {\url{https://amplify.fixstars.com/en/docs/amplify-bbopt/v0/}}\BibitemShut {NoStop}%
\bibitem [{\citenamefont {Kim}\ \emph {et~al.}(2024{\natexlab{a}})\citenamefont {Kim}, \citenamefont {Park}, \citenamefont {Moon}, \citenamefont {Zhang}, \citenamefont {Hwang}, \citenamefont {Kim}, \citenamefont {Luo},\ and\ \citenamefont {Lee}}]{Kim2024}%
  \BibitemOpen
  \bibfield  {author} {\bibinfo {author} {\bibfnamefont {S.}~\bibnamefont {Kim}}, \bibinfo {author} {\bibfnamefont {S.-J.}\ \bibnamefont {Park}}, \bibinfo {author} {\bibfnamefont {S.}~\bibnamefont {Moon}}, \bibinfo {author} {\bibfnamefont {Q.}~\bibnamefont {Zhang}}, \bibinfo {author} {\bibfnamefont {S.}~\bibnamefont {Hwang}}, \bibinfo {author} {\bibfnamefont {S.-K.}\ \bibnamefont {Kim}}, \bibinfo {author} {\bibfnamefont {T.}~\bibnamefont {Luo}}, \ and\ \bibinfo {author} {\bibfnamefont {E.}~\bibnamefont {Lee}},\ }\bibfield  {title} {\enquote {\bibinfo {title} {Quantum annealing-aided design of an ultrathin-metamaterial optical diode},}\ }\href {http://dx.doi.org/10.1186/s40580-024-00425-6} {\bibfield  {journal} {\bibinfo  {journal} {Nano Convergence}\ }\textbf {\bibinfo {volume} {11}} (\bibinfo {year} {2024}{\natexlab{a}})}\BibitemShut {NoStop}%
\bibitem [{\citenamefont {Kim}\ \emph {et~al.}(2023)\citenamefont {Kim}, \citenamefont {Wu}, \citenamefont {Jian}, \citenamefont {Xiong},\ and\ \citenamefont {Luo}}]{Kim:2023aa}%
  \BibitemOpen
  \bibfield  {author} {\bibinfo {author} {\bibfnamefont {S.}~\bibnamefont {Kim}}, \bibinfo {author} {\bibfnamefont {S.}~\bibnamefont {Wu}}, \bibinfo {author} {\bibfnamefont {R.}~\bibnamefont {Jian}}, \bibinfo {author} {\bibfnamefont {G.}~\bibnamefont {Xiong}}, \ and\ \bibinfo {author} {\bibfnamefont {T.}~\bibnamefont {Luo}},\ }\bibfield  {title} {\enquote {\bibinfo {title} {Design of a high-performance titanium nitride metastructure-based solar absorber using quantum computing-assisted optimization},}\ }\href {\doibase 10.1021/acsami.3c08214} {\bibfield  {journal} {\bibinfo  {journal} {ACS Applied Materials \& Interfaces}\ }\textbf {\bibinfo {volume} {15}},\ \bibinfo {pages} {40606--40613} (\bibinfo {year} {2023})}\BibitemShut {NoStop}%
\bibitem [{\citenamefont {Kim}\ \emph {et~al.}(2024{\natexlab{b}})\citenamefont {Kim}, \citenamefont {Jung}, \citenamefont {Bobbitt}, \citenamefont {Lee},\ and\ \citenamefont {Luo}}]{KIM2024101847}%
  \BibitemOpen
  \bibfield  {author} {\bibinfo {author} {\bibfnamefont {S.}~\bibnamefont {Kim}}, \bibinfo {author} {\bibfnamefont {S.}~\bibnamefont {Jung}}, \bibinfo {author} {\bibfnamefont {A.}~\bibnamefont {Bobbitt}}, \bibinfo {author} {\bibfnamefont {E.}~\bibnamefont {Lee}}, \ and\ \bibinfo {author} {\bibfnamefont {T.}~\bibnamefont {Luo}},\ }\bibfield  {title} {\enquote {\bibinfo {title} {Wide-angle spectral filter for energy-saving windows designed by quantum annealing-enhanced active learning},}\ }\href {\doibase https://doi.org/10.1016/j.xcrp.2024.101847} {\bibfield  {journal} {\bibinfo  {journal} {Cell Reports Physical Science}\ }\textbf {\bibinfo {volume} {5}},\ \bibinfo {pages} {101847} (\bibinfo {year} {2024}{\natexlab{b}})}\BibitemShut {NoStop}%
\bibitem [{\citenamefont {Guo}\ \emph {et~al.}(2024)\citenamefont {Guo}, \citenamefont {Kitai}, \citenamefont {Jippo},\ and\ \citenamefont {Shiomi}}]{arXiv:2408.05799}%
  \BibitemOpen
  \bibfield  {author} {\bibinfo {author} {\bibfnamefont {J.}~\bibnamefont {Guo}}, \bibinfo {author} {\bibfnamefont {K.}~\bibnamefont {Kitai}}, \bibinfo {author} {\bibfnamefont {H.}~\bibnamefont {Jippo}}, \ and\ \bibinfo {author} {\bibfnamefont {J.}~\bibnamefont {Shiomi}},\ }\bibfield  {title} {\enquote {\bibinfo {title} {Boosting the quality factor of tamm structures to millions by quantum inspired classical annealer with factorization machine},}\ }\href@noop {} {\bibfield  {journal} {\bibinfo  {journal} {arXiv:2408.05799}\ } (\bibinfo {year} {2024})}\BibitemShut {NoStop}%
\bibitem [{\citenamefont {Urushihara}\ \emph {et~al.}(2023)\citenamefont {Urushihara}, \citenamefont {Karube}, \citenamefont {Yamaguchi},\ and\ \citenamefont {Tamura}}]{https://doi.org/10.1002/adpr.202300226}%
  \BibitemOpen
  \bibfield  {author} {\bibinfo {author} {\bibfnamefont {M.}~\bibnamefont {Urushihara}}, \bibinfo {author} {\bibfnamefont {M.}~\bibnamefont {Karube}}, \bibinfo {author} {\bibfnamefont {K.}~\bibnamefont {Yamaguchi}}, \ and\ \bibinfo {author} {\bibfnamefont {R.}~\bibnamefont {Tamura}},\ }\bibfield  {title} {\enquote {\bibinfo {title} {Optimization of core--shell nanoparticles using a combination of machine learning and ising machine},}\ }\href {\doibase https://doi.org/10.1002/adpr.202300226} {\bibfield  {journal} {\bibinfo  {journal} {Advanced Photonics Research}\ }\textbf {\bibinfo {volume} {4}},\ \bibinfo {pages} {2300226} (\bibinfo {year} {2023})}\BibitemShut {NoStop}%
\bibitem [{\citenamefont {Tamura}\ \emph {et~al.}(2024)\citenamefont {Tamura}, \citenamefont {Nagata}, \citenamefont {Sodeyama}, \citenamefont {Nakamura}, \citenamefont {Tokuhira}, \citenamefont {Shibata}, \citenamefont {Hammura}, \citenamefont {Sugisawa}, \citenamefont {Kawamura}, \citenamefont {Tsurimoto}, \citenamefont {Naito}, \citenamefont {Demura},\ and\ \citenamefont {Nakanishi}}]{doi:10.1080/14686996.2024.2388016}%
  \BibitemOpen
  \bibfield  {author} {\bibinfo {author} {\bibfnamefont {R.}~\bibnamefont {Tamura}}, \bibinfo {author} {\bibfnamefont {K.}~\bibnamefont {Nagata}}, \bibinfo {author} {\bibfnamefont {K.}~\bibnamefont {Sodeyama}}, \bibinfo {author} {\bibfnamefont {K.}~\bibnamefont {Nakamura}}, \bibinfo {author} {\bibfnamefont {T.}~\bibnamefont {Tokuhira}}, \bibinfo {author} {\bibfnamefont {S.}~\bibnamefont {Shibata}}, \bibinfo {author} {\bibfnamefont {K.}~\bibnamefont {Hammura}}, \bibinfo {author} {\bibfnamefont {H.}~\bibnamefont {Sugisawa}}, \bibinfo {author} {\bibfnamefont {M.}~\bibnamefont {Kawamura}}, \bibinfo {author} {\bibfnamefont {T.}~\bibnamefont {Tsurimoto}}, \bibinfo {author} {\bibfnamefont {M.}~\bibnamefont {Naito}}, \bibinfo {author} {\bibfnamefont {M.}~\bibnamefont {Demura}}, \ and\ \bibinfo {author} {\bibfnamefont {T.}~\bibnamefont {Nakanishi}},\ }\bibfield  {title} {\enquote {\bibinfo {title} {Machine learning prediction of the mechanical properties of injection-molded polypropylene through {X}-ray diffraction
  analysis},}\ }\href {\doibase 10.1080/14686996.2024.2388016} {\bibfield  {journal} {\bibinfo  {journal} {Science and Technology of Advanced Materials}\ }\textbf {\bibinfo {volume} {25}},\ \bibinfo {pages} {2388016} (\bibinfo {year} {2024})}\BibitemShut {NoStop}%
\bibitem [{\citenamefont {Huang}\ and\ \citenamefont {Ju}(2024)}]{Huang2024}%
  \BibitemOpen
  \bibfield  {author} {\bibinfo {author} {\bibfnamefont {X.}~\bibnamefont {Huang}}\ and\ \bibinfo {author} {\bibfnamefont {S.}~\bibnamefont {Ju}},\ }\bibfield  {title} {\enquote {\bibinfo {title} {Tutorial: {AI}-assisted exploration and active design of polymers with high intrinsic thermal conductivity},}\ }\href {http://dx.doi.org/10.1063/5.0201522} {\bibfield  {journal} {\bibinfo  {journal} {Journal of Applied Physics}\ }\textbf {\bibinfo {volume} {135}} (\bibinfo {year} {2024})}\BibitemShut {NoStop}%
\bibitem [{\citenamefont {Morgan}(1965)}]{doi:10.1021/c160017a018}%
  \BibitemOpen
  \bibfield  {author} {\bibinfo {author} {\bibfnamefont {H.~L.}\ \bibnamefont {Morgan}},\ }\bibfield  {title} {\enquote {\bibinfo {title} {The generation of a unique machine description for chemical structures-{A} technique developed at chemical abstracts service.}}\ }\href {\doibase 10.1021/c160017a018} {\bibfield  {journal} {\bibinfo  {journal} {Journal of Chemical Documentation}\ }\textbf {\bibinfo {volume} {5}},\ \bibinfo {pages} {107--113} (\bibinfo {year} {1965})}\BibitemShut {NoStop}%
\bibitem [{dim()}]{dimod}%
  \BibitemOpen
  \href@noop {} {\enquote {\bibinfo {title} {{GitHub}, dimod},}\ }\bibinfo {note} {\url{https://github.com/dwavesystems/dimod}}\BibitemShut {NoStop}%
\bibitem [{\citenamefont {Hida}\ \emph {et~al.}(2024)\citenamefont {Hida}, \citenamefont {Ikeda}, \citenamefont {Maruo}, \citenamefont {Sato},\ and\ \citenamefont {Yamazaki}}]{HIDA2024}%
  \BibitemOpen
  \bibfield  {author} {\bibinfo {author} {\bibfnamefont {M.}~\bibnamefont {Hida}}, \bibinfo {author} {\bibfnamefont {H.}~\bibnamefont {Ikeda}}, \bibinfo {author} {\bibfnamefont {A.}~\bibnamefont {Maruo}}, \bibinfo {author} {\bibfnamefont {M.}~\bibnamefont {Sato}}, \ and\ \bibinfo {author} {\bibfnamefont {T.}~\bibnamefont {Yamazaki}},\ }\bibfield  {title} {\enquote {\bibinfo {title} {Topology optimization of analog circuit design via global optimization using factorization machines with digital annealer},}\ }\href {\doibase 10.1299/jamdsm.2024jamdsm0076} {\bibfield  {journal} {\bibinfo  {journal} {Journal of Advanced Mechanical Design, Systems, and Manufacturing}\ }\textbf {\bibinfo {volume} {18}},\ \bibinfo {pages} {JAMDSM0076} (\bibinfo {year} {2024})}\BibitemShut {NoStop}%
\bibitem [{\citenamefont {Tibshirani}(1996)}]{tibshirani_regression_1996}%
  \BibitemOpen
  \bibfield  {author} {\bibinfo {author} {\bibfnamefont {R.}~\bibnamefont {Tibshirani}},\ }\bibfield  {title} {\enquote {\bibinfo {title} {Regression {Shrinkage} and {Selection} via the {Lasso}},}\ }\href {http://www.jstor.org/stable/2346178} {\bibfield  {journal} {\bibinfo  {journal} {Journal of the Royal Statistical Society. Series B (Methodological)}\ }\textbf {\bibinfo {volume} {58}} (\bibinfo {year} {1996})}\BibitemShut {NoStop}%
\bibitem [{\citenamefont {Nagata}\ \emph {et~al.}(2015)\citenamefont {Nagata}, \citenamefont {Kitazono}, \citenamefont {Nakajima}, \citenamefont {Eifuku}, \citenamefont {Tamura},\ and\ \citenamefont {Okada}}]{Nagata:2015aa}%
  \BibitemOpen
  \bibfield  {author} {\bibinfo {author} {\bibfnamefont {K.}~\bibnamefont {Nagata}}, \bibinfo {author} {\bibfnamefont {J.}~\bibnamefont {Kitazono}}, \bibinfo {author} {\bibfnamefont {S.}~\bibnamefont {Nakajima}}, \bibinfo {author} {\bibfnamefont {S.}~\bibnamefont {Eifuku}}, \bibinfo {author} {\bibfnamefont {R.}~\bibnamefont {Tamura}}, \ and\ \bibinfo {author} {\bibfnamefont {M.}~\bibnamefont {Okada}},\ }\bibfield  {title} {\enquote {\bibinfo {title} {An exhaustive search and stability of sparse estimation for feature selection problem},}\ }\href {\doibase 10.2197/ipsjtrans.8.25} {\bibfield  {journal} {\bibinfo  {journal} {IPSJ Online Transactions}\ }\textbf {\bibinfo {volume} {8}},\ \bibinfo {pages} {25--32} (\bibinfo {year} {2015})}\BibitemShut {NoStop}%
\bibitem [{\citenamefont {Igarashi}\ \emph {et~al.}(2018)\citenamefont {Igarashi}, \citenamefont {Takenaka}, \citenamefont {Nakanishi-Ohno}, \citenamefont {Uemura}, \citenamefont {Ikeda},\ and\ \citenamefont {Okada}}]{doi:10.7566/JPSJ.87.044802}%
  \BibitemOpen
  \bibfield  {author} {\bibinfo {author} {\bibfnamefont {Y.}~\bibnamefont {Igarashi}}, \bibinfo {author} {\bibfnamefont {H.}~\bibnamefont {Takenaka}}, \bibinfo {author} {\bibfnamefont {Y.}~\bibnamefont {Nakanishi-Ohno}}, \bibinfo {author} {\bibfnamefont {M.}~\bibnamefont {Uemura}}, \bibinfo {author} {\bibfnamefont {S.}~\bibnamefont {Ikeda}}, \ and\ \bibinfo {author} {\bibfnamefont {M.}~\bibnamefont {Okada}},\ }\bibfield  {title} {\enquote {\bibinfo {title} {Exhaustive search for sparse variable selection in linear regression},}\ }\href {\doibase 10.7566/JPSJ.87.044802} {\bibfield  {journal} {\bibinfo  {journal} {Journal of the Physical Society of Japan}\ }\textbf {\bibinfo {volume} {87}},\ \bibinfo {pages} {044802} (\bibinfo {year} {2018})}\BibitemShut {NoStop}%
\bibitem [{\citenamefont {Nagata}, \citenamefont {Sugita},\ and\ \citenamefont {Okada}(2012)}]{NAGATA201282}%
  \BibitemOpen
  \bibfield  {author} {\bibinfo {author} {\bibfnamefont {K.}~\bibnamefont {Nagata}}, \bibinfo {author} {\bibfnamefont {S.}~\bibnamefont {Sugita}}, \ and\ \bibinfo {author} {\bibfnamefont {M.}~\bibnamefont {Okada}},\ }\bibfield  {title} {\enquote {\bibinfo {title} {Bayesian spectral deconvolution with the exchange {Monte Carlo} method},}\ }\href {\doibase https://doi.org/10.1016/j.neunet.2011.12.001} {\bibfield  {journal} {\bibinfo  {journal} {Neural Networks}\ }\textbf {\bibinfo {volume} {28}},\ \bibinfo {pages} {82--89} (\bibinfo {year} {2012})}\BibitemShut {NoStop}%
\bibitem [{\citenamefont {Matsumura}\ \emph {et~al.}(2023)\citenamefont {Matsumura}, \citenamefont {Nagamura}, \citenamefont {Akaho}, \citenamefont {Nagata},\ and\ \citenamefont {Ando}}]{doi:10.1080/27660400.2022.2159753}%
  \BibitemOpen
  \bibfield  {author} {\bibinfo {author} {\bibfnamefont {T.}~\bibnamefont {Matsumura}}, \bibinfo {author} {\bibfnamefont {N.}~\bibnamefont {Nagamura}}, \bibinfo {author} {\bibfnamefont {S.}~\bibnamefont {Akaho}}, \bibinfo {author} {\bibfnamefont {K.}~\bibnamefont {Nagata}}, \ and\ \bibinfo {author} {\bibfnamefont {Y.}~\bibnamefont {Ando}},\ }\bibfield  {title} {\enquote {\bibinfo {title} {High-throughput {XPS} spectrum modeling with autonomous background subtraction for 3d$_{5/2}$ peak mapping of {SnS}},}\ }\href {\doibase 10.1080/27660400.2022.2159753} {\bibfield  {journal} {\bibinfo  {journal} {Science and Technology of Advanced Materials: Methods}\ }\textbf {\bibinfo {volume} {3}},\ \bibinfo {pages} {2159753} (\bibinfo {year} {2023})}\BibitemShut {NoStop}%
\bibitem [{\citenamefont {Sukegawa}\ \emph {et~al.}(2017)\citenamefont {Sukegawa}, \citenamefont {Kato}, \citenamefont {Belmoubarik}, \citenamefont {Cheng}, \citenamefont {Daibou}, \citenamefont {Shimomura}, \citenamefont {Kamiguchi}, \citenamefont {Ito}, \citenamefont {Yoda}, \citenamefont {Ohkubo}, \citenamefont {Mitani},\ and\ \citenamefont {Hono}}]{10.1063/1.4977946}%
  \BibitemOpen
  \bibfield  {author} {\bibinfo {author} {\bibfnamefont {H.}~\bibnamefont {Sukegawa}}, \bibinfo {author} {\bibfnamefont {Y.}~\bibnamefont {Kato}}, \bibinfo {author} {\bibfnamefont {M.}~\bibnamefont {Belmoubarik}}, \bibinfo {author} {\bibfnamefont {P.-H.}\ \bibnamefont {Cheng}}, \bibinfo {author} {\bibfnamefont {T.}~\bibnamefont {Daibou}}, \bibinfo {author} {\bibfnamefont {N.}~\bibnamefont {Shimomura}}, \bibinfo {author} {\bibfnamefont {Y.}~\bibnamefont {Kamiguchi}}, \bibinfo {author} {\bibfnamefont {J.}~\bibnamefont {Ito}}, \bibinfo {author} {\bibfnamefont {H.}~\bibnamefont {Yoda}}, \bibinfo {author} {\bibfnamefont {T.}~\bibnamefont {Ohkubo}}, \bibinfo {author} {\bibfnamefont {S.}~\bibnamefont {Mitani}}, \ and\ \bibinfo {author} {\bibfnamefont {K.}~\bibnamefont {Hono}},\ }\bibfield  {title} {\enquote {\bibinfo {title} {{MgGa$_2$O$_4$ spinel barrier for magnetic tunnel junctions: Coherent tunneling and low barrier height}},}\ }\href {\doibase 10.1063/1.4977946} {\bibfield  {journal} {\bibinfo  {journal}
  {Applied Physics Letters}\ }\textbf {\bibinfo {volume} {110}},\ \bibinfo {pages} {122404} (\bibinfo {year} {2017})}\BibitemShut {NoStop}%
\bibitem [{\citenamefont {Suga}, \citenamefont {Maruo},\ and\ \citenamefont {Jippo}(2024)}]{202420244237}%
  \BibitemOpen
  \bibfield  {author} {\bibinfo {author} {\bibfnamefont {Y.}~\bibnamefont {Suga}}, \bibinfo {author} {\bibfnamefont {A.}~\bibnamefont {Maruo}}, \ and\ \bibinfo {author} {\bibfnamefont {H.}~\bibnamefont {Jippo}},\ }\bibfield  {title} {\enquote {\bibinfo {title} {A feasibility study for quantum computing methodologies in automotive advanced material investigation},}\ }\href {\doibase 10.11351/jsaeronbun.55.621} {\bibfield  {journal} {\bibinfo  {journal} {Transactions of Society of Automotive Engineers of Japan}\ }\textbf {\bibinfo {volume} {55}},\ \bibinfo {pages} {621--627} (\bibinfo {year} {2024})}\BibitemShut {NoStop}%
\bibitem [{\citenamefont {Matsumori}, \citenamefont {Taki},\ and\ \citenamefont {Kadowaki}(2022)}]{Matsumori:2022aa}%
  \BibitemOpen
  \bibfield  {author} {\bibinfo {author} {\bibfnamefont {T.}~\bibnamefont {Matsumori}}, \bibinfo {author} {\bibfnamefont {M.}~\bibnamefont {Taki}}, \ and\ \bibinfo {author} {\bibfnamefont {T.}~\bibnamefont {Kadowaki}},\ }\bibfield  {title} {\enquote {\bibinfo {title} {Application of qubo solver using black-box optimization to structural design for resonance avoidance},}\ }\href {\doibase 10.1038/s41598-022-16149-8} {\bibfield  {journal} {\bibinfo  {journal} {Scientific Reports}\ }\textbf {\bibinfo {volume} {12}},\ \bibinfo {pages} {12143} (\bibinfo {year} {2022})}\BibitemShut {NoStop}%
\bibitem [{fix({\natexlab{d}})}]{fixwing}%
  \BibitemOpen
  \href@noop {} {\enquote {\bibinfo {title} {Black-box optimization of the airfoil geometry with fluid flow simulation},}\ } ({\natexlab{d}}),\ \bibinfo {note} {\url{https://amplify.fixstars.com/en/demo/fmqa_3_aerofoil}}\BibitemShut {NoStop}%
\bibitem [{\citenamefont {Lin}\ \emph {et~al.}(2025{\natexlab{b}})\citenamefont {Lin}, \citenamefont {Tada}, \citenamefont {Koizumi}, \citenamefont {Sumita}, \citenamefont {Tsuda},\ and\ \citenamefont {Tamura}}]{Lin2025}%
  \BibitemOpen
  \bibfield  {author} {\bibinfo {author} {\bibfnamefont {J.}~\bibnamefont {Lin}}, \bibinfo {author} {\bibfnamefont {T.}~\bibnamefont {Tada}}, \bibinfo {author} {\bibfnamefont {A.}~\bibnamefont {Koizumi}}, \bibinfo {author} {\bibfnamefont {M.}~\bibnamefont {Sumita}}, \bibinfo {author} {\bibfnamefont {K.}~\bibnamefont {Tsuda}}, \ and\ \bibinfo {author} {\bibfnamefont {R.}~\bibnamefont {Tamura}},\ }\bibfield  {title} {\enquote {\bibinfo {title} {Determination of stable proton configurations by black-box optimization using an {Ising} machine},}\ }\href {\doibase 10.1021/acs.jpcc.4c07104} {\bibfield  {journal} {\bibinfo  {journal} {The Journal of Physical Chemistry C}\ }\textbf {\bibinfo {volume} {129}},\ \bibinfo {pages} {2332–2340} (\bibinfo {year} {2025}{\natexlab{b}})}\BibitemShut {NoStop}%
\bibitem [{\citenamefont {Inoue}\ \emph {et~al.}(2022)\citenamefont {Inoue}, \citenamefont {Seki}, \citenamefont {Tanaka}, \citenamefont {Togawa}, \citenamefont {Ishizaki},\ and\ \citenamefont {Noda}}]{Inoue:22}%
  \BibitemOpen
  \bibfield  {author} {\bibinfo {author} {\bibfnamefont {T.}~\bibnamefont {Inoue}}, \bibinfo {author} {\bibfnamefont {Y.}~\bibnamefont {Seki}}, \bibinfo {author} {\bibfnamefont {S.}~\bibnamefont {Tanaka}}, \bibinfo {author} {\bibfnamefont {N.}~\bibnamefont {Togawa}}, \bibinfo {author} {\bibfnamefont {K.}~\bibnamefont {Ishizaki}}, \ and\ \bibinfo {author} {\bibfnamefont {S.}~\bibnamefont {Noda}},\ }\bibfield  {title} {\enquote {\bibinfo {title} {Towards optimization of photonic-crystal surface-emitting lasers via quantum annealing},}\ }\href {\doibase 10.1364/OE.476839} {\bibfield  {journal} {\bibinfo  {journal} {Opt. Express}\ }\textbf {\bibinfo {volume} {30}},\ \bibinfo {pages} {43503--43512} (\bibinfo {year} {2022})}\BibitemShut {NoStop}%
\bibitem [{\citenamefont {Kondo}, \citenamefont {Kohira},\ and\ \citenamefont {Minamoto}(2025)}]{kondo2024}%
  \BibitemOpen
  \bibfield  {author} {\bibinfo {author} {\bibfnamefont {T.}~\bibnamefont {Kondo}}, \bibinfo {author} {\bibfnamefont {T.}~\bibnamefont {Kohira}}, \ and\ \bibinfo {author} {\bibfnamefont {Y.}~\bibnamefont {Minamoto}},\ }\bibfield  {title} {\enquote {\bibinfo {title} {Simultaneous structure design optimization of multiple car models using {FMQA}},}\ }\href@noop {} {\bibfield  {journal} {\bibinfo  {journal} {Transactions of the Society of Automotive Engineers of Japan}\ }\textbf {\bibinfo {volume} {56}},\ \bibinfo {pages} {229--236} (\bibinfo {year} {2025})}\BibitemShut {NoStop}%
\bibitem [{\citenamefont {Samuel}\ and\ \citenamefont {Turnbull}(2007)}]{Samuel:2007aa}%
  \BibitemOpen
  \bibfield  {author} {\bibinfo {author} {\bibfnamefont {I.~D.~W.}\ \bibnamefont {Samuel}}\ and\ \bibinfo {author} {\bibfnamefont {G.~A.}\ \bibnamefont {Turnbull}},\ }\bibfield  {title} {\enquote {\bibinfo {title} {Organic semiconductor lasers},}\ }\href {\doibase 10.1021/cr050152i} {\bibfield  {journal} {\bibinfo  {journal} {Chemical Reviews}\ }\textbf {\bibinfo {volume} {107}},\ \bibinfo {pages} {1272--1295} (\bibinfo {year} {2007})}\BibitemShut {NoStop}%
\bibitem [{\citenamefont {Kennedy}\ and\ \citenamefont {Eberhart}(1995)}]{488968}%
  \BibitemOpen
  \bibfield  {author} {\bibinfo {author} {\bibfnamefont {J.}~\bibnamefont {Kennedy}}\ and\ \bibinfo {author} {\bibfnamefont {R.}~\bibnamefont {Eberhart}},\ }\bibfield  {title} {\enquote {\bibinfo {title} {Particle swarm optimization},}\ }in\ \href {\doibase 10.1109/ICNN.1995.488968} {\emph {\bibinfo {booktitle} {Proceedings of ICNN'95 - International Conference on Neural Networks}}},\ Vol.~\bibinfo {volume} {4}\ (\bibinfo {year} {1995})\ pp.\ \bibinfo {pages} {1942--1948 vol.4}\BibitemShut {NoStop}%
\bibitem [{\citenamefont {Wang}, \citenamefont {Tan},\ and\ \citenamefont {Liu}(2018)}]{Wang:2018aa}%
  \BibitemOpen
  \bibfield  {author} {\bibinfo {author} {\bibfnamefont {D.}~\bibnamefont {Wang}}, \bibinfo {author} {\bibfnamefont {D.}~\bibnamefont {Tan}}, \ and\ \bibinfo {author} {\bibfnamefont {L.}~\bibnamefont {Liu}},\ }\bibfield  {title} {\enquote {\bibinfo {title} {Particle swarm optimization algorithm: an overview},}\ }\href {\doibase 10.1007/s00500-016-2474-6} {\bibfield  {journal} {\bibinfo  {journal} {Soft Computing}\ }\textbf {\bibinfo {volume} {22}},\ \bibinfo {pages} {387--408} (\bibinfo {year} {2018})}\BibitemShut {NoStop}%
\bibitem [{\citenamefont {Whitley}(1994)}]{Whitley:1994aa}%
  \BibitemOpen
  \bibfield  {author} {\bibinfo {author} {\bibfnamefont {D.}~\bibnamefont {Whitley}},\ }\bibfield  {title} {\enquote {\bibinfo {title} {A genetic algorithm tutorial},}\ }\href {\doibase 10.1007/BF00175354} {\bibfield  {journal} {\bibinfo  {journal} {Statistics and Computing}\ }\textbf {\bibinfo {volume} {4}},\ \bibinfo {pages} {65--85} (\bibinfo {year} {1994})}\BibitemShut {NoStop}%
\bibitem [{\citenamefont {Suzuki}, \citenamefont {Minami},\ and\ \citenamefont {Inamuro}(2015)}]{Suzuki_Minami_Inamuro_2015}%
  \BibitemOpen
  \bibfield  {author} {\bibinfo {author} {\bibfnamefont {K.}~\bibnamefont {Suzuki}}, \bibinfo {author} {\bibfnamefont {K.}~\bibnamefont {Minami}}, \ and\ \bibinfo {author} {\bibfnamefont {T.}~\bibnamefont {Inamuro}},\ }\bibfield  {title} {\enquote {\bibinfo {title} {Lift and thrust generation by a butterfly-like flapping wing--body model: immersed boundary--lattice boltzmann simulations},}\ }\href {\doibase 10.1017/jfm.2015.57} {\bibfield  {journal} {\bibinfo  {journal} {Journal of Fluid Mechanics}\ }\textbf {\bibinfo {volume} {767}},\ \bibinfo {pages} {659--695} (\bibinfo {year} {2015})}\BibitemShut {NoStop}%
\bibitem [{\citenamefont {Bando}\ \emph {et~al.}(1994)\citenamefont {Bando}, \citenamefont {Hasebe}, \citenamefont {Nakayama}, \citenamefont {Shibata},\ and\ \citenamefont {Sugiyama}}]{Bando:1994aa}%
  \BibitemOpen
  \bibfield  {author} {\bibinfo {author} {\bibfnamefont {M.}~\bibnamefont {Bando}}, \bibinfo {author} {\bibfnamefont {K.}~\bibnamefont {Hasebe}}, \bibinfo {author} {\bibfnamefont {A.}~\bibnamefont {Nakayama}}, \bibinfo {author} {\bibfnamefont {A.}~\bibnamefont {Shibata}}, \ and\ \bibinfo {author} {\bibfnamefont {Y.}~\bibnamefont {Sugiyama}},\ }\bibfield  {title} {\enquote {\bibinfo {title} {Structure stability of congestion in traffic dynamics},}\ }\href {\doibase 10.1007/BF03167222} {\bibfield  {journal} {\bibinfo  {journal} {Japan Journal of Industrial and Applied Mathematics}\ }\textbf {\bibinfo {volume} {11}},\ \bibinfo {pages} {203--223} (\bibinfo {year} {1994})}\BibitemShut {NoStop}%
\bibitem [{\citenamefont {Bando}\ \emph {et~al.}(1995)\citenamefont {Bando}, \citenamefont {Hasebe}, \citenamefont {Nakayama}, \citenamefont {Shibata},\ and\ \citenamefont {Sugiyama}}]{PhysRevE.51.1035}%
  \BibitemOpen
  \bibfield  {author} {\bibinfo {author} {\bibfnamefont {M.}~\bibnamefont {Bando}}, \bibinfo {author} {\bibfnamefont {K.}~\bibnamefont {Hasebe}}, \bibinfo {author} {\bibfnamefont {A.}~\bibnamefont {Nakayama}}, \bibinfo {author} {\bibfnamefont {A.}~\bibnamefont {Shibata}}, \ and\ \bibinfo {author} {\bibfnamefont {Y.}~\bibnamefont {Sugiyama}},\ }\bibfield  {title} {\enquote {\bibinfo {title} {Dynamical model of traffic congestion and numerical simulation},}\ }\href {\doibase 10.1103/PhysRevE.51.1035} {\bibfield  {journal} {\bibinfo  {journal} {Physical Review E}\ }\textbf {\bibinfo {volume} {51}},\ \bibinfo {pages} {1035--1042} (\bibinfo {year} {1995})}\BibitemShut {NoStop}%
\bibitem [{\citenamefont {Kohira}\ \emph {et~al.}(2018)\citenamefont {Kohira}, \citenamefont {Kemmotsu}, \citenamefont {Akira},\ and\ \citenamefont {Tatsukawa}}]{kohira2018}%
  \BibitemOpen
  \bibfield  {author} {\bibinfo {author} {\bibfnamefont {T.}~\bibnamefont {Kohira}}, \bibinfo {author} {\bibfnamefont {H.}~\bibnamefont {Kemmotsu}}, \bibinfo {author} {\bibfnamefont {O.}~\bibnamefont {Akira}}, \ and\ \bibinfo {author} {\bibfnamefont {T.}~\bibnamefont {Tatsukawa}},\ }\bibfield  {title} {\enquote {\bibinfo {title} {Proposal of benchmark problem based on real-world car structure design optimization},}\ }\href {\doibase 10.1145/3205651.320570} {\bibfield  {journal} {\bibinfo  {journal} {Proceedings of the Genetic and Evolutionary Computation Conference Companion}\ ,\ \bibinfo {pages} {183--184}} (\bibinfo {year} {2018})}\BibitemShut {NoStop}%
\bibitem [{\citenamefont {Ootomo}, \citenamefont {Harada},\ and\ \citenamefont {Thawonmas}(2018)}]{Ootomo2018}%
  \BibitemOpen
  \bibfield  {author} {\bibinfo {author} {\bibfnamefont {S.}~\bibnamefont {Ootomo}}, \bibinfo {author} {\bibfnamefont {T.}~\bibnamefont {Harada}}, \ and\ \bibinfo {author} {\bibfnamefont {R.}~\bibnamefont {Thawonmas}},\ }\bibfield  {title} {\enquote {\bibinfo {title} {Proposal of optimization method using common parts information and virtual parent in simultaneous design optimization problem of multiple car structures},}\ }\href {\doibase 10.11394/tjpnsec.9.41} {\bibfield  {journal} {\bibinfo  {journal} {Transaction of the Japanese Society for Evolutionary Computation}\ }\textbf {\bibinfo {volume} {9}},\ \bibinfo {pages} {41--52} (\bibinfo {year} {2018})}\BibitemShut {NoStop}%
\bibitem [{\citenamefont {Oyama}(2018)}]{Oyama2018}%
  \BibitemOpen
  \bibfield  {author} {\bibinfo {author} {\bibfnamefont {A.}~\bibnamefont {Oyama}},\ }\bibfield  {title} {\enquote {\bibinfo {title} {Report of evolutionary computation competition 2017},}\ }\href {\doibase 10.11394/tjpnsec.9.86} {\bibfield  {journal} {\bibinfo  {journal} {Transaction of the Japanese Society for Evolutionary Computation}\ }\textbf {\bibinfo {volume} {9}},\ \bibinfo {pages} {86--92} (\bibinfo {year} {2018})}\BibitemShut {NoStop}%
\bibitem [{\citenamefont {Daulton}\ \emph {et~al.}(2022)\citenamefont {Daulton}, \citenamefont {Eriksson}, \citenamefont {Balandat},\ and\ \citenamefont {Bakshy}}]{Daulton2022}%
  \BibitemOpen
  \bibfield  {author} {\bibinfo {author} {\bibfnamefont {S.}~\bibnamefont {Daulton}}, \bibinfo {author} {\bibfnamefont {D.}~\bibnamefont {Eriksson}}, \bibinfo {author} {\bibfnamefont {M.}~\bibnamefont {Balandat}}, \ and\ \bibinfo {author} {\bibfnamefont {E.}~\bibnamefont {Bakshy}},\ }\bibfield  {title} {\enquote {\bibinfo {title} {Multi-objective {Bayesian} optimization over high-dimensional search spaces},}\ }\href@noop {} {\bibfield  {journal} {\bibinfo  {journal} {Proceedings of Machine Learning Research}\ }\textbf {\bibinfo {volume} {180}},\ \bibinfo {pages} {507--517} (\bibinfo {year} {2022})}\BibitemShut {NoStop}%
\bibitem [{\citenamefont {Wilson}\ \emph {et~al.}(2021{\natexlab{a}})\citenamefont {Wilson}, \citenamefont {Kudyshev}, \citenamefont {Kildishev}, \citenamefont {Kais}, \citenamefont {Shalaev},\ and\ \citenamefont {Boltasseva}}]{Wilson:21}%
  \BibitemOpen
  \bibfield  {author} {\bibinfo {author} {\bibfnamefont {B.~A.}\ \bibnamefont {Wilson}}, \bibinfo {author} {\bibfnamefont {Z.~A.}\ \bibnamefont {Kudyshev}}, \bibinfo {author} {\bibfnamefont {A.~V.}\ \bibnamefont {Kildishev}}, \bibinfo {author} {\bibfnamefont {S.}~\bibnamefont {Kais}}, \bibinfo {author} {\bibfnamefont {V.~M.}\ \bibnamefont {Shalaev}}, \ and\ \bibinfo {author} {\bibfnamefont {A.}~\bibnamefont {Boltasseva}},\ }\bibfield  {title} {\enquote {\bibinfo {title} {Metasurface design optimization via {D-Wave} based sampling},}\ }in\ \href {\doibase 10.1364/CLEO_QELS.2021.FTh2M.2} {\emph {\bibinfo {booktitle} {Conference on Lasers and Electro-Optics}}}\ (\bibinfo  {publisher} {Optica Publishing Group},\ \bibinfo {year} {2021})\ p.\ \bibinfo {pages} {FTh2M.2}\BibitemShut {NoStop}%
\bibitem [{\citenamefont {Wilson}\ \emph {et~al.}(2021{\natexlab{b}})\citenamefont {Wilson}, \citenamefont {Kudyshev}, \citenamefont {Kildishev}, \citenamefont {Kais}, \citenamefont {Shalaev},\ and\ \citenamefont {Boltasseva}}]{10.1063/5.0060481}%
  \BibitemOpen
  \bibfield  {author} {\bibinfo {author} {\bibfnamefont {B.~A.}\ \bibnamefont {Wilson}}, \bibinfo {author} {\bibfnamefont {Z.~A.}\ \bibnamefont {Kudyshev}}, \bibinfo {author} {\bibfnamefont {A.~V.}\ \bibnamefont {Kildishev}}, \bibinfo {author} {\bibfnamefont {S.}~\bibnamefont {Kais}}, \bibinfo {author} {\bibfnamefont {V.~M.}\ \bibnamefont {Shalaev}}, \ and\ \bibinfo {author} {\bibfnamefont {A.}~\bibnamefont {Boltasseva}},\ }\bibfield  {title} {\enquote {\bibinfo {title} {{Machine learning framework for quantum sampling of highly constrained, continuous optimization problems}},}\ }\href {\doibase 10.1063/5.0060481} {\bibfield  {journal} {\bibinfo  {journal} {Applied Physics Reviews}\ }\textbf {\bibinfo {volume} {8}},\ \bibinfo {pages} {041418} (\bibinfo {year} {2021}{\natexlab{b}})}\BibitemShut {NoStop}%
\bibitem [{\citenamefont {Mao}\ \emph {et~al.}(2023)\citenamefont {Mao}, \citenamefont {Matsuda}, \citenamefont {Tamura},\ and\ \citenamefont {Tsuda}}]{D3DD00047H}%
  \BibitemOpen
  \bibfield  {author} {\bibinfo {author} {\bibfnamefont {Z.}~\bibnamefont {Mao}}, \bibinfo {author} {\bibfnamefont {Y.}~\bibnamefont {Matsuda}}, \bibinfo {author} {\bibfnamefont {R.}~\bibnamefont {Tamura}}, \ and\ \bibinfo {author} {\bibfnamefont {K.}~\bibnamefont {Tsuda}},\ }\bibfield  {title} {\enquote {\bibinfo {title} {Chemical design with {GPU}-based {Ising} machines},}\ }\href {\doibase 10.1039/D3DD00047H} {\bibfield  {journal} {\bibinfo  {journal} {Digital Discovery}\ }\textbf {\bibinfo {volume} {2}},\ \bibinfo {pages} {1098--1103} (\bibinfo {year} {2023})}\BibitemShut {NoStop}%
\bibitem [{\citenamefont {Weininger}(1988)}]{Weininger:1988aa}%
  \BibitemOpen
  \bibfield  {author} {\bibinfo {author} {\bibfnamefont {D.}~\bibnamefont {Weininger}},\ }\bibfield  {title} {\enquote {\bibinfo {title} {{SMILES}, a chemical language and information system. 1. introduction to methodology and encoding rules},}\ }\href {\doibase 10.1021/ci00057a005} {\bibfield  {journal} {\bibinfo  {journal} {Journal of Chemical Information and Computer Sciences}\ }\textbf {\bibinfo {volume} {28}},\ \bibinfo {pages} {31--36} (\bibinfo {year} {1988})}\BibitemShut {NoStop}%
\bibitem [{\citenamefont {Jin}, \citenamefont {Barzilay},\ and\ \citenamefont {Jaakkola}(2018)}]{pmlr-v80-jin18a}%
  \BibitemOpen
  \bibfield  {author} {\bibinfo {author} {\bibfnamefont {W.}~\bibnamefont {Jin}}, \bibinfo {author} {\bibfnamefont {R.}~\bibnamefont {Barzilay}}, \ and\ \bibinfo {author} {\bibfnamefont {T.}~\bibnamefont {Jaakkola}},\ }\bibfield  {title} {\enquote {\bibinfo {title} {Junction tree variational autoencoder for molecular graph generation},}\ }in\ \href {https://proceedings.mlr.press/v80/jin18a.html} {\emph {\bibinfo {booktitle} {Proceedings of the 35th International Conference on Machine Learning}}},\ \bibinfo {series} {Proceedings of Machine Learning Research}, Vol.~\bibinfo {volume} {80},\ \bibinfo {editor} {edited by\ \bibinfo {editor} {\bibfnamefont {J.}~\bibnamefont {Dy}}\ and\ \bibinfo {editor} {\bibfnamefont {A.}~\bibnamefont {Krause}}}\ (\bibinfo  {publisher} {PMLR},\ \bibinfo {year} {2018})\ pp.\ \bibinfo {pages} {2323--2332}\BibitemShut {NoStop}%
\bibitem [{\citenamefont {Jang}, \citenamefont {Gu},\ and\ \citenamefont {Poole}(2017)}]{jang2017categoricalreparameterizationgumbelsoftmax}%
  \BibitemOpen
  \bibfield  {author} {\bibinfo {author} {\bibfnamefont {E.}~\bibnamefont {Jang}}, \bibinfo {author} {\bibfnamefont {S.}~\bibnamefont {Gu}}, \ and\ \bibinfo {author} {\bibfnamefont {B.}~\bibnamefont {Poole}},\ }\href {https://arxiv.org/abs/1611.01144} {\enquote {\bibinfo {title} {Categorical reparameterization with {Gumbel}-softmax},}\ } (\bibinfo {year} {2017}),\ \Eprint {http://arxiv.org/abs/1611.01144} {arXiv:1611.01144 [stat.ML]} \BibitemShut {NoStop}%
\bibitem [{\citenamefont {Baynazarov}\ and\ \citenamefont {Piontkovskaya}(2019)}]{10.1007/978-3-030-34518-1_10}%
  \BibitemOpen
  \bibfield  {author} {\bibinfo {author} {\bibfnamefont {R.}~\bibnamefont {Baynazarov}}\ and\ \bibinfo {author} {\bibfnamefont {I.}~\bibnamefont {Piontkovskaya}},\ }\bibfield  {title} {\enquote {\bibinfo {title} {Binary autoencoder for text modeling},}\ }in\ \href@noop {} {\emph {\bibinfo {booktitle} {Artificial Intelligence and Natural Language}}},\ \bibinfo {editor} {edited by\ \bibinfo {editor} {\bibfnamefont {D.}~\bibnamefont {Ustalov}}, \bibinfo {editor} {\bibfnamefont {A.}~\bibnamefont {Filchenkov}}, \ and\ \bibinfo {editor} {\bibfnamefont {L.}~\bibnamefont {Pivovarova}}}\ (\bibinfo  {publisher} {Springer International Publishing},\ \bibinfo {address} {Cham},\ \bibinfo {year} {2019})\ pp.\ \bibinfo {pages} {139--150}\BibitemShut {NoStop}%
\bibitem [{\citenamefont {Gao}\ \emph {et~al.}(2023)\citenamefont {Gao}, \citenamefont {Jones}, \citenamefont {Kobayashi}, \citenamefont {Sugawara}, \citenamefont {Yamashita}, \citenamefont {Kawaguchi}, \citenamefont {Tanaka},\ and\ \citenamefont {Yamamoto}}]{doi:10.34133/icomputing.0037}%
  \BibitemOpen
  \bibfield  {author} {\bibinfo {author} {\bibfnamefont {Q.}~\bibnamefont {Gao}}, \bibinfo {author} {\bibfnamefont {G.~O.}\ \bibnamefont {Jones}}, \bibinfo {author} {\bibfnamefont {T.}~\bibnamefont {Kobayashi}}, \bibinfo {author} {\bibfnamefont {M.}~\bibnamefont {Sugawara}}, \bibinfo {author} {\bibfnamefont {H.}~\bibnamefont {Yamashita}}, \bibinfo {author} {\bibfnamefont {H.}~\bibnamefont {Kawaguchi}}, \bibinfo {author} {\bibfnamefont {S.}~\bibnamefont {Tanaka}}, \ and\ \bibinfo {author} {\bibfnamefont {N.}~\bibnamefont {Yamamoto}},\ }\bibfield  {title} {\enquote {\bibinfo {title} {Quantum-classical computational molecular design of deuterated high-efficiency {OLED} emitters},}\ }\href {\doibase 10.34133/icomputing.0037} {\bibfield  {journal} {\bibinfo  {journal} {Intelligent Computing}\ }\textbf {\bibinfo {volume} {2}},\ \bibinfo {pages} {0037} (\bibinfo {year} {2023})}\BibitemShut {NoStop}%
\bibitem [{IBM()}]{IBMQ}%
  \BibitemOpen
  \href@noop {} {\enquote {\bibinfo {title} {{IBMQ}},}\ }\bibinfo {note} {\url{https://www.ibm.com/quantum/technology}}\BibitemShut {NoStop}%
\bibitem [{\citenamefont {Schneider}(2018)}]{Schneider:2018aa}%
  \BibitemOpen
  \bibfield  {author} {\bibinfo {author} {\bibfnamefont {G.}~\bibnamefont {Schneider}},\ }\bibfield  {title} {\enquote {\bibinfo {title} {Automating drug discovery},}\ }\href {\doibase 10.1038/nrd.2017.232} {\bibfield  {journal} {\bibinfo  {journal} {Nature Reviews Drug Discovery}\ }\textbf {\bibinfo {volume} {17}},\ \bibinfo {pages} {97--113} (\bibinfo {year} {2018})}\BibitemShut {NoStop}%
\bibitem [{\citenamefont {Burger}\ \emph {et~al.}(2020)\citenamefont {Burger}, \citenamefont {Maffettone}, \citenamefont {Gusev}, \citenamefont {Aitchison}, \citenamefont {Bai}, \citenamefont {Wang}, \citenamefont {Li}, \citenamefont {Alston}, \citenamefont {Li}, \citenamefont {Clowes}, \citenamefont {Rankin}, \citenamefont {Harris}, \citenamefont {Sprick},\ and\ \citenamefont {Cooper}}]{Burger:2020aa}%
  \BibitemOpen
  \bibfield  {author} {\bibinfo {author} {\bibfnamefont {B.}~\bibnamefont {Burger}}, \bibinfo {author} {\bibfnamefont {P.~M.}\ \bibnamefont {Maffettone}}, \bibinfo {author} {\bibfnamefont {V.~V.}\ \bibnamefont {Gusev}}, \bibinfo {author} {\bibfnamefont {C.~M.}\ \bibnamefont {Aitchison}}, \bibinfo {author} {\bibfnamefont {Y.}~\bibnamefont {Bai}}, \bibinfo {author} {\bibfnamefont {X.}~\bibnamefont {Wang}}, \bibinfo {author} {\bibfnamefont {X.}~\bibnamefont {Li}}, \bibinfo {author} {\bibfnamefont {B.~M.}\ \bibnamefont {Alston}}, \bibinfo {author} {\bibfnamefont {B.}~\bibnamefont {Li}}, \bibinfo {author} {\bibfnamefont {R.}~\bibnamefont {Clowes}}, \bibinfo {author} {\bibfnamefont {N.}~\bibnamefont {Rankin}}, \bibinfo {author} {\bibfnamefont {B.}~\bibnamefont {Harris}}, \bibinfo {author} {\bibfnamefont {R.~S.}\ \bibnamefont {Sprick}}, \ and\ \bibinfo {author} {\bibfnamefont {A.~I.}\ \bibnamefont {Cooper}},\ }\bibfield  {title} {\enquote {\bibinfo {title} {A mobile robotic chemist},}\ }\href {\doibase
  10.1038/s41586-020-2442-2} {\bibfield  {journal} {\bibinfo  {journal} {Nature}\ }\textbf {\bibinfo {volume} {583}},\ \bibinfo {pages} {237--241} (\bibinfo {year} {2020})}\BibitemShut {NoStop}%
\bibitem [{\citenamefont {MacLeod}\ \emph {et~al.}(2020)\citenamefont {MacLeod}, \citenamefont {Parlane}, \citenamefont {Morrissey}, \citenamefont {H{\"a}se}, \citenamefont {Roch}, \citenamefont {Dettelbach}, \citenamefont {Moreira}, \citenamefont {Yunker}, \citenamefont {Rooney}, \citenamefont {Deeth}, \citenamefont {Lai}, \citenamefont {Ng}, \citenamefont {Situ}, \citenamefont {Zhang}, \citenamefont {Elliott}, \citenamefont {Haley}, \citenamefont {Dvorak}, \citenamefont {Aspuru-Guzik}, \citenamefont {Hein},\ and\ \citenamefont {Berlinguette}}]{doi:10.1126/sciadv.aaz8867}%
  \BibitemOpen
  \bibfield  {author} {\bibinfo {author} {\bibfnamefont {B.~P.}\ \bibnamefont {MacLeod}}, \bibinfo {author} {\bibfnamefont {F.~G.~L.}\ \bibnamefont {Parlane}}, \bibinfo {author} {\bibfnamefont {T.~D.}\ \bibnamefont {Morrissey}}, \bibinfo {author} {\bibfnamefont {F.}~\bibnamefont {H{\"a}se}}, \bibinfo {author} {\bibfnamefont {L.~M.}\ \bibnamefont {Roch}}, \bibinfo {author} {\bibfnamefont {K.~E.}\ \bibnamefont {Dettelbach}}, \bibinfo {author} {\bibfnamefont {R.}~\bibnamefont {Moreira}}, \bibinfo {author} {\bibfnamefont {L.~P.~E.}\ \bibnamefont {Yunker}}, \bibinfo {author} {\bibfnamefont {M.~B.}\ \bibnamefont {Rooney}}, \bibinfo {author} {\bibfnamefont {J.~R.}\ \bibnamefont {Deeth}}, \bibinfo {author} {\bibfnamefont {V.}~\bibnamefont {Lai}}, \bibinfo {author} {\bibfnamefont {G.~J.}\ \bibnamefont {Ng}}, \bibinfo {author} {\bibfnamefont {H.}~\bibnamefont {Situ}}, \bibinfo {author} {\bibfnamefont {R.~H.}\ \bibnamefont {Zhang}}, \bibinfo {author} {\bibfnamefont {M.~S.}\ \bibnamefont {Elliott}}, \bibinfo {author}
  {\bibfnamefont {T.~H.}\ \bibnamefont {Haley}}, \bibinfo {author} {\bibfnamefont {D.~J.}\ \bibnamefont {Dvorak}}, \bibinfo {author} {\bibfnamefont {A.}~\bibnamefont {Aspuru-Guzik}}, \bibinfo {author} {\bibfnamefont {J.~E.}\ \bibnamefont {Hein}}, \ and\ \bibinfo {author} {\bibfnamefont {C.~P.}\ \bibnamefont {Berlinguette}},\ }\bibfield  {title} {\enquote {\bibinfo {title} {Self-driving laboratory for accelerated discovery of thin-film materials},}\ }\href {\doibase 10.1126/sciadv.aaz8867} {\bibfield  {journal} {\bibinfo  {journal} {Science Advances}\ }\textbf {\bibinfo {volume} {6}},\ \bibinfo {pages} {eaaz8867} (\bibinfo {year} {2020})}\BibitemShut {NoStop}%
\bibitem [{\citenamefont {Tamura}, \citenamefont {Tsuda},\ and\ \citenamefont {Matsuda}(2023)}]{doi:10.1080/27660400.2023.2232297}%
  \BibitemOpen
  \bibfield  {author} {\bibinfo {author} {\bibfnamefont {R.}~\bibnamefont {Tamura}}, \bibinfo {author} {\bibfnamefont {K.}~\bibnamefont {Tsuda}}, \ and\ \bibinfo {author} {\bibfnamefont {S.}~\bibnamefont {Matsuda}},\ }\bibfield  {title} {\enquote {\bibinfo {title} {{NIMS-OS}: an automation software to implement a closed loop between artificial intelligence and robotic experiments in materials science},}\ }\href {\doibase 10.1080/27660400.2023.2232297} {\bibfield  {journal} {\bibinfo  {journal} {Science and Technology of Advanced Materials: Methods}\ }\textbf {\bibinfo {volume} {3}},\ \bibinfo {pages} {2232297} (\bibinfo {year} {2023})}\BibitemShut {NoStop}%
\bibitem [{\citenamefont {Tom}\ \emph {et~al.}(2024)\citenamefont {Tom}, \citenamefont {Schmid}, \citenamefont {Baird}, \citenamefont {Cao}, \citenamefont {Darvish}, \citenamefont {Hao}, \citenamefont {Lo}, \citenamefont {Pablo-García}, \citenamefont {Rajaonson}, \citenamefont {Skreta}, \citenamefont {Yoshikawa}, \citenamefont {Corapi}, \citenamefont {Akkoc}, \citenamefont {Strieth-Kalthoff}, \citenamefont {Seifrid},\ and\ \citenamefont {Aspuru-Guzik}}]{Tom2024}%
  \BibitemOpen
  \bibfield  {author} {\bibinfo {author} {\bibfnamefont {G.}~\bibnamefont {Tom}}, \bibinfo {author} {\bibfnamefont {S.~P.}\ \bibnamefont {Schmid}}, \bibinfo {author} {\bibfnamefont {S.~G.}\ \bibnamefont {Baird}}, \bibinfo {author} {\bibfnamefont {Y.}~\bibnamefont {Cao}}, \bibinfo {author} {\bibfnamefont {K.}~\bibnamefont {Darvish}}, \bibinfo {author} {\bibfnamefont {H.}~\bibnamefont {Hao}}, \bibinfo {author} {\bibfnamefont {S.}~\bibnamefont {Lo}}, \bibinfo {author} {\bibfnamefont {S.}~\bibnamefont {Pablo-García}}, \bibinfo {author} {\bibfnamefont {E.~M.}\ \bibnamefont {Rajaonson}}, \bibinfo {author} {\bibfnamefont {M.}~\bibnamefont {Skreta}}, \bibinfo {author} {\bibfnamefont {N.}~\bibnamefont {Yoshikawa}}, \bibinfo {author} {\bibfnamefont {S.}~\bibnamefont {Corapi}}, \bibinfo {author} {\bibfnamefont {G.~D.}\ \bibnamefont {Akkoc}}, \bibinfo {author} {\bibfnamefont {F.}~\bibnamefont {Strieth-Kalthoff}}, \bibinfo {author} {\bibfnamefont {M.}~\bibnamefont {Seifrid}}, \ and\ \bibinfo {author} {\bibfnamefont
  {A.}~\bibnamefont {Aspuru-Guzik}},\ }\bibfield  {title} {\enquote {\bibinfo {title} {Self-driving laboratories for chemistry and materials science},}\ }\href {\doibase 10.1021/acs.chemrev.4c00055} {\bibfield  {journal} {\bibinfo  {journal} {Chemical Reviews}\ }\textbf {\bibinfo {volume} {124}},\ \bibinfo {pages} {9633–9732} (\bibinfo {year} {2024})}\BibitemShut {NoStop}%
\bibitem [{\citenamefont {Yoshikawa}\ \emph {et~al.}(2025)\citenamefont {Yoshikawa}, \citenamefont {Asano}, \citenamefont {Futaba}, \citenamefont {Harada}, \citenamefont {Hitosugi}, \citenamefont {Kanda}, \citenamefont {Matsuda}, \citenamefont {Nagata}, \citenamefont {Nagato}, \citenamefont {Naito}, \citenamefont {Natsume}, \citenamefont {Nishio}, \citenamefont {Ono}, \citenamefont {Ozaki}, \citenamefont {Shin}, \citenamefont {Shiomi}, \citenamefont {Shizume}, \citenamefont {Takahashi}, \citenamefont {Takeda}, \citenamefont {Takeuchi}, \citenamefont {Tamura}, \citenamefont {Tsuda},\ and\ \citenamefont {Ushiku}}]{Yoshikawa2025}%
  \BibitemOpen
  \bibfield  {author} {\bibinfo {author} {\bibfnamefont {N.}~\bibnamefont {Yoshikawa}}, \bibinfo {author} {\bibfnamefont {Y.}~\bibnamefont {Asano}}, \bibinfo {author} {\bibfnamefont {D.~N.}\ \bibnamefont {Futaba}}, \bibinfo {author} {\bibfnamefont {K.}~\bibnamefont {Harada}}, \bibinfo {author} {\bibfnamefont {T.}~\bibnamefont {Hitosugi}}, \bibinfo {author} {\bibfnamefont {G.~N.}\ \bibnamefont {Kanda}}, \bibinfo {author} {\bibfnamefont {S.}~\bibnamefont {Matsuda}}, \bibinfo {author} {\bibfnamefont {Y.}~\bibnamefont {Nagata}}, \bibinfo {author} {\bibfnamefont {K.}~\bibnamefont {Nagato}}, \bibinfo {author} {\bibfnamefont {M.}~\bibnamefont {Naito}}, \bibinfo {author} {\bibfnamefont {T.}~\bibnamefont {Natsume}}, \bibinfo {author} {\bibfnamefont {K.}~\bibnamefont {Nishio}}, \bibinfo {author} {\bibfnamefont {K.}~\bibnamefont {Ono}}, \bibinfo {author} {\bibfnamefont {H.}~\bibnamefont {Ozaki}}, \bibinfo {author} {\bibfnamefont {W.}~\bibnamefont {Shin}}, \bibinfo {author} {\bibfnamefont {J.}~\bibnamefont {Shiomi}},
  \bibinfo {author} {\bibfnamefont {K.}~\bibnamefont {Shizume}}, \bibinfo {author} {\bibfnamefont {K.}~\bibnamefont {Takahashi}}, \bibinfo {author} {\bibfnamefont {S.}~\bibnamefont {Takeda}}, \bibinfo {author} {\bibfnamefont {I.}~\bibnamefont {Takeuchi}}, \bibinfo {author} {\bibfnamefont {R.}~\bibnamefont {Tamura}}, \bibinfo {author} {\bibfnamefont {K.}~\bibnamefont {Tsuda}}, \ and\ \bibinfo {author} {\bibfnamefont {Y.}~\bibnamefont {Ushiku}},\ }\bibfield  {title} {\enquote {\bibinfo {title} {Self-driving laboratories in {Japan}},}\ }\href {\doibase 10.1039/d4dd00387j} {\bibfield  {journal} {\bibinfo  {journal} {Digital Discovery}\ }\textbf {\bibinfo {volume} {4}},\ \bibinfo {pages} {1384–1403} (\bibinfo {year} {2025})}\BibitemShut {NoStop}%
\bibitem [{\citenamefont {Gonzalez}\ \emph {et~al.}(2016)\citenamefont {Gonzalez}, \citenamefont {Dai}, \citenamefont {Hennig},\ and\ \citenamefont {Lawrence}}]{pmlr-v51-gonzalez16a}%
  \BibitemOpen
  \bibfield  {author} {\bibinfo {author} {\bibfnamefont {J.}~\bibnamefont {Gonzalez}}, \bibinfo {author} {\bibfnamefont {Z.}~\bibnamefont {Dai}}, \bibinfo {author} {\bibfnamefont {P.}~\bibnamefont {Hennig}}, \ and\ \bibinfo {author} {\bibfnamefont {N.}~\bibnamefont {Lawrence}},\ }\bibfield  {title} {\enquote {\bibinfo {title} {Batch {Bayesian} optimization via local penalization},}\ }in\ \href {https://proceedings.mlr.press/v51/gonzalez16a.html} {\emph {\bibinfo {booktitle} {Proceedings of the 19th International Conference on Artificial Intelligence and Statistics}}},\ \bibinfo {series} {Proceedings of Machine Learning Research}, Vol.~\bibinfo {volume} {51},\ \bibinfo {editor} {edited by\ \bibinfo {editor} {\bibfnamefont {A.}~\bibnamefont {Gretton}}\ and\ \bibinfo {editor} {\bibfnamefont {C.~C.}\ \bibnamefont {Robert}}}\ (\bibinfo  {publisher} {PMLR},\ \bibinfo {address} {Cadiz, Spain},\ \bibinfo {year} {2016})\ pp.\ \bibinfo {pages} {648--657}\BibitemShut {NoStop}%
\bibitem [{\citenamefont {Azimi}, \citenamefont {Fern},\ and\ \citenamefont {Fern}(2010)}]{NIPS2010_e702e51d}%
  \BibitemOpen
  \bibfield  {author} {\bibinfo {author} {\bibfnamefont {J.}~\bibnamefont {Azimi}}, \bibinfo {author} {\bibfnamefont {A.}~\bibnamefont {Fern}}, \ and\ \bibinfo {author} {\bibfnamefont {X.}~\bibnamefont {Fern}},\ }\bibfield  {title} {\enquote {\bibinfo {title} {Batch {Bayesian} optimization via simulation matching},}\ }in\ \href {https://proceedings.neurips.cc/paper_files/paper/2010/file/e702e51da2c0f5be4dd354bb3e295d37-Paper.pdf} {\emph {\bibinfo {booktitle} {Advances in Neural Information Processing Systems}}},\ Vol.~\bibinfo {volume} {23},\ \bibinfo {editor} {edited by\ \bibinfo {editor} {\bibfnamefont {J.}~\bibnamefont {Lafferty}}, \bibinfo {editor} {\bibfnamefont {C.}~\bibnamefont {Williams}}, \bibinfo {editor} {\bibfnamefont {J.}~\bibnamefont {Shawe-Taylor}}, \bibinfo {editor} {\bibfnamefont {R.}~\bibnamefont {Zemel}}, \ and\ \bibinfo {editor} {\bibfnamefont {A.}~\bibnamefont {Culotta}}}\ (\bibinfo  {publisher} {Curran Associates, Inc.},\ \bibinfo {year} {2010})\BibitemShut {NoStop}%
\bibitem [{\citenamefont {Dai}\ \emph {et~al.}(2023)\citenamefont {Dai}, \citenamefont {Nguyen}, \citenamefont {Tay}, \citenamefont {Urano}, \citenamefont {Leong}, \citenamefont {Low},\ and\ \citenamefont {Jaillet}}]{NEURIPS2023_727a5a5c}%
  \BibitemOpen
  \bibfield  {author} {\bibinfo {author} {\bibfnamefont {Z.}~\bibnamefont {Dai}}, \bibinfo {author} {\bibfnamefont {Q.~P.}\ \bibnamefont {Nguyen}}, \bibinfo {author} {\bibfnamefont {S.}~\bibnamefont {Tay}}, \bibinfo {author} {\bibfnamefont {D.}~\bibnamefont {Urano}}, \bibinfo {author} {\bibfnamefont {R.}~\bibnamefont {Leong}}, \bibinfo {author} {\bibfnamefont {B.~K.~H.}\ \bibnamefont {Low}}, \ and\ \bibinfo {author} {\bibfnamefont {P.}~\bibnamefont {Jaillet}},\ }\bibfield  {title} {\enquote {\bibinfo {title} {Batch {Bayesian} optimization for replicable experimental design},}\ }in\ \href {https://proceedings.neurips.cc/paper_files/paper/2023/file/727a5a5c77be15d053b47b7c391800c2-Paper-Conference.pdf} {\emph {\bibinfo {booktitle} {Advances in Neural Information Processing Systems}}},\ Vol.~\bibinfo {volume} {36},\ \bibinfo {editor} {edited by\ \bibinfo {editor} {\bibfnamefont {A.}~\bibnamefont {Oh}}, \bibinfo {editor} {\bibfnamefont {T.}~\bibnamefont {Naumann}}, \bibinfo {editor} {\bibfnamefont
  {A.}~\bibnamefont {Globerson}}, \bibinfo {editor} {\bibfnamefont {K.}~\bibnamefont {Saenko}}, \bibinfo {editor} {\bibfnamefont {M.}~\bibnamefont {Hardt}}, \ and\ \bibinfo {editor} {\bibfnamefont {S.}~\bibnamefont {Levine}}}\ (\bibinfo  {publisher} {Curran Associates, Inc.},\ \bibinfo {year} {2023})\ pp.\ \bibinfo {pages} {36476--36506}\BibitemShut {NoStop}%
\bibitem [{\citenamefont {Glos}, \citenamefont {Krawiec},\ and\ \citenamefont {Zimbor{\'a}s}(2022)}]{Glos:2022aa}%
  \BibitemOpen
  \bibfield  {author} {\bibinfo {author} {\bibfnamefont {A.}~\bibnamefont {Glos}}, \bibinfo {author} {\bibfnamefont {A.}~\bibnamefont {Krawiec}}, \ and\ \bibinfo {author} {\bibfnamefont {Z.}~\bibnamefont {Zimbor{\'a}s}},\ }\bibfield  {title} {\enquote {\bibinfo {title} {Space-efficient binary optimization for variational quantum computing},}\ }\href {\doibase 10.1038/s41534-022-00546-y} {\bibfield  {journal} {\bibinfo  {journal} {npj Quantum Information}\ }\textbf {\bibinfo {volume} {8}},\ \bibinfo {pages} {39} (\bibinfo {year} {2022})}\BibitemShut {NoStop}%
\bibitem [{\citenamefont {Domino}\ \emph {et~al.}(2022)\citenamefont {Domino}, \citenamefont {Kundu}, \citenamefont {Salehi},\ and\ \citenamefont {Krawiec}}]{Domino:2022aa}%
  \BibitemOpen
  \bibfield  {author} {\bibinfo {author} {\bibfnamefont {K.}~\bibnamefont {Domino}}, \bibinfo {author} {\bibfnamefont {A.}~\bibnamefont {Kundu}}, \bibinfo {author} {\bibfnamefont {{\"O}.}~\bibnamefont {Salehi}}, \ and\ \bibinfo {author} {\bibfnamefont {K.}~\bibnamefont {Krawiec}},\ }\bibfield  {title} {\enquote {\bibinfo {title} {Quadratic and higher-order unconstrained binary optimization of railway rescheduling for quantum computing},}\ }\href {\doibase 10.1007/s11128-022-03670-y} {\bibfield  {journal} {\bibinfo  {journal} {Quantum Information Processing}\ }\textbf {\bibinfo {volume} {21}},\ \bibinfo {pages} {337} (\bibinfo {year} {2022})}\BibitemShut {NoStop}%
\end{thebibliography}%

\end{document}